\documentclass[orange]{TesisAndes}
 \usepackage[latin1]{inputenc}
 \usepackage{epsfig}
 \usepackage{amsmath}
 \usepackage{amssymb,color,graphpap}
 \usepackage{ae,graphics,enumerate}
 \title{CLASSIFICATION OF QUANTUM SYMMETRIC NONZERO-SUM 2X2 GAMES IN THE EISERT SCHEME}
 \author{Álvaro Francisco Huertas Rosero}
 \advisor{Dr. Alonso Botero}
 \degreeyear{2004}
 \degreemonth{February}
 \degreename{Magister in Physical Sciences}
\begin{document}
 \begin{frontmatter}
 \begin{quote}
  \textit{\large
 "Now, there is a law written in the darkest of the Books of Life, and it is this: If you look at a thing nine hundred and ninety-nine times, you are perfectly safe; if you look at it the thousandth time, you are in frightful danger of seeing it for the first time."}\\
  \begin{flushright}{\large Gilbert Keith Chesterton}\\
   in ``The Napoleon of Notting Hill''(1906)\end{flushright}
 \end{quote}
 \newpage
   \begin{quote}
    \makeatletter
     \addcontentsline{toc}{chapter}{Aknowledgements}
    \makeatother
    \centerline{\Large AKNOWLEDGEMENTS}\vspace{1cm}

This was a very rewarding work by itself, but I certainly have to thank some people that made the experience even more interesting and pleasant.

First, I would like to express my gratitude for the great amount (and variety) of knowledge I received from my advisor, Alonso Botero, as well as for his auspicious and patient orientation.

Special thanks to Víctor Tapia, for his valuable advice in the writting and general presentation of the document.

Thanks also for Liliana Martín, who, besides of her pleasant company, gave me necessary opinions from a different point of view about some aspects of the work and the document .

And last but not least, I am indebted to the Department of Physics of Universidad de los Andes, which under the efficacious direction of Dr. Bernardo Gómez has been a wonderful place to develop my work. 

   \end{quote}
 \cleardoublepage
 \begin{thesisabstract}
    
\begin{center}
{\large CLASSIFICATION OF QUANTUM SYMMETRIC NONZERO-SUM 2X2 GAMES IN THE EISERT SCHEME}\vspace{1cm}\\
Álvaro Francisco Huertas Rosero\vspace{0.5cm}\\
Universidad de los Andes\vspace{1cm}\\
{\large \textbf{Abstract}}\vspace{1cm}
\end{center}
  A quantum game in the Eisert scheme is defined by the payoff matrix, plus some quantum entanglement parameters.   In the symmetric nonzero-sum 2x2 games, the relevant features of the game are given by two parameters in the payoff matrix, and only one extra entanglement parameter is introduced by quantizing it in the Eisert scheme.

  The criteria adopted in this work to classify quantum games are the amount and characteristics of Nash equilibria and Pareto-optimal positions.  A new classification based on them is developed for symmetric nonzero-sum classical 2x2 games, as well as classifications for quantum games with different restricted subsets of the total strategy set.   Finally, a classification is presented taking the whole set of strategies into account, both unitary strategies and nonunitary strategies studied as convex mixures of unitary strategies.

  The classification reproduces features which have been previously found in other works, like appearance of multiple equilibria, changes in the character of equilibria, and entanglement regime transitions.

 \end{thesisabstract}
 \end{frontmatter}
 \bibliographystyle{plain}
 \chapter{ABOUT THIS WORK}
  This is a work on a very specific subject in the newborn field of Quantum Game Theory, a field that is familiar only to a restricted group of people (at least, at the date of the publication of this dissertation).

The work is presented here mainly for physicists, specially those working on Quantum Mechanics.  Therefore, the physical concepts will be explained to a minor extent than those of the mathematical Theory of Games.

The dissertation is divided in 10 chapters (this is chapter \thechapter), for which a little reading guide is presented here:
\begin{enumerate}
 \setcounter{enumi}{1}
 \item Why study Quantum Games?\\
  In this chapter, an introduction to some of the concepts of both Game Theory and Quantum Information are given with an emphasis on their potential practical application. 
 \item Classical Games, Quantum Games\\
  An introduction to the relevant concepts from ``Classical'' game theory, and a description of the Eisert scheme for defining Quantum Games.
 \item What we know so far\\
  In this chapter a brief description is presented of the relevant work developed on the subject of Quantum games, as well as the questions posed by earlier studys, with an emphasis on the questions that have inspired this work
 \item Classification of Symmetric 2x2 Classical Games\\
  A classification is developed for this kind of games, that will be the basis to devise a similar classification of quantum games.   The results obtained can have some interest in some developing fields of Game Theory as well as in Quantum Game Theory.
 \item The Quantum Game with Deterministic Strategies\\
  The \textbf{main results} of the entire work can be found in this chapter.
 \item A Geometrical approach to Unitary Strategies\\
  In this chapter a methodology for studying a larger set of strategies is developed.  It can be interesting to the reader that is interested in the transformations of qubits, but is not essential to the work, except as a basis for further results.
 \item Critical Responses in Quantum 2x2 Games\\
  Here a methodology is presented to characterize games taking  all the possible strategies into account, and some further partial results are presented.
 \item Exploration of the Strategy Space\\
  Here a final exploration of the characteristics of the quantum games is presented, that completes the sought results.
 \item Conclusions and Perspectives
\end{enumerate}
  To guide the reader, here is also a little map:
  \footnote{The map shows the main concepts treated in this work, with their relations as arrows.  The numbers within the parentheses are the chapters where they are developed.}  
  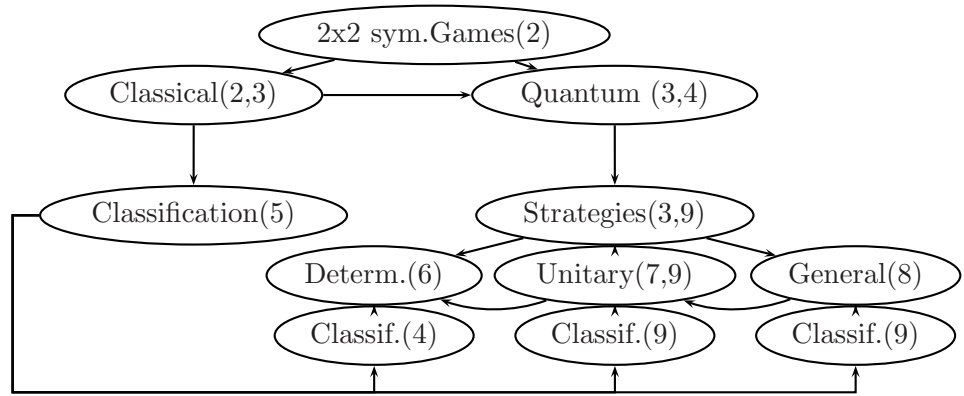
\begin{figure}[bht]
   \begin{center}
   \psset{unit=0.8cm}
   \begin{pspicture}(-4,-1)(12,5)
    \rput[cc](3,5){\ovalnode{Games}{\small 2x2 sym.Games(2)}}
    \rput[cc](-1,4){\ovalnode{Cla}{\small Classical(2,3)}}
    \rput[cc](6,4){\ovalnode{Quant}{\small Quantum (3,4)}}
    \rput[cc](-1,2){\ovalnode{ClassC}{\small Classification(5)}}
    \rput[cc](6,2){\ovalnode{Strategies}{\small Strategies(3,9)}}
    \rput[cc](2,1){\ovalnode{Determ}{\small Determ.(6)}}
    \rput[cc](6,1){\ovalnode{Unit}{\small Unitary(7,9)}}
    \rput[cc](10,1){\ovalnode{General}{\small General(8)}}
    \rput[cc](2,0){\ovalnode{Class1}{\small Classif.(4)}}
    \rput[cc](6,0){\ovalnode{Class2}{\small Classif.(9)}}
    \rput[cc](10,0){\ovalnode{Class3}{\small Classif.(9)}}
    \ncline{->}{Games}{Cla}
    \ncline{->}{Games}{Quant}
    \ncline{->}{Cla}{ClassC}
    \ncline{->}{Cla}{Quant}
    \ncline{->}{Quant}{Strategies}
    \ncline{->}{Strategies}{Determ}
    \ncline{->}{Strategies}{Unit}
    \ncline{->}{Strategies}{General}
    \ncline{->}{Determ}{Class1}
    \ncline{->}{Unit}{Class2}
    \ncline{->}{General}{Class3}
    \ncangles[angleA=180,angleB=-90]{->}{ClassC}{Class1}
    \ncangles[angleA=180,angleB=-90]{->}{ClassC}{Class2}
    \ncangles[angleA=180,angleB=-90]{->}{ClassC}{Class3}
    \ncarc[arcangle=20]{->}{General}{Unit}
    \ncarc[arcangle=20]{->}{Unit}{Determ}
   \end{pspicture}
    \end{center}
   \caption{A road map of the work}
  \end{figure}

 \chapter{WHY STUDY QUANTUM GAMES?}
   \section{WHY STUDY GAMES}
   Games are situations where several deciding agents get a certain individual payoff according to what \textit{all of them} have decided.  A certain rationality is assumed from the agents that forces all of them to choose with a payoff maximizing criterion.

   There is a number of situations in life where the best thing to do, or what can be expected to happen,  can be computed by game theoretical tools, and some of these situations have been portrayed in the movies and TV.   An example:\\
  \textit{In the famous movie ``\textbf{Rebel Without a Cause}'', a simple game is stated, called ``\textbf{The Chicken Game}''.   Two rebel young man drive on a highway in opposite directions, each approaching the other at high speed. If one of them get scared and avoid the collision, he will lose control of his car and be ridiculed by his friends.  Obviously, the other driver will gain a great prestige among these risk-loving young men. But, on the other hand, if neither driver turns to avoid the collision right on time, both will probably die or be severely hurt.}\cite{GamesTV}

   We can ask several questions about this game... Is there a ``winning'' strategy?  How can one predict which player is going to win?  Is it more likely to have a happy ending or a tragic one?   Game Theory is a systematic attempt to answer this kinds of questions.

   Game Theory is, indeed, part of the heart of what is called by Aumann ``rational side of social sciences'' \cite{RationalSocial}.  Any social situation consists in a number of deciding agents whose decisions affect the whole group.  Game theory is, then, very useful to study the behaviour of extremely complex systems in a complex environment (individuals in a society), provided that there is some way of defining the payoff to ensure rationality.

   Evolution \cite{Evolution}, population dynamics \cite{Population1} and a variety of biological phenomena\cite{Animals} have been studied with game models, obtaining in this way interesting predictions and suggestive interpretations.  It is possible, for example, to give a definition for \textit{altruism} as a way to perceive the payoff in a social situation \cite{Altruism}.

   Game theory has even been proposed for modeling complex physical phenomena such as decoherence and irreversibility \cite{Parrondo}, \cite{Decoherence}.

  \subsection{RATIONALITY AND UNILATERAL OPTIMIZATION}
   The concept of instrumental rationality adopted in Game Theory is sometimes problematic when applied to human individuals \cite{Rational} but it can be applied comfortably to other kind of agents.  It consists simply in the tendency to maximize a certain kind of \textbf{utility}.  Here is an example given by Rapoport in \cite{Utility}:
   \begin{quote}
   ``The utility scale, as it has been defined, implies that of two \textit{risky}\footnote{The italics were put by the author} outcomes, the one with the greater expected utility is always preferred.  A risky outcome can be thought of essentially as a lottery ticket entitling the owner to any several prizes depending on the outcome of a chance event. Such a lottery ticket carries a list of the prizes with the probability attached to each, namely the probability of the event that entitles the holder of the ticket to the prize in question.''
   \end{quote}
   \begin{definition}
    \textbf{Expected Payoff} $(\hat{\$})$:= The average payoff obtained in  a big number of attempts.
   \end{definition}
   This average (expected) payoff gives us a good criterion to decide in situations where there is no complete certainty of what can happen.   There is an utility, but there is also a risk that arises from the lack of information.

   Now suppose we have to choose between two different lotteries.  Each has different prizes, and different probabilities to win.   The information we have about the prizes and probabilities is in table \ref{Lotteries}.
   \begin{table}
    \center
    \begin{tabular}{|c|c||c|c|}\hline
     \multicolumn{2}{|c||}{Lottery 1} & \multicolumn{2}{|c|}{Lottery 2} \\ \hline\hline
     {\small Probability} & {\small Prize} & {\small Probability} & {\small Prize}\\\hline
     0.010 & 1000 & 0.001 & 10000\\
     0.040 & 250 & 0.005 & 2000\\
     0.150 & 40 & 0.104 & 45\\
     0.800 & 0 & 0.890 & 0\\ \hline
    \end{tabular}
    \caption{Comparison of two lotteries}\label{Lotteries}
   \end{table}
   The expected payoff of the first lottery can be computed as an average, that is, the sum of each probability times the corresponding prize:
   \begin{equation*}\begin{aligned}
    \boxed{\hat{\$}_1 = 0.01\times 1000 + 0.040\times250 + 0.150\times 40 + 0.800\times0 = 10 + 10 + 6 + 0 = 26}\\
    \boxed{\hat{\$}_2 = 0.001\times 10000 + 0.005\times2500 + 0.104\times40 + 0.800\times0 = 10 + 10 + 4.16 = 24.16}
   \end{aligned}\end{equation*}
   The expected payoff analysis tells us that the better lottery is the first one, which give us an expected payoff of \$26, even when the second one promises us \$10000 if we win the big prize.

   This is a very simple way of taking risk into account, but there are more elaborate ones.   There is, then, a way of predicting the behaviour of a decision-making agent accurately in terms of maximization of a certain payoff with a certain use of risk statistical parameters.  This is the base of \textbf{decision theory} \cite{Decision}.

   The maximization of quantities, on the other hand, has been used as a cornerstone to construct very elegant theoretic buildings, not only in physics but in all the so called ``hard sciences''.  The best example is probably Lagrangian Mechanics, based on the Maupertuis least action principle \cite{ExtremalPrinc}.  Nowadays, as it has always happened, this still motivates physicists and other theoretical scientists to untiringly seek \textit{extremal principles} as a foundation for their theories.  (An example of this is \cite{ExtremalPhysics}, where three extremal principles are claimed to give rise to the entire formulation of General and Special Relativity and Quantum Mechanics)

   Rationality, then, seems to be a very useful assumption for studying both living and non-living systems, and therefore Game Theory arises as a powerful tool in those fields.

  \subsection{MULTIAGENT OPTIMIZATION}\label{Multiagent}
   The lottery we have seen can be bought by a large amount of people or by none at all, and this does not affect either the probabilities or the prizes.  But this is not always the case.

  For example: The administrators of a lottery decide to do something to incentive the people to buy tickets.  If a certain prize is not reclaimed, then it is given to the winner of another game, at random.   Then the payoffs \textit{depend} now on what other lottery players do: if they don't buy, they increase the payoffs of the game.

  The theory of multi-agent decision is \textbf{game theory}.   A number of interesting new concepts arise when we consider the interplay between many agents.  A typical example of strange things that can happen is the game called \textbf{Prisoner's dilemma}.

  Two partners in crime are held by the police in two separate rooms at the police station and given a similar deal. If one implicates the other, he will go free while the other receives a life in prison. If neither implicates the other, both will be given moderate sentences, and if both implicate the other, the sentences for both will be severe \cite{Prisoner}.

  The payoff in this case would be the avoided  years in prison.  A maximizes his payoff choosing the upper row no matter what player B does, and B maximizes his payoff choosing the left column no matter what player A does.  They reach an equilibrium with \textit{severe} sentences, while if they maximize \textit{the other player's payoff} both get \textit{moderate} sentences.
  \begin{table}[hbt]
   \center
   \begin{tabular}{|p{6.4cm} p{6.4cm}|}\hline
    \textbf{Payoff for A} & \textbf{Payoff for B}\\
    \begin{tabular}{|c|c|c|}\hline
      & B {\tiny confesses} & B {\tiny implicates A}\\ \hline
      A {\tiny confesses} & {\small moderate}  & {\small lifetime}\\ \hline
      A {\tiny implicates B} & {\small free} & {\small severe} \\ \hline
    \end{tabular}
    &
    \begin{tabular}{|c|c|c|}\hline
      & B {\tiny confesses} & B {\tiny implicates A}\\ \hline
      A {\tiny confesses} & {\small moderate} & {\small free} \\ \hline
      A {\tiny implicates B} & {\small lifetime}  & {\small severe}\\ \hline
    \end{tabular}\\ \hline
   \end{tabular}
  \caption{Qualitative Payoffs in Prisoner's Dilemma}\label{TablePrisoner}
  \end{table}

  In situations involving interacting agents like those described by games,  overall payoff optimization can give a different result than unilateral optimization.   In prisoner's dilemma, for example, an overall optimization of payoff would lead us to the \textit{both confess} situation, with a moderate sentence for both, while the unilateral optimization lead us to the \textit{both implicate one another} situation, where both get severe sentences.

  \subsection{MODELS OF BEHAVIOUR}
   Game theory have also proved to be a very useful tool to devise models of learning and, in general, models of behaviour for interacting individuals within a population \cite{Population2}.   But this requires a careful choice of the payoffs to be assigned to the players according each state of affairs.

   The concept of payoff function has been applied successfully as an objective tendency as well as a subjective preference.  Repeated versions of 2x2 games like Prisoner's Dilemma are now extensively used to model evolution of behavior within populations \cite{Behaviour}.   In some works, it is assumed that the perceived payoff function varies with time when the game is played repeatedly, and sometimes it arrives to a stationary state where a very stable equilibrium is established.  This is, for example, an explanation proposed by Castillo and Salazar for the mad persistence of the guerrilla conflict in Colombia \cite{WarGame}.

   There are also interesting approaches to a ``rational'' (in a game-theoretical sense) definition of altruistic, aggressive, and other kins of behaviour, as learning schemes used along a game \cite{Population2}, or even as subjective perceptions of the payoff \cite{Subjective}.  This last approach \textit{requires a complete knowledge of the relation between the payoff matrix and the relevant features in the game}, because the characteristics of a game can change dramatically when certain thresholds are reached in the variation of the payoff matrix. It can be then necessary to have a certain \textbf{cartographic knowledge of the space of payoff matrices}, to predict the qualitative changes of behavior arising from changes in the perceived payoff matrix.
   
 \section{THE APPEAL OF QUANTUM INFORMATION}
  All we have said is about theory of (classical) games.  But, what about \textit{quantum} games?

  In quantum mechanics, the description of the state of the system is said to be incomplete \cite{EPR}, because the values of some observables are sharply determined, and the values of some others are not.    In that sense, we can think of a quantum state as \textit{less informative} than a classical one.  This is a consequence of the existence of incompatible observables:
   \begin{definition}
    Two observables are \textbf{incompatible} when measuring one implies losing information about the other.
   \end{definition}

   This can lead us to formulate a question:  Which, then, is the advantage of quantum information?  There are two ways to make the laws of quantum mechanics be advantageously used for information processing:
   \begin{enumerate}
    \item Coherent Superposition
    \item Entanglement
   \end{enumerate}

  \begin{definition}
    \textbf{COHERENT SUPERPOSITION:}

    When an observable does not have a definite value in a state, the system behaves as if there are several superposed states with definite observable values.  This state is said to be a \textbf{coherent superposition} of states.
  \end{definition}
    The states in the superposition  evolve independently, but when the observable is measured, all disappear except one, thus destroying the information about \textit{other} observables in the system.

  A superposition is represented in the following way:
    \begin{equation}
    \psi_{superposition}= c_1\phi_1 + c_2\phi_2 + ... + c_n\phi_n
    \end{equation}
    where the coefficients $c_i$ are complex numbers, and $\phi_i$ are the states for which the value of observable $\hat{O}$ is $\lambda_i$.   The probabilities that, when measuring, we get the value $o_i$ are the square of the magnitude of these numbers
    \begin{equation}
     Probability(o_i\text{ in state }\psi) = |c_i|^2.
    \end{equation}

    This feature of quantum states led Richard P. Feynman\cite{FeynmanRules} and, independently, to Paul Benioff in 1982 to suggest that this can be used to perform parallel computation \cite{BeginingQuantComp}.  If such a superposition is processed in a certain way, then every component is processed \textit{independently and simultaneously}.  This could be used, they suggested, to accelerate the processing of data.

    There is still another potential use of coherent superposition.   If a measurement on quantum systems destroys information of the state, why not encode a secret message in that way, a message that is destroyed when not properly read?  That question well may have been the beginning of \textbf{quantum cryptography} \cite{BeginingQuantComp}.

    \begin{definition}
     \textbf{ENTANGLEMENT}

    An \textbf{entangled} state is a state of several quantum subsystems where no observable of the individual subsystems is sharply determinate, but some global (nonlocal) observables are.
   \end{definition}
    Now suppose we prepare two quantum systems in an entangled state, and take them to places that are far apart from each other.  The systems will still share an entangled state.   The results of measuring an observable on any of them is unpredictable, because its value is not determinate.   But once we measure an observable in one, the other is \textbf{instantly} put in a state where a certain local observable is also determinate.

  \subsection{INFORMATION AND THE QUANTUM WORLD}
   There are not only pragmatic motivations to study quantum information, but also philosophical ones.   The understanding of how a system \textit{can be described} has an outstanding role in the ontology\footnote{the set of the things that are supposed to exist} of a theory.   And information is, now, part of the ontology of physics.   When speaking in information in physics, it is customary to cite the visionary statement of Rolf Landauer:
   \begin{quote}
    ``Information is physical''\\
     \flushright{Rolf Landauer, cited by\cite{RLandauer}}
   \end{quote}
   Information was given a precise mathematical meaning by Shannon in 1944 \cite{TheoryInf}, and since this work physics has used it extensively in the field of statistical mechanics, and others related.   This put information in the line for the general conceptual scrutiny quantum mechanics has caused on other physical concepts.

   One of the first remarks done on the problematic character of the quantum description of the world, and one of the most important, was made in 1935 by Einstein, Podolsky and Rosen.    They showed that the new theory included essentially problematic non-local aspects \cite{EPR};  which suggest new ways to think the \textit{information} concepts regarding our knowledge of a physical system \cite{BellPhysics}.

   \subsubsection{THE EPR PARADOX}
   A quantum state will have a determinate value for some observables, but not for some others that are incompatible with the former.   Let us call a state that has a determinate value $a_i$ for observable $A$, the state $\psi_i$.  Now suppose $B$ is an observable that is incompatible with $A$, and $\phi_j$ is a state where it gives a value $b_j$.

   Now suppose we have two quantum systems, both in the state $\psi_i$. The total state is $\psi_i\psi_i$.  But if they interact, a new state must be formed, and it can be an \textbf{entangled} state.

   Suppose that the interaction leave the  entire system the entangled state  $\psi_{ent}c_1\psi_1\psi_2 + c_2\psi_2\psi_1$.  According to quantum mechanics, if observable $A$ is measured now on subsystem 1 and the result is $a_1$, then the result of measuring $A$ on subsystem 2 is instantly determinate (is $a_2$) no matter how far apart the two subsystems are.  However, there is no certainty of getting $a_1$ when measuring $A$ on system 1, because individual observables are not sharply defined.

   When we measured $a_1$ in system 1, then we know that the result of measuring on $A$ on system 2 would be $a_2$.  There is no point in measuring $A$. Then, let us measure B.  We get, for example $b_1$.  But alas!  It looks like we fooled nature!! We know that the value of A for subsystem 2 is $a_2$, and the value of B for subsystem 2 is $b_2$ \textbf{even thought they are incompatible}!

   That is the EPR paradox.  The bad news are that if we want to check $A$ on system 2, just to be sure, it is perfectly possible that the result \textbf{is not} $a_2$ as we expected, because \textit{when we measured B on system 2, we destroyed the information about observable A in that subsystem.}

   Nature is not so easy to fool.   The strange properties of entanglement were indeed used to test the very fitness of Quantum Theory itself experimentally, with favourable results \cite{AspectExperiment}.

  \subsection{QUANTUM ENTANGLEMENT AND ITS USES}
   
   John Stuart Bell shown in 1966 that systems in entangled states exhibit correlations beyond those explainable by local ``hidden'' properties\footnote{Hidden properties are properties that are determinate in the system, but cannot be measured.} \cite{BellInequalities}.

   Suppose now that we have a  black cube, a white cube, a black ball and a white ball to be divided randomly to Alice and Bob: one cube and one ball to each.  But only one of them can be taken out of the bag (the hole is too narrow), so each one must decide whether he (she) is going to see the cube or the ball (they can touch the things inside the bag to decide).

   So far, this is equivalent to an entangled state.  But now, let us consider the case when Alice saw a black ball and Bob saw a black cube.  Everything is now determinate: a further measurement from Alice should give a ``white'' result, and a further measurement by Bob should give a ``black'' result.   But the quantum case is not like that.

   In the quantum case, the hidden Alice's cube not only must be hidden when Alice sees the ball, but must be able to \textbf{change its color} with a probability of $\frac{1}{2}$ when she does!   If the objects were quantum, we may well end up with two black balls and two black cubes in the considered case.   Once a local measurement is done, the quantum correlations (unlike the classical correlations) must disappear.

   This correlating power of entanglement can be used to speed up the communication, but not to send a signal faster than light.   This is called \textbf{superdense coding}\cite{SDcoding1}, because several bits are encoded in one particle.    Suppose Alice and Bob have particles sharing the following entangled state:
   \begin{equation}
    \frac{1}{\sqrt{2}}\left(\phi_0\phi_0 + \phi_1\phi_1\right).
   \end{equation}
   Alice can apply a transformation to her particle according to the message she wants to transmit, according to table \ref{SDcoding}:
   \begin{table}[hbt]
    \center
    \begin{tabular}{|c|l|r|}\hline
     Message & Operation & Final state \\\hline
     00 & she does nothing & $\phi_0\phi_0 + \phi_1\phi_1$ \\
     01 & she turns $\phi_0$ into $\phi_1$ and vice versa & $\phi_0\phi_1 + \phi_1\phi_0$\\
     10 & she turns $\phi_0$ into -$\phi_1$ and $\phi_0$ into $\phi_1$ & $\phi_0\phi_1 - \phi_1\phi_0$ \\
     11 & she turns $\phi_0$ into -$\phi_0$ & $\phi_0\phi_0 - \phi_1\phi_1$ \\\hline
    \end{tabular}
    \caption{Operations for super dense coding}\label{SDcoding}
   \end{table}
   Then Alice sends her particle to Bob, and he measures on the whole two-particle system an observable whose value gives him the message\footnote{this is possible because the final states given in table \ref{SDcoding} are completely distinguishable}.   Then,Alice transmitted  \textbf{two bits} with \textbf{one} two-state particle.   This has been tested experimentally using photons with partially successful results \cite{SDcoding2}.

 \section{THE FIRST QUANTUM GAME}
  In 1999 a version of the ``penny flipover'' game appeared in a major Physics journal \cite{QuantumPenny}.  This was the first quantum game that attracted the attention of the physicists community.

  This game was presented as played by Captain Picard from Enterprise starship and a quantum player \textit{Q}.   Q gives captain Picard a penny in a closed box, then Picard decides whether to turn it over or not, and gives it back to Q.  This procedure is repeated a number of times. Then Q decides to turn it over or not.  The penny is initially head up, and Picard wins if the penny is tail up after Q's second turn.

    \begin{figure}[h!]
      \center  \psset{unit=20pt}
     \begin{pspicture}(0,2)(18,5)
      \rput(2,4){\ovalnode{H0}{HEADS}}
      \rput(6,4){\ovalnode{H1}{HEADS}}
      \rput(10,4){\ovalnode{H2}{HEADS}}
      \rput(14,4){\ovalnode{H3}{HEADS}}
      \rput(6,3){\ovalnode{T1}{TAILS}}
      \rput(10,3){\ovalnode{T2}{TAILS}}
      \rput(14,3){\ovalnode{T3}{TAILS}}
      \uput[l](18,4){P wins}
      \uput[l](18,3){Q wins}
      \uput[l](5,2){Q plays}
      \uput[l](9,2){P plays}
      \uput[l](13,2){Q plays}
      \ncline{->}{H0}{H1}
      \ncline{->}{H0}{T1}
      \ncline{->}{H1}{H2}
      \ncline{->}{H1}{T2}
      \ncline{->}{T1}{H2}
      \ncline{->}{T1}{T2}
      \ncline{->}{H2}{H3}
      \ncline{->}{H2}{T3}
      \ncline{->}{T2}{H3}
      \ncline{->}{T2}{T3}
     \end{pspicture}
     \caption{\small Penny Flipover Game}
    \end{figure}
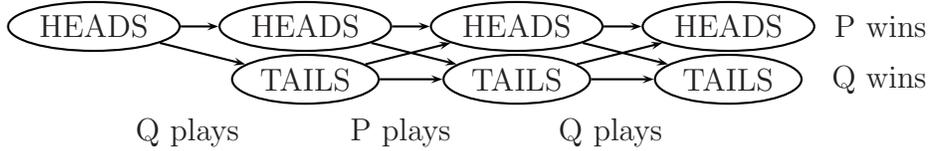
   The quantum feature of this game is that Q can perform a ``half turn'' on the penny, leaving it at first in a quantum state which is a superposition of head up and tail up, and recovering the initial state with the opposite turn at the end, whatever P has done.  Giving Q this advantage, there is no way P can win.\vspace{6pt}

   The state generated by Q can be represented as a quantum superposition of the two classical possibilities:
   \begin{equation}
    \vert STATE \rangle = \frac{1}{\sqrt{2}}\left(\vert HEAD-UP \rangle \pm \vert TAIL-UP \rangle \right).
   \end{equation}
   A classical turnover turns head-up into tail-up, and tail-up into head-up.  The effect on the state generated by Q is none for the symmetric superposition (that with a plus sign) or a global change of sign for the antisymmetric superposition (that with a minus sign):
   {\small \begin{multline}
    \hat{\boxed{turnover}}\vert STATE \rangle =\\ \pm\frac{1}{\sqrt{2}}\left(\vert TAIL-UP \rangle \pm \vert HEAD-UP \rangle \right)\\
    = \pm \vert STATE \rangle.
   \end{multline}
   This overall sign is irrelevant when measuring.

  \subsection{QUANTUM COINS}
   This first quantum game deals with a ``quantum coin'', that is, a coin that can be put in stable enough quantum states.  This quantum coin can be thought as a two-level quantum system, or \textit{qubit}.  A qubit is a quantum extension of the classical concept of the bit, or, in game theory,  a binary choice.

   There is a number of physical realizations of quantum coins.   The conditions necessary to get a good quantum coin can be grouped in two categories:
   \begin{enumerate}
    \item Only two states can be involved in the manipulations.  Normally quantum systems have a very big number of states, but if we want to use the system as a coin, it only must show heads, coins, or superposition of both.  Transitions to other states must be avoided
    \item The quantum state of the system, whatever it is, must be well defined.  Interaction with large systems as environment causes a loss of determinacy in the quantum state, that is called \textit{decoherence} \cite{Decoherence}.  That is probably the reason why we have no ``quantum casino'' so far.
   \end{enumerate}

   Each category of conditions encompasses several technical requirements, which are different for each experimental realization.   Some of these realizations are \cite{ExpQuantComp}:
   \begin{itemize}
    \item \textbf{Quantum wells :} An electron restricted to move within a small region of semiconductive material can have definite states of energy just like one spinning around a nucleus.   The technology to limit its transitions to only some of the lower energy states is now developed enough to produce reliable microscopic quantum coins.  These coins can be designed with an outstanding freedom, to suit the necessities of a particular use.  
    \item \textbf{Light states :} LASER light is a collective quantum state of photons (called coherent state) where the number of photons is not sharply determinate, but its average value can be decreased to nearly one.  Then, we would have states that are superposition of states with no photon and one photon, for example, and can act as good quantum coins.
    \item \textbf{Oscillators :} Currents in microcircuits and vibrating nuclei in molecules are fairly harmonic oscillators.  This systems, when small enough, have clearly defined discrete energy states.   These states have nearly equal energy differences, and it is possible to control the transitions between them using lasers or modulated electromagnetic fields. If we are able to limit the system to two of these states (``heads'' and ``tails'') or two sets of states, we will have a quantum coin.
    \item \textbf{Atoms in cavities :} The one-electron states in an atom can be very efficiently controlled when the atom is in an optical cavity\footnote{A cavity with mirrored walls}.   The states of this system can be controlled with laser light with a high accuracy, and designed with a great freedom to fit a wide range of applications.   This is in fact one of the most developed fields in experimental quantum computation.
    \item \textbf{Spinning nuclei: } Atomic nuclei have a property that has only sense in the frame of quantum mechanics: spin.   This is a vector property, and its components are mutually incompatible observables.   The values that this components can have are always naturally limited to a few multiples of a basic quantity, so it is not necessary to limit transitions to other states.  The evolution of these states is controlled with modulated magnetic fields, that can be controlled with precision.  The drawback of this realization of quantum coins is that a great degree of control at a very small length scale is needed.

   This particular realization was indeed used by Du, et al. to actually play a quantum game \cite{ExpQG}.
   \end{itemize}
 \section{MULTIPARTY CONTROL AND THE EISERT SCHEME}
   The game between Picard and Q discussed above would turn rather dull if captain Picard were allowed to perform quantum operations as well.  In fact, the ``both quantum'' version of the game is as unsolvable as the classical game, in the sense that there is no winning strategy.  But some months after Meyer's paper, another proposal for quantum games appeared, where the game was still very interesting when both players play on equal terms.  This scheme was proposed by Jens Eisert, from Potsdam University,  and named after him \cite{Eisert}.

  \subsection{EISERT'S IDEA}
   It has been noticed by John Bell \cite{BellInequalities} that two person sharing entangled qubits have a correlation resource beyond any classical possibility, and, even when they are not able to communicate faster than light, they can make use of that resource to act coordinately \cite{QuantumCoordination}.

   Coordination can make a great difference in some two-person games like Prisoner's Dilemma and Stag-Hunt (Chicken) game \cite{Miracle}.  Why not try to play these games using entangled qubits?

   This was probably the motivation of Eisert.   His scheme (explained later in detail) is obtained by the classical game changing some elements:
   \begin{enumerate}
    \item ENTANGLEMENT\\
     A referee takes a composite system and prepares an entangled state.   The subsystems have as much levels as choices are in the classical game
    \item QUANTUM STRATEGIES\\
     The players are allowed to perform quantum operations on the entangled subsystems.  These operations are the quantum analogue to the classical strategies
    \item PAYOFF FUNCTION\\
      States of the composite system are defined as eigenstates of the payoff operator.\par
      There must be a ``classical subset'' in each player's strategy set, and players can always choose strategies within this subset that are equivalent to probabilistic strategies in the classical game.  Otherwise, we would not have a quantum version of a classical game, but rather a different game.
   \end{enumerate}
    The formulation of this game is, by itself, very interesting.   There is, for example \textit{only one way} to entangle the subsystems for 2 players with 2 choices that is coherent with the objectives pursued.  The space of local operations on a 2-level quantum systems becomes here the strategy space, and is a key concept in the study of quantum computation; here we study it from a new point of view, that of the maximization of an observable.

    The description of this scheme for quantum games, described in chapter \ref{EisertScheme} is, therefore,  the natural next step in this thesis.
  \subsection{THE SIMPLEST GAMES}
   The obvious way to approach the study of quantum games is choosing first the simplest possible games that can be interesting.  In the current stage, quantum game theory is involved mainly in the study of 2x2 games, that is, games with two players that (in the classical game) have two possible choices.  The quantum games are then 2-qubit games.

   The most studied game is perhaps the Prisoner's Dilemma; a game where both players are completely equivalent (a symmetric game).  The same occurs for Chicken Game, which is perhaps the second most studied game \cite{Miracle}.

\section{THE OBJECTIVES OF THIS WORK}
 The current research on 2x2 classical games shows that in most cases a \textit{complete knowledge of the relation between the payoff matrix and the main characteristics of the game} is a necessary point of departure.    The payoff of a game can vary within certain limits without changing the properties of the game, but outside these limits new features can appear, and old features can disappear.

   In some methodologies the perception of the payoff matrix or even the payoff matrix itself is subject of variations \cite{Behaviour} \cite{Subjective}, thus increasing the importance of the qualitative changes generated by continuous variations of the payoffs.

   There is then a number of situations where it is necessary to have a good cartography of the space of possible payoff matrices.  And we can expect that this is also the case in the field of quantum games.

   In this respect, Rappoport developed a classification for 2x2 games with strictly ordered payoff\footnote{A strictly ordered payoff matrix is one whose elements have relations of strict order in each column and each row}, and found 78 classes of them \cite{Rapoport0},\cite{Rapoport1}.  If he restricts to symmetric 2x2 games, the number of classes is only 9 \cite{Robinson}.  It can be advisable, as a first step, to study this simple classification (that of symmetric games) in their quantum versions.

 The main purpose of this work can be formulated in the following way:
 \begin{center}
 \textbf{Find how quantization affects the classification of symmetric nonzero-sum 2x2 games}
 \end{center}
  There are many ways to classify games.  In this work two criteria are chosen:
  \begin{enumerate}
   \item Number and nature of Nash Equilibria
   \item Number and nature of Pareto Optimal positions
  \end{enumerate}

 \chapter{CLASSICAL GAMES, QUANTUM GAMES}
    Before starting to develop the work, it can be useful to introduce and define some concepts of Game Theory (both classical and quantum) that are going to be used along the next chapters.
 \section{CLASSICAL GAMES}
  A classical game is defined as three elements:\cite{TheoryGamesEB}
  \begin{enumerate}
   \item Two or more agents (players)
   \item A set of choices for each player at each point of the game. (in this work, the only relevant games are those where each player chooses once only)
   \item A payoff function that depends on the choices taken by all the players.
   \item A set of rules governing the influence of one player over the others, like information restrictions.
  \end{enumerate}
   The only element among these that must have some special properties is the payoff function.  To be able to define some operative player's rationality the values of the payoff function must belong to an \textit{ordered set}.  That means that it must be possible to order all the possible values of the payoff in a consistent way, thus allowing the player to choose always a best result.  The existence and characteristics of payoffs that can be considered as utilities is a rather cumbersome philosophical problem\cite{Rational}, that we can avoid by simply working with payoffs which are real finite numbers.

  According to the set of rules about the influences among the players, the games can be divided in two classes:

  \begin{definition}
   \textbf{Games with Imperfect Information : }In this games the players are not influenced by the choices of the others.
  \end{definition}
  This is the kind of games we will study.

  \begin{definition}
   \textbf{Games with Perfect Information : }In this games all players know what the others have decided in all the other steps of the game
  \end{definition}

  \subsection{EXTENSIVE FORM}\label{ExtensiveForm}
   A popular way to represent games (specially complex, multi-stage games) is the extensive form.   It consists in a tree where each inner node is a point where an agent chooses, and a payoff is assigned to each terminal node.

   \begin{definition}
    The \textbf{extensive form of a game} is a tree representation, where each decision of each player is represented by branching nodes, and a payoff is assigned to every terminal branch.
   \end{definition}

   Let's see an example taken from Jim Ratliff's Game Theory Course \cite{GTCourse}.  It is called ``Cholesterol: friend or foe''

   Suppose that there are two companies planing to launch a new low-cholesterol product, but each can choose not to.  But after the launching date, a medical association releases a study about cholesterol.   This study can give two results:
   \begin{itemize}
    \item Cholesterol is injurious for health.  Good news for the companies.
    \item Cholesterol is health promoting.  Bad news for the companies. (But good news for gourmets)
   \end{itemize}
   This result will certainly affect the payoffs of the companies, but nature (the agent that determines whether cholesterol is injurious or not) is not affected by the result of the game, and will not maximize anything.  Nature simply will choose ``unhealthy'' with a probability $p$ or ``healthy'' with a probability $1-p$.\\
    \begin{figure}[htb]
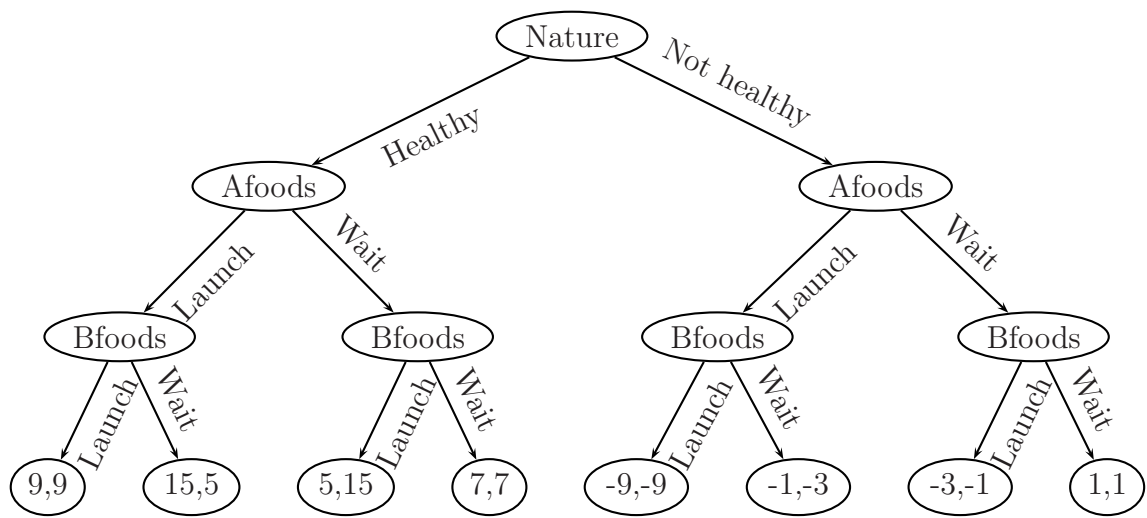

    \center
     \pstree[arrows=->]{
       \Toval{Nature}}{ 
       \pstree[arrows=->]{
       \Toval{Afoods}\naput[nrot=:D]{Healthy}}{ 
        \pstree[arrows=->]{
        \Toval{Bfoods}\naput[nrot=:D]{Launch}}{ 
         {\Toval{9,9}\naput[nrot=:D]{Launch}} 
         {\Toval{15,5}\naput[nrot=:U]{Wait}}} 
        \pstree[arrows=->]{
        \Toval{Bfoods}\naput[nrot=:U]{Wait}}{ 
         {\Toval{5,15}\naput[nrot=:D]{Launch}} 
         {\Toval{7,7}\naput[nrot=:U]{Wait}}}} 
       \pstree[arrows=->]{
       \Toval{Afoods}\naput[nrot=:U]{Not healthy}}{ 
        \pstree[arrows=->]{
        \Toval{Bfoods}\naput[nrot=:D]{Launch}}{ 
         {\Toval{-9,-9}\naput[nrot=:D]{Launch}} 
         {\Toval{-1,-3}\naput[nrot=:U]{Wait}}} 
        \pstree[arrows=->]{
        \Toval{Bfoods}\naput[nrot=:U]{Wait}}{ 
         {\Toval{-3,-1}\naput[nrot=:D]{Launch}} 
         {\Toval{1,1}\naput[nrot=:U]{Wait}}}}} 
     \caption{Game Tree for ``Cholesterol: friend or foe''}
   \end{figure}
   This way of representing games is appropriate to include some rules about available information (example: a curve can enclose the set of nodes that a certain player have information about, defining an ``information set'' of the player in a certain step) \cite{InformationGames}.

  \subsection{STRATEGIC FORM}\label{StrategicForm}
   Most of the games are much more complicated than the one described in the last section.  The tree game for that game is somewhat big, and can be too big to be useful for a number of games having more steps, more players, or both.   Besides, there is a number of different trees that corresponds to the very same game (for complex games, a huge number of them), and it is desirable to remove as much as possible that multiplicity of representations.   A simpler representation would be very useful in those cases.

   It was proved by von Neuman and Morgenstern for games where each player is ignorant of what the others do, there is a simpler representation that is completely equivalent.  It is called the \textbf{strategic form} \cite{TheoryGamesEB}.

   To generate the strategic-form representation of a game, we first define \textbf{strategy}
   \begin{definition}
    A \textbf{strategy} is a certain set of decisions taken by a player all along the game.
   \end{definition}
   A strategy, then, describes all the performance of a player in the game.  With this concept, we can define the strategic-form representation
   \begin{definition}
    The \textbf{strategic-form} representation of a game is a table where a certain payoff for each player is assigned to each possible set of players' strategies.
   \end{definition}
   To show one example of this representation, let us define a ($n_1 \times n_2 \times n_3 \times ... \times n_N$) game:
   \begin{definition}
    \textbf{A (}$\mathbf{n_1 \times n_2 \times n_3 \times ... \times n_N}$\textbf{) game} is a N-player game where player 1 can use $n_1$ strategies, player 2 can use $n_2$ strategies, and, in general, player i has $n_i$ strategies (where i is an integer between 1 and N)
   \end{definition}

   The strategic representation of such game can be made with N hypermatrices, each with N indexes and dimensions $n_1, n_2, n_3, ..., n_N$.

   The games to be studied here are 2x2 games.  For this games the strategic representation can be made with two 2x2 matrices, one for each player, where the payoffs are tabulated, just like in table \ref{TablePrisoner}.

   \subsubsection{TWO-PLAYER GAMES}
   It is necessary at this point to make some definitions about two-player games that will be important later:

   The first concept is \textbf{zero-sum} games.  This kind of games are not studied in this work, but they are tangentially important, because some studies about them are important precedents to this one.
   \begin{definition}
    A \textbf{zero-sum game} is a game where the payoffs of both players sum up to zero for any position.
   \end{definition}
   In these games, what one player wins, the other player lose it.  The reason why they are not as studied in their quantum versions is that in this games the coordination between players is not important.  And new possibilities of coordination are one of the main reasons to study quantum games.

   The other concept is \textbf{symmetric} games.
   \begin{definition}
    A \textbf{symmetric game} is one where all the players are exactly in the same conditions, both in terms of payoff, number of choices, and information rules.
   \end{definition}
    This concept is important, of course, because this is the kind of games we have chosen for this study.  The reason why this kind of games was chosen is simplicity.  The symmetry imposes certain conditions on the payoff matrices (or poly-hypermatrices for games with more players), that make the number of free parameters lower, and allows to study the variations of the defining payoffs more easily.

    This will be noticed in  chapter 5, where a complete classification of the 2x2 symmetric games is made, that heavily relies on the symmetry condition.   A similar study of the non-symmetric games can be made with the same methodology, but the results would be much more complex.

  \subsection{REDUCTION OF THE GAMES}
   Even when the strategic representation is much simpler than the extensive representation, it can sometimes be still too complex (or, better, too big).  It can be necessary to simplify it further to make it operational.

   An example of this simplification (reduction) of the strategic form can be performed on the three-player game ``Cholesterol: friend or foe''.   The payoff can be tabulated in three hypermatrices with three indexes, one for Nature, other por Afoods, and other for Bfoods.  To present the payoffs in matrices, we can consider the two choices of nature separately:
  \begin{table}[hbt]
   \center
   \textbf{CHOLESTEROL IS HEALTHY}\vspace{6pt}
   \begin{tabular}{|p{4.2cm} p{4.2cm} p{4.2cm}|}\hline
    \textbf{Nature} &\textbf{Afoods} & \textbf{Bfoods} \\
    \begin{tabular}{|c|c|c|}\hline
      & BL & BW\\ \hline
     AL & X  & X \\ \hline
     AW & X & X \\ \hline
    \end{tabular}
    &
    \begin{tabular}{|c|c|c|}\hline
      &BL  & BW\\ \hline
      AL & -9  & -3 \\ \hline
      AW & -1 & 1 \\ \hline
    \end{tabular}
    &
    \begin{tabular}{|c|c|c|}\hline
      & BL & B BW\\ \hline
      AL & -9  & -1 \\ \hline
      AW & -3 & 1 \\ \hline
    \end{tabular}\\\hline
   \end{tabular} \vspace{12pt}
   \textbf{CHOLESTEROL IS UNHEALTHY}\vspace{6pt}
   \begin{tabular}{|p{4cm} p{4cm} p{4cm}|}\hline
    \textbf{Nature} &\textbf{Afoods} & \textbf{Bfoods} \\
    \begin{tabular}{|c|c|c|}\hline
      &BL & B BW\\ \hline
      AL & Y & Y \\ \hline
      AW & Y & Y \\ \hline
    \end{tabular}
    &
    \begin{tabular}{|c|c|c|}\hline
      & BL & BW\\ \hline
      AL & 9  & 15 \\ \hline
      AW & 5 & 7 \\ \hline
    \end{tabular}
    &
    \begin{tabular}{|c|c|c|}\hline
      & BL & BW\\ \hline
      AL & 9  & 5 \\ \hline
      AW & 15 & 7 \\ \hline
    \end{tabular}\\\hline
   \end{tabular}
  \caption{Strategic ``Cholesterol: friend or foe''}\label{TableCFF}
  \end{table}
   In table \label{TableCFF} we called the strategies of launching the product AL and BL respectively, and the strategies of waiting AW and BW.   The variables X and Y are random variables which allow us to model the behaviour of Nature in this game.  With a probability $p$ Y>X and nature will select ``not healthy'', and with a probability ($1-p$) X>Y and nature will select ``healthy''.

   The information we have about nature allows us to reduce the strategic form of the game.  According to their statistical knowledge of nature, players should work with \textbf{average payoffs} to decide their actions.  This puts nature out of the picture, and allow us to convert the game in a 2x2 game, with a parameter $p$ in the payoff:
  \begin{table}[hbt]
   \center
   \begin{tabular}{|p{6.4cm} p{6.4cm}|}\hline
    \textbf{Payoff for A} & \textbf{Payoff for B}\\
    \begin{tabular}{|c|c|c|}\hline
      & B launches & B waits \\ \hline
      A launches & 9(2p-1)  & 18p-3  \\ \hline
      A waits & 6p-1  & 6p + 1 \\ \hline
    \end{tabular}
    &
    \begin{tabular}{|c|c|c|}\hline
      & B launches & B waits \\ \hline
      A launches & 9(2p-1)  & 6p-1  \\ \hline
      A waits & 18p-3  & 6p + 1 \\ \hline
    \end{tabular}\\ \hline
   \end{tabular}
  \caption{Reduced ``Cholesterol: friend or foe''}\label{ReducedCFF}
  \end{table}

  \subsection{OPTIMAL RESPONSES AND NASH EQUILIBRIUM}\label{OptimalNash}
   The strategic representation of games makes very easy to forecast what a rational player would do if he knew the strategies of all the others.  He will probably react with an optimal response.
   \begin{definition}\label{DefOptResp}
    \textbf{Optimal response} is the strategy that gives the player the highest payoff if the other players' strategies are known and fixed.
   \end{definition}
   Each player, of course, will not know the strategies of the others in the games that we are studying.   But he can assume that they are rational too.  So, the other player will want to play optimal strategies also.

   That means that all players will tend to be at a certain position\footnote{We will call \textbf{position} a certain choice of strategies for all players} that is optimal for all the players.  That is called a Nash Equilibrium.
   \begin{definition}\label{DefNashEq}
    \textbf{Nash Equilibrium} is a position where all the strategies are mutually optimal responses.  That means that no unilateral change of strategy will give a higher payoff to the corresponding player.
   \end{definition}

  \subsection{PARETO OPTIMAL STRATEGIES}
   In Prisoner's Dilemma (section \ref{Multiagent}), we saw that a better result can sometimes be obtained in two person games if we maximize \textit{the other player's payoff}.  Thus we could define a concept related to Nash Equilibrium, called Pareto Optimal Position: \cite{Pareto}
   \begin{definition}\label{DefParetoOp}
    A position is \textbf{Pareto Optimal} if the only way each player has to increase her payoff is decreasing someone else's payoff.
   \end{definition}
   This concept is an interesting approach to include the advantages of a limited sympathy towards the other players.

 \section{QUANTUM GAMES}
  \subsection{INTRODUCTION}\label{EisertScheme}
   The puzzling properties of \textit{entanglement} are currently intensely studied with a number of motivations, and its applicability in communication technology is certainly among the most popular.   Entangled quantum systems behave in a somehow coordinated way, and even though we cannot fool relativity using entanglement, it has been proved possible to enhance communication speeds to a large amount by this resource \cite{QuantumCommunication}.

   But, what about coordination through entanglement?  Even if there is no way to communicate, it can be possible for two people to act in a coordinated way if they share entanglement.   A natural arena to check this possible coordination is \textbf{game theory}. (an example can be found in \cite{QuantumCoordination})

   The most popular scheme for quantum games is perhaps the \textbf{Eisert scheme} \cite{Eisert}.  A quantum game in this scheme is played in the following way:
   \begin{itemize}
    \item A referee produces an entangled state of N \textit{qunits} (A \textit{qunit} is a n-level quantum system), applying a nonlocal unitary transformation to an initial known separable state
    \item Each player gets one of the entangled qunits
    \item Each player acts on her qunit in any way she wants
    \item The qunits are returned to the referee
    \item The referee applies the inverse nonlocal transformation, so that if neither of the players did anything, the result state is the initial known state
    \item The referee makes a measurement on each of the qunits, and gives a payoff to each player.
    \begin{figure}[h!]
     \center
     \includegraphics{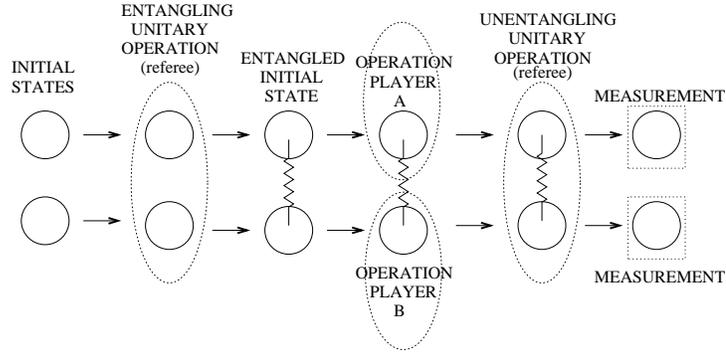}
     \caption{The Eisert Protocol for a Quantum Game}
    \end{figure}
   \end{itemize}
   Let us discuss some of the concepts involved in the quantum game:

   \subsection{THE INITIAL STATE}
    This state must be known by all the players, at least to some level of accuracy, because of the rationality condition \cite{LinearUtility}.   This condition requires from the players some criterion to choose the strategy, which cannot exist without knowledge of the initial state.

    This state must be separable (at least to some extent), because otherwise the players would have uncertainty in the \textit{local} state of their subsystem. In the ideal case (total certainty) the initial state is given by the state vector $\vert 00 \rangle$, or by the density operator:
    \begin{equation}
     \rho_{ini} = (\vert 0 \rangle \langle 0 \vert)\otimes(\vert 0 \rangle \langle 0 \vert).
    \end{equation}

   \subsection{THE STRATEGIES OF THE PLAYERS}
    In the Eisert scheme, each player is given an initial state, and the strategies of a quantum player correspond to quantum operations to be performed on that state.

    Whenever the game is played, the players get one of the payoff values tabulated in the payoff matrix, but there is some randomness only in the result.  To apply some kind of rationality in these games, it is necessary to work with the \textbf{expected payoff} rather than the actual earned payoff, as it is done in games involving random moves \cite{TheoryGamesEB}.

    We will make a distinction, however subtle, between \textbf{quantum operations} and \textbf{quantum strategies}, the former being physical operations carried on a physical system (a qubit) and the latter being a particular choice of parameters that leads to some expected payoff.

   \subsection{A CLASSIFICATION OF QUANTUM STRATEGIES}
    In classical games, there are only two kinds of strategies: pure strategies (those who define the space of positions) and mixed strategies, defined as distributions of probability.

    The set of quantum operations  on a qubit, on the other hand, is very big.  A general quantum operation can be defined by means of the $\chi$ matrix representation, \cite{Chi} which consists in:
    \begin{equation}\label{EstratChi}
     \rho_{transformed}=\hat{\hat{O}}\rho = \sum_{i,j} (E_i)^\dagger \rho E_j \chi_{i,j}.
    \end{equation}
    where $\hat{\hat{O}}$ is the superoperator describing the quantum operation, $\rho$ is a density matrix, $\{E_i\}$ is a complete basis for the hermitian operators in the Hilbert-Schmidt space, and $\chi$ is an hermitian matrix with the dimension of the space.  The parameters in the $\chi$ matrix define a quantum strategy.
    A physically meaningful quantum operation fulfills the positiveness condition: \cite{Completeness}
    \begin{equation}\label{Completeness}
     \sum_{i,j} \chi_{i,j} (E_i)^\dagger E_j \leqslant Identity.
    \end{equation}
    The number of parameters necessary to describe an operation on n-level systems is $n^4-n^2$.   For a qubit, this number is 12. (the space of quantum strategies is 12-fold!)

    From a game theoretical point of view, we may use a determinacy criterion to classify quantum strategies (operations), as we did in classical games to distinguish pure and mixed strategies.  Therefore we can define two kinds of strategies:
    \begin{enumerate}
      \item DETERMINISTIC STRATEGIES\\
       These are strategies that ensure a known payoff, without randomness.  They consist simply on acting on the density matrix in the following way:
      \begin{equation}\label{Deterministic}
       \rho_{final} = \rho_{initial} \hspace{16pt}or\hspace{16pt} \rho_{final} = \sigma_x \rho_{initial} \sigma_x.
      \end{equation}
      The reason for using $\sigma_x$ as a deterministic strategy is merely conventional.
      \item WEAKLY DETERMINISTIC STRATEGIES\\
       These are strategies that are deterministic in the totally nonentangled and totally entangled game, but can be otherwise in intermediate entangled games.  They consist simply on acting on the density matrix in the following way:
      \begin{equation}
       \rho_{final} = \sigma_y \rho_{initial}\sigma_y \hspace{16pt}or\hspace{16pt} \rho_{final} = \sigma_z \rho_{initial} \sigma_z.
      \end{equation}
       In what follows, we will refer to both deterministic and weakly deterministic strategies as \textbf{semideterministic strategies}
      \item NONDETERMINISTIC STRATEGIES\\
       These are strategies that involve randomness in the payoff for any entanglement condition.\\
       Among these we can still distinguish two cases:
      \begin{enumerate}
       \item UNITARY STRATEGIES
       In these strategies the state of the entangled pair of qubits remains pure, that is, there is no coherence loss.   That means that there is no randomness in the \textit{quantum state} of the entangled pair of qubits.  The source of the randomness in the payoff is for this cases is the indeterminacy of the payoff in the quantum state of the system.  They consist simply on acting on the density matrix in the following way:
     \begin{equation}
      \rho_{final} = U^\dagger \rho_{initial}U.
     \end{equation}
     The form of the $chi$ matrix corresponding to an unitary strategy is:
     \begin{equation}\label{ChiUnitary}
      \chi = \begin{pmatrix}\sqrt{1-\vec{a}\centerdot\vec{a}} & -ia_z & -ia_y & -ia_x\end{pmatrix}
      \begin{pmatrix}\sqrt{1-\vec{a}\centerdot\vec{a}} \\ ia_z \\ ia_y \\ ia_x\end{pmatrix}.
     \end{equation}
      \item MIXED UNITARY STRATEGIES
       These consist in a number of unitary strategies used at random with a distribution of probabilities:
       \begin{equation}
        \rho_{final} =  \sum_j \lambda_j\left((U_j)^\dagger \rho_{qubit} U_j\right)
       \end{equation}
       where $G_j$ is some hermitian operator, like  $\hat{r}\centerdot\vec{\sigma}$ (for $\hat{r}$ real and unitary), $0\leqslant \lambda_j \leqslant 1 $ and $\sum_j \lambda_j = 1$.

   Unitary interaction with other relatively small systems in indeterminate states is a one-step physical realization of this kind of strategies.  An important characteristic of this kind of quantum operations is that if the initial density operator is identity, it remains the identity after applying the operation.  The operations that behave in that way are called \textit{unital}.   For single qubits, \textbf{all unital operations can be represented by mixtures of unitaries} \cite{UnitalQubit}.

       Another important characteristic is that this operation \textit{preserve the entanglement}.  In the Eisert scheme, the consequence is that the final state (the one on which the payoff is measured) is unentangled.\\
       This means that if both players act in this way on their entangled qubits, the referee ``unentangles'' them with the inverse of the unitary he used to entangle them before being transformed.
       \begin{multline}
        J \left(\sum_{j,k}\lambda_j \lambda_k (U_j\otimes U_k)^\dagger \left[J^\dagger (\rho_A\otimes\rho_B)J\right] (U_j\otimes U_k)\right) J^\dagger =\\
        = \sum_{j,k}\left(\lambda_j \lambda_k(\tilde{\rho}_A)_j \otimes (\tilde{\rho}_B)_k \right) =\\
       = \sum_l \left(\tilde{\lambda}_l(\tilde{\rho}_A)_l \otimes (\tilde{\rho}_B)_l\right)
       \end{multline}
       where J is the entangling operation. The final density matrix is separable, but \textbf{classically correlated}, thus allowing coordination.

      \item INFORMATION WITHDRAWING STRATEGIES\\
      Non-unital operations can be regarded as information withdrawing strategies, because it is not possible to recover the initial state from the knowledge of the operation and the final state.\\
      They are associated with a measurement, and consist simply on acting on the density matrix in the following way:
      \begin{equation}
       \rho_{final} = \sum_j (F_j)^\dagger \rho_{initial} F_j.
      \end{equation}
      Where $R_j$ are positive operators such that $\sum_j (F_j)^\dagger F_j <\ Identity$\\
      This kind of operations, even being in a sense local, \textit{do not preserve entanglement}.   That means that the final state, the one on which payoff is going to be measured, is entangled.   The payoff operator is separable (even when, in general, payoffs are classically correlated), and the entanglement in the state to be measured becomes an \textbf{additional source of randomness}.

     Since $\chi$ is an hermitian matrix, which is, besides, positive if the basis operators are positive, it can be decomposed in terms of projectors:
     \begin{equation}\label{ChiDecomposed}
      \chi_A = \sum_k \lambda_k \begin{pmatrix}\sqrt{1-\vec{(a_k)^*}\centerdot\vec{a_k}} \\ ((a_k)_z)^* \\ ((a_k)_y)^*  \\ ((a_k)_x)^*\end{pmatrix}
           \begin{pmatrix}\sqrt{1-\vec{(a_k)^*}\centerdot\vec{a_k}} & (a_k)_z)^* & (a_k)_y  & (a_k)_x\end{pmatrix}
     \end{equation}
     where the $\lambda_k$ are real numbers from 0 to 1 and $\sum_k \lambda_k \leqslant 1$.

     This expression is very similar to that of a unital map (a convex sum of (\ref{ChiUnitary})), except that the unit vectors $a_k$ are not necessarily imaginary in $x,y,z$.
     This expansion of the general map will be useful later.
     \end{enumerate}
    \end{enumerate}
   \subsection{THE PAYOFF FUNCTION}
    The payoff function is in quantum games an observable, and it is, of course, represented by an operator.    The payoff matrix gives a definite payoff for the positions defined by deterministic strategies (\ref{Deterministic}).  Each of the 4 numbers of the 2x2 payoff matrix is then an eigenvalue of the payoff operator.   The eigenvectors are simply direct products of the initial state $\vert 0 \rangle$ transformed by the deterministic operation of A, and the one transformed by the deterministic operation of B.

    If the payoff matrix for the classical game is {\small$\begin{pmatrix}a & b\\c&d\end{pmatrix}$} then the payoff operator is:
    \begin{equation}
     \hat{\$} = a\vert 00 \rangle \langle 00 \vert + b\vert 01 \rangle \langle 01 \vert + c\vert 10 \rangle \langle 10 \vert + d\vert 10 \rangle \langle 10 \vert.
    \end{equation}
    The expected payoff will have then the following form:
    \begin{equation}\label{ChiPayoff}
     \langle \$ \rangle = \sum_{i,j,k,l}(\chi_A)_{i,j} (\chi_B)_{k,l} Tr \left(J\left((F_i\otimes F_k)^\dagger J^\dagger (\vert 00 \rangle \langle 00 \vert)J(F_j\otimes F_l)\right)J^\dagger \ \hat{\$} \right).
    \end{equation}
    The trace in the expression is a bitensor of rank 4 and dimension 12, that we can call $P^{i,j,k,l}$.  The payoff will then be an scalar:
    \begin{equation}
     \langle \$ \rangle = \sum_{i,j,k,l}(\chi_A)_{i,j} (\chi_B)_{k,l} P^{i,j,k,l}.
    \end{equation}

   \subsection{THE CLASSICAL GAME WITHIN THE QUANTUM GAME}
    One of the guidelines in the quantization of games is that the classical game must be embedded in the quantum game \cite{QGTheory}.  Therefore, there must be a classical strategy subset within the total quantum strategy set.

    In the Eisert scheme, the classical strategies are those constructed with generators commuting with the entangling unitary transformation \cite{Eisert}.   As a convention, the chosen entangling unitary is:
    \begin{equation}\label{entangling}
     J = \sqrt{1-\gamma^2}\mathbb{I} + i\gamma(\sigma_x \otimes \sigma_x)
    \end{equation}
   where $\mathbb{I}$ is the 2x2 identity and $\gamma$ is a real entanglement parameter.

    Any operation $\hat{\hat{O}}$ defined with  $E_i=\alpha\mathbb{I} + \beta\sigma_x$ according to equation \ref{EstratChi}, will commute with the entangling unitary $J$, and will therefore be a \textit{classical strategy}.

    Restricting the elements of the quantum operations to the identity and the flip Pauli matrix, a classical strategy would be represented by a 2x2 hermitian matrix.  The payoff tensor computed in this subspace is very simple:
    \begin{multline}
     P^{i,j,k,l} = \frac{1}{4}((a+b+c+d) + \delta_{ij}(-1)^{1+ij}(a+b-c-d) \\ + \delta_{kl}(-1)^{1+kl}(a-b-c+d) + \delta_{ij}\delta_{kl}(-1)^{1+ij+kl}(a-b-c+d) )
    \end{multline}
    where the indexes $i,j,k,l$ take values of 0 for the identity and 1 for $\sigma_x$.

    The expected payoff is simply:
    \begin{equation}
    \langle \$ \rangle = a(\chi_A)_{00}(\chi_B)_{00} + b(\chi_A)_{00}(\chi_B)_{11} + c(\chi_A)_{11}(\chi_B)_{00}+d(\chi_A)_{11}(\chi_B)_{11}.
    \end{equation}
    The condition of completeness (\ref{Completeness}) forces the relation $\chi_{00}+\chi_{11} = 1$ for both players, and we can see that the expected value of the payoff is exactly equivalent to that of the expected value for mixed strategies in the classical game:
    \begin{equation}
     \langle \$_{classical} \rangle = (1-P_A)(1-P_B)a + (1-P_A)P_Bb + P_A(1-P_B)c + P_AP_bd.
    \end{equation}
    It must be noted that the nondiagonal elements of the $\chi$ matrix in the classical subspace are irrelevant for the expected payoff.

   \subsection{THE SPACE OF PLAYER'S PARAMETERS}
    We have seen that the strategy of each player can be represented by an hermitian matrix fulfilling the completeness condition, and that not all the elements of the matrix are relevant to the expected payoff ($\chi_{0X}$ and $\chi_{0X}$ are not relevant).

    As the parameters are 24 (12 from each player) but the payoff is only one number.  That means that we can find 23 functions of the parameters given by the players that are not relevant for the payoff (we can call them redundant parameters).  The two irrelevant parameters mentioned are not independent, they are related by the completeness relation to each other and to other elements. (If all the elements outside the classical subspace are zero, the relation is $\chi_{0X}=-\chi_{0X}$ The hermiticity of the matrix forces each to be the conjugate of the other, so they must be plus and minus an imaginary number.\\
 We can only reduce the space of parameters if some of these relevant parameters depend only on one player's parameters.  In that case, some unilateral variations would not affect the payoff.

 \chapter{WHAT WE KNOW SO FAR}
  
  These work is restricted to quantum games in the Eisert scheme.  There are interesting proposals outside this scheme, but these are not discussed here. However, in table \ref{OtherGames} a little description is presented:
   \begin{table}[htb]
    \center
    \begin{tabular}{|p{10cm}|c|}\hline
     {\centering \textbf{Description}} & \textbf{Reference} \\\hline
     ``\textbf{Quantum solutions for coordination problems}'' and ``\textbf{Correlated Equilibria of Classical Strategic Games with Quantum Signals}''

      The use of quantum communication (communication through entanglement) is explored for players playing quantum games  & \cite{QuantumCoordination} and \cite{QCorrEquilibr}\\ \hline
     ``\textbf{Probability amplitude in quantum like games}''

      Classical games that emule quantum situations are explored & \cite{QuantumLike}\\ \hline
      ``\textbf{Generalized quantum games with Nash Equilibrium}''

      Generalized quantum games are defined relaxing some features of the Eisert scheme, like locality of initial state, to give the probability amplitudes a protagonic role.  They call these games ``fully-quantum game theory''. & \cite{GeneralizedQG}\\ \hline
      ``\textbf{Quantum Games of asymmetric information}''

      Asymmetric information is introduced in the Eisert Scheme. & \cite{Asymmetric}\\\hline
      ``\textbf{Continuous-variable Quantum Games}''

     &  \cite{QContinuousG}\\ \hline
    \end{tabular}
    \caption{Other quantum games} \center \label{OtherGames}
   \end{table}
  However, the quantum versions of symmetric 2x2 nonzero-sum quantum games are in some cases complex enough to constitute a broad field of research, which is not yet depleted.

  The first important results to discuss are a quantum version of the Nash Theorem, and some criticisms about the very ``quantumness'' of the quantum games.

  \section{NASH EQUILIBRIA AND PARETO OPTIMAL POSITIONS}
   A \textbf{Nash Equilibrium} is a position in which the payoff of all players can only remain equal or decrease by unilateral changes of strategy.   The Nash Theorem states that \textbf{every finite game has at least one equilibrium point} \cite{NashTheorem}.

   This Nash equilibrium, however, can correspond to a \textbf{mixed strategy}, a strate.gy that consists in playing one pure strategy with a certain probability, and playing another with another probability.   For symmetric a stronger theorem is fulfilled: Every symmetric game has at least one \textbf{pure strategy Nash Equilibrium} \cite{NashSym},\cite{Samaritan}.

   The quantum version of Nash theorem was proved by Lee and Johnson \cite{QGTheory} for quantum games in the Eisert scheme (Quantum games defined in other schemes can fail to fulfill this theorem, for example those studied in \cite{QmatrixStrat}).

  The Pareto-optimal positions are positions where \textbf{each player can increase his own payoff only decreasing the other player payoff} \cite{Pareto}. These have not been studied as extensively as a central feature of a game like the Nash Equilibria, but rather as a stability additional test.  A Nash Equilibria that is Pareto Optimal at the same time, is more stable than one that is not.

 \subsection{EQUIVALENT CLASSICAL GAMES?}
  How ``quantum'' are quantum games?  van Enk and Pike show in \cite{ClassicalRules} that some quantum games are equivalent to a classical game, in particular Quantum Games with a restricted set of strategies.   They criticize the assumption that the quantum version of a game is a quantum version of \textit{that} game, even when can have the characteristics of \textit{another} classical game.

 \section{THE PRISONER'S DILEMMA}
  The preferred example for studying quantum games is the Prisoner Dilemma.  The most exhaustive study about this game can be found in \cite{Eisert}.  A popular quantitative assignation of payoff for this game is:
  \begin{equation}
   \begin{pmatrix} 3 & 5 \\ 0 & 1 \end{pmatrix}
  \end{equation}

  \subsection{ONE-PARAMETER SUBSPACE}
   The first study done was the exploration of a one-parameter subspace of the strategy space, that of the Unitary transformations generated by $\sigma_y$.
   In this subspace, Nash Equilibrium in the completely entangled game is the same as in the classical one. (position(1,1), payoff 1 for both players) 

  \subsection{TWO-PARAMETER SUBSPACE}
   The second step was the study of the following  parametrization for a subset of SU(2):
   \begin{equation}\label{TwoParam}
    U(\theta,\phi)=
    \begin{pmatrix}
     \cos(\theta/2)e^{i\phi} & \sin(\theta/2)\\
    -\sin(\theta/2) & \cos(\theta/2)e^{-i\phi}
    \end{pmatrix}
   \end{equation}
   A new equilibrium was found for this subspace in $\theta=0$, $\phi=\pi/2$, with the same payoff that the Pareto Optimal position of the classical game, which is 3 for both players.  This is the best possible payoff for any symmetrical position.

  \subsection{THREE PARAMETER SUBSPACE}
   In the whole unitary group SU(2) it was found that the perfect Nash Equilibrium found in the last subspace loses its character, and the Nash equilibrium is again the (0,0) position, the same as in the classical game.

  \subsection{THE GENERAL CASE}
   In the general case, there appears once again an efficient Nash Equilibrium, which can be put as a mixed unitary strategy that is different for each player (a nonsymmetric mixed strategy)

   For player A:
  \begin{equation}
   U_1= \begin{pmatrix} 1 & 0 \\ 0 & 1 \end{pmatrix}\hspace{16pt}\lambda_1=\frac{1}{2}\hspace{1.5cm} U_2= \begin{pmatrix} -i & 0 \\ 0 & i \end{pmatrix}\hspace{16pt}\lambda_2=\frac{1}{2}
  \end{equation}
   For player B:
  \begin{equation}
   U_1= \begin{pmatrix} 0 & 1 \\ -1  & 0 \end{pmatrix}\hspace{16pt}\lambda_1=\frac{1}{2}\hspace{1.5cm} U_2= \begin{pmatrix} 0 & -i \\ -i & 0 \end{pmatrix}\hspace{16pt}\lambda_2=\frac{1}{2}
  \end{equation}

   In the total strategy space, the only extra operation is a measurement that, according to Nielsen \cite{Equivalence} is equivalent to the mixture of unitaries.

 \section{THE CHICKEN GAME}
  The Chicken Game (Also called Stag-Hunt) is another 2x2 symmetric game which is popular among the quantum game theoreticians.   This game exhibits a dilemma of a different kind, based on the nonsymmetric character of their Nash Equilibria.  This produces, however, a rather stable mixed Nash Equilibria.   In this game it is interesting to find out which quantum strategy can mimic the mixed classical strategy.  A quantitative payoff matrix for this game is:
  \begin{equation}
   \begin{pmatrix} 6 & 2 \\ 8 & 0 \end{pmatrix}
  \end{equation}

  \subsection{ONE-PARAMETER SUBSPACE}
   Again, in the one-parameter subspace the game has the same Nash Equilibrium in its classical version and in its quantum version. (the classical Nash equilibria are the positions (0,1) and (1,0), with payoff 8 for both players)

  \subsection{TWO-PARAMETER SUBSPACE}
   Using the same parametrization (\ref{TwoParam}), a unique diagonal Nash Equilibrium in $\theta=0,\phi=\pi/2$ with a payoff of 6 is found .

  \subsection{THE WHOLE SPACE}
   In the whole space the focal point\footnote{A \textbf{focal point} is a Nash Equilibrium which is stable under unilateral departures from rationality} turns out to be the same as the mixed strategy in the classical game, with a payoff of 4 for both players.

 \section{OTHER GAMES}
  A number of nonsymmetric games, or even zero-sum games have been studied.  Here are some references to them, even though they are not going to be discussed here.
  \begin{enumerate}
   \item The Battle of Sexes: Some common properties of this game and Stag-Hunt (Chicken) are described \cite{QBattleSexes}.   This can be a result of the fact that relabeling of strategies of the Battle of the Sexes give a game that is very similar to Stag-Hunt.
   \item Samaritan Dilemma and Welfare Game:  This non-symmetric game has no pure strategy Nash Equilibrium; but it is found that in the completely entangled case several equilibrium arise, converting the game in a coordination game.\cite{Samaritan}
  \end{enumerate}

 \section{TRANSITIONS OF ENTANGLEMENT REGIME}
  Eisert's work, and others, are focused on studying completely entangled games.  But, as we have seen, in the Eisert scheme there is the possibility to \textit{modulate} entanglement.

  Then it is interesting to look for \textit{critical amount of entanglement} in a game, where the qualitative properties of it change.  This was studied in \cite{QTransition}, where Du et. Al. found precisely this for games like the Prisoner's Dilemma.

   They studied games with a ordered payoff matrix
   \begin{equation}
    \begin{pmatrix}r & s \\ t & p\end{pmatrix}
   \end{equation}
   where t>r>p>s. (Ordered payoff matrix means there is an order relation between the entries, like this).  They found that using the entangling unitary
   \begin{equation}
    J = \sqrt{1-\gamma^2}\mathbb{I}\otimes\mathbb{I} + i\gamma \sigma_x\otimes\sigma_x
   \end{equation}
    where $\gamma$ is a real parameter between 0 and 1.
    There are two transitions in the entanglement regime:
    \begin{align}
     \vert 1-2\gamma^2 \vert = \sqrt{\frac{p-s}{t-s}}\\
     \vert 1-2\gamma^2 \vert = \sqrt{\frac{t-s}{r-s}}
    \end{align}
     \begin{figure}[h!]
      \center
      \includegraphics{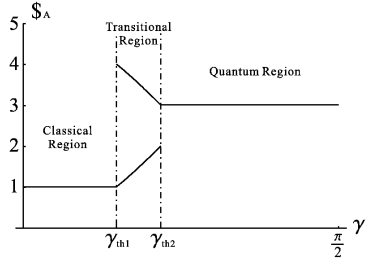}
      \caption{Variation of Nash Equilibrium Payoff with Entanglement (Reproduced from \cite{QTransition})}
     \end{figure}
    All this was computed, however, for the two parameter subspace mentioned above.

    For the total space, they find that there is only one transition (the first) and that after it there is  only one symmetric mixed Nash Equilibrium, whose payoff does not change.  (Essentially, the same found by Eisert)

    All these studies have taken payoff matrix from games that were known to be interesting in their classical version.  But what has \textit{not} been studied suggest a direction for further studies:

   {\large
    \begin{enumerate}
     \item  How to choose a payoff matrix to get a game that is interesting in its \textbf{quantum version}?
     \item How to avoid the choice of rather arbitrary parameters to define a set of quantum strategies?  (a more general classification of quantum strategies is perhaps needed)
     \item If there are non-unitary strategies, why unitary strategies seem to show all the important features we can expect?
    \end{enumerate}}

   In later chapters we will try to make some aportations to answer these questions.

 \chapter[CLASSIFICATION OF CLASSICAL GAMES]{CLASSIFICATION OF NONZERO-SUM SYMMETRIC 2x2 GAMES}
   \section{INTRODUCTION}
  There is a large amount of investigation on dynamic models based on 2x2 symmetric nonzero-sum games \cite{Behaviour},\cite{Selection}, where a certain payoff, and some times some uncertainty on it \cite{Subjective}, is chosen to construct the model.

   The features of the static games defined by the possible payoff matrices are very important in these dynamic models \cite{Suitable}, but so far, we lack a complete cartography of the space of possible payoff matrices to guide our selection for a certain model.   It is then very important to know at which moment a change in the assigned payoffs will lead to a different kind of game, and if so, which kind of game.  The aim of this work is to provide a guide to ``move'' in the space of possible 2x2 symmetric nonzero-sum games in the strategic form.
 
 \section{PAYOFF MATRIX FOR 2x2 GAMES}
  Among the nonzero sum games, the most simple kind of game is probably the 2x2 symmetric with perfect information.  And the most simple way to represent it is the strategic form presented in \ref{StrategicForm}
 \begin{equation} 
  \Game = (\{A,B\},\{0,1\} \otimes \{0,1\},\$_{A}:\{0,1\}\otimes\{0,1\}\to \mathbb{R},\$_{B}:\{0,1\}\otimes\{0,1\}\to \mathbb{R})
\end{equation}
  where $\Game$ represents the game, $\{A,B\}$ is the set of players, $\{0,1\}$ is the space of strategies of any of the players, and the composite space $\{0,1\}\otimes\{0,1\}$ is the space of \textit{positions}.  $\$_A$ and $\$_B$ are the payoff function for the two players. This is a function that goes from the space of positions to some ordered set, which is generally $\mathbb{Z}$, the set of the integer numbers.  These payoff functions are sometimes referred as \textit{payoff matrices}, taking the labels of the strategies as indexes
  \begin{equation}
   (\mathbb{P}_A)_{ij} =  (\$_A(i,j)).
  \end{equation}
  In symmetric games, there is a symmetry condition on the payoffs, which ensures equality of conditions for the two players \cite{Simetricos}.  If we interchange players, both the indexes $i$, $j$ are interchanged, and the payoff matrices $\mathbb{P}_{A}$, $\mathbb{P}_{B}$ are interchanged.
  \begin{equation}
   (\mathbb{P}_{A})_{i,j}=(\mathbb{P}_{B})_{j,i}.
  \end{equation}

 \section[FEATURES OF THE PAYOFF MATRIX]{ESSENTIAL AND UNESSENTIAL FEATURES IN THE PAYOFF BIMATRIX}\label{ClasParameters}
  An acceptable payoff can be defined in any set with an order relation \cite{OrdinalUtility}, but it proves to be very useful to define it within a compact set.   If we assign a numerical payoff to both players in every choice situation, we need 8 numbers.  The condition of symmetry reduces them to 4.  But as it was stated in \cite{LinearInvariance} and \cite{LinearUtility}, the properties of the game cannot depend either on the value of payoff represented by number 0 (additive invariance), or an overall factor in the scale of payoffs shared by both players (scale invariance).  Then, there are two parameters that can be ruled out, and we are left with only \textbf{two numbers}.

  Let us represent the payoff matrices for A and B in the following way:
  \begin{equation} \label{Gclasica}
   \mathbb{P}_{A}=
    \begin{pmatrix}a & b \\ c & d \end{pmatrix}\hspace{1cm}
   \mathbb{P}_{B}=
    \begin{pmatrix}a & c \\ b & d \end{pmatrix}.
   \end{equation}
  The mean of this numbers is irrelevant (additive invariance \cite{LinearUtility}), so can subtract it from the four parameters (that is $a'=a-\frac{1}{4}(a+b+c+d)$) and the properties of the game are preserved.

  We can also use the scale invariance to get rid of the overall scale considering (a',b',c',d') as a vector in $\mathbb{R}^4$ and normalizing. ($a'' = a'/norm(a',b',c',d')$).

  To make the two relevant parameters explicit, we choose a SO(4) transformation which takes $a''$ into $\frac{1}{2}(a''+b''+c''+d'')$ which we know is always zero, and get, for example,these new parameters:
  {\small \begin{equation}\label{transformation}
   (G_0,G_A,G_B,G_{AB})=(a,b,c,d)
   \frac{1}{2}\begin{pmatrix}
    1 & 1 & 1 & 1\\
    1 & 1 & -1 & -1\\
    1 & -1 & 1 & -1\\
    1 & -1 & -1 & 1
    \end{pmatrix}.
   \end{equation}}
   $G_{0}$ is always zero, but $G_{A}$ is the difference between the expected payoffs of each player if \textit{the other} player chooses 0 y 1 with equal probability.  $G_{B}$ is the difference in \textit{the other player}'s payoffs if each player plays an uniform distribution of probabilities 0 and 1.   $G_{AB}$ is the payoff difference between the two symmetric situations (A choosing the same as B).

   The parameter $G_{0}$ is always zero, so we can ignore it. The rest of the parameters define a unit vector ($G_{A}$,$G_{B}$,$G_{AB}$), which allows us to devise a geometric representation of the space of games as the surface of a 3D sphere.

 \section{GEOMETRIC REPRESENTATION}
   The conditions for the existence of Nash equilibria and Pareto Optimal conditions are sets of inequalities involving these three parameters.  These conditions cut the surface of the sphere at planes, so it is possible to view the different kinds of 2x2 non-zero symmetric games as portions of the unit sphere.

   Every possible game in the considered type, defined by a payoff matrix, has a ``normalized'' representative in the surface of the sphere (except for the trivial one with the same payoff for any situation).  \textbf{The fraction of surface of the unit sphere enclosed by the planes corresponding to the conditions is exactly the fraction of the possible games set which fulfills those conditions}.
 \section{NASH EQUILIBRIA (NE)}
  The definition for a Nash Equilibrium (NE from now on) is simple: if both player decrease their payoff departing individually from a certain choice condition, then this condition is a NE \cite{NashTheorem}.

  For example, the position($i^*$,$j^*$) is a NE iff:
  \begin{equation}
   \forall (k,l)\epsilon \{positions\}\hspace{16pt}  \mathbb{P}_{i^*j^*} \geqslant \mathbb{P}_{kj^*}\hspace{16pt}and\hspace{16pt}
   \mathbb{P}_{i^*j^*} \geqslant \mathbb{P}_{i^*l}.
  \end{equation}
  In a symmetric game, that means that we only compare elements in the same column of the player A payoff matrix, or in the same row of the player B payoff matrix.  In terms of our payoff parameters, all these conditions have the form:
   \begin{equation}\label{NashConditions} \begin{aligned}
    (-1)^i(G_{A}+ (-1)^j G_{AB}) &\geqslant 0\\
    (-1)^j(G_{A}+ (-1)^i G_{AB}) &\geqslant 0.
   \end{aligned} \end{equation}
   Where the indexes $i$ and $j$ can take values within the set of strategies $\{0,1\}$.

   A remarkable feature of these conditions is that they only involve the parameters $G_{A}$ and $G_{AB}$.  They are completely insensitive to $G_{B}$, the average difference in payoff owed to a change in the \textit{other player} strategy. 

   \subsection{Symmetric positions}
    For the symmetric positions (i=j) it is easily seen that both conditions are the same: 
    \begin{equation}
    (-1)^i(G_{A}+(-1)^i G_{AB}) \geqslant 0.
    \end{equation}
   This condition will be fulfilled for half the games, for it defines a plane that cuts the sphere in two halves, both for $(1-i)(1-j)=1$ and $ij=1$.  The planes are orthogonal, and that allows us to infer that $\frac{1}{4}$ of the games will have (0,0) as a Nash equilibrium,  $\frac{1}{4}$ of the games will have (1,1) as a Nash equilibrium and $\frac{1}{4}$ of the games will have both positions as Nash equilibria.

   \subsection{Nonsymmetric positions}
    For the nonsymmetric positions , the two conditions are different ($s_1=1, s_2=-s_3$). The explicit conditions are:
    \begin{equation}\begin{aligned}
    (-1)^iG_{A}-G_{AB} &\geqslant 0\\
    (-1)^jG_{A}-G_{AB} &\geqslant 0.
    \end{aligned}\end{equation}
   These conditions, again,  define two perpendicular planes.  In fact, the planes defined are the same as in the conditions for the symmetric positions, but the conditions are fulfilled precisely in the zone excluded by the conditions for (0,0) and (1,1).

   A fraction of $\frac{1}{4}$ of the possible games will then have two non-symmetric Nash equilibria, and none will have only one.

  \subsection{A plane representation}
   So far, the conditions do not involve the $G_{B}$ parameter at all, and therefore we can restrict ourselves to the $(G_{A},G_{AB})$ plane to consider NE alone.  The NE conditions for 00 and 11 define two orthogonal planes perpendicular to $G_{B}=0$.  Each divides the sphere in two halves, and the halves so defined overlap in $\frac{1}{4}$ of the sphere.

   The NE conditions for 01 and 10 define the same two planes mentioned above.   The region of the sphere they enclose is the quarter of the sphere where there is no symmetric NE.  Summing up, the fractions are:
   \begin{itemize}
    \item One symmetric NE: 1/2
    \item Two symmetric NE: 1/4
    \item Two nonsymmetric NE: 1/4
   \end{itemize}
 \section{PARETO OPTIMA (PO)}
  A situation is Pareto-optimum (PO from now on) when each player can improve his payoff, but doing so, he diminishes the other player's payoff \cite{Pareto}.

  One of the conditions for a certain choice situation to be a PO is clearly that \textbf{it is not a NE}. The other, concerning the other player's payoff maximization, can be viewed as the NE condition on the \textit{transposed} payoff matrix.  Transposing payoff matrix amounts to interchanging $G_{A}$ and $G_{B}$, so we already have them:
   \begin{equation}\label{ParetoConditions} \begin{aligned}
    s_1 G_{B}+ s_2 G_{AB} &\geqslant 0\\
    s_1 G_{B}+ s_3 G_{AB} &\geqslant 0.
   \end{aligned} \end{equation}
   These conditions share the same form with conditions \ref{NashConditions}, but refer to a different subspace.

  \subsection{The geometric complete picture}
   NE conditions generates planes defined by $(G_{A}\pm G_{AB}=0)$, and OP conditions generate planes defined by $(G_{B}\pm G_{AB}=0)$.  To achieve some symmetry in our partition of the sphere we need to cut the sphere with the plane $(G_{A}\pm G_{B}=0)$ as well.  This involves comparing one diagonal element of the payoff matrix with the other diagonal element, and comparing one nondiagonal element with the other nondiagonal element\vspace{8pt}

   Now we divided the sphere with these three pairs of orthogonal planes, getting 24 identical regions, which will allow us to compute easily the fraction of the sphere corresponding to certain set of such NE or PO conditions. Every region of the sphere characterized by NE and OP conditions will be conformed by some of these 24 small curve ``triangles'', which we will call \textit{elementary regions} from now on.
    \begin{figure}[htb]
     \center
     \includegraphics{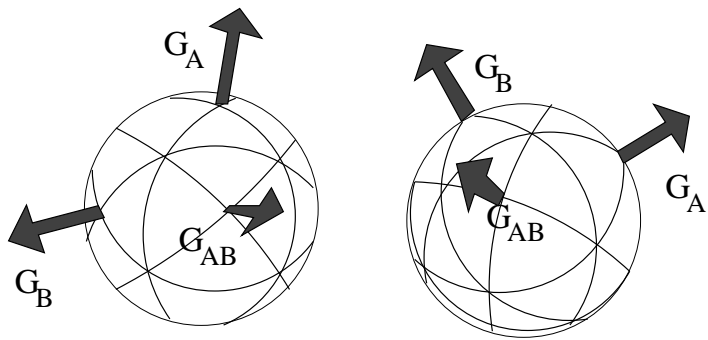}
     \caption{Partition of the Sphere}
    \end{figure}

  We can also use another norm to choose the set of game representatives which is going to give us a clearer picture.   If we take the higher absolute value of the elements of the payoff matrix as a norm, we are projecting the space of possible games into the \textbf{surface of the unit cube}.\vspace{8pt}

  The Nash and Pareto conditions for this cube will be all the diagonal planes cutting the cube in two equal halves. Then, it becomes apparent that all the areas are equal.  In figure\ref{Cube} this partition of the cube is shown, together with the unfolded surface. In the central square $G_{AB}>0$, while in the square formed by the four extreme triangles of the unfolded cube $G_{AB}<0$
  \begin{figure}[hbt]
   \center
   \includegraphics{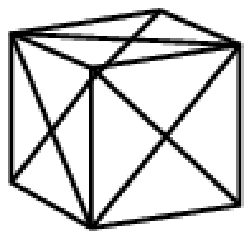}
   \includegraphics{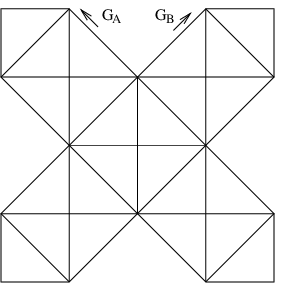}
   \caption{Partition of the Cube}\label{Cube}
  \end{figure}

  The projection on the cube works very well for strictly-ordered payoffs, but we have to be cautious to use with non-strictly ordered payoff.  If we impose a boundary to the differences of payoffs within a matrix, we are restricting the space of parameter to a cube.  This makes no difference for the fraction of strictly ordered payoffs games in a class, but for non-strictly-ordered games, the difference mentioned above is important.  Non-strictly-ordered games lie, in our representation, at the lines between triangles, and the fractions are given by the length of a straight line (for the cube) or by arc lengths (for the sphere).  There is, then, a slight difference between the representations.

  \section{A classification of 2x2 nonzero sum symmetric games}
   We can see the EN conditions  affecting $G_{AB}$ and $G_{A}$ and the PO conditions affecting $G_{AB}$ and $G_{B}$ as classification criteria, meaning the following for a certain position:
   \begin{itemize}
    \item If the position is NE it means that each player gets a maximum payoff \textbf{for himself}.
    \item If the position is  PO it means that each player produces a maximum payoff \textbf{for the other player}.
   \end{itemize}
    The importance of the first criteria is evident by itself, but the other criteria becomes important in cooperative game \cite{Threats} where players are allow to interact before choosing, or in dynamic games, where in some cases players learn that it can be advantageous to behave ``generously'' \cite{BehaviouralStrategies}\vspace{12pt}

    Normally the PO definition include the inverse of the NE condition (the PO payoff must be susceptible to be raised by any player) \cite{Pareto}.  But here we are going to try a less restrictive approach, taking the conditions on the transpose of the payoff matrices as \textit{independent} from the conditions of the actual payoff matrix.\vspace{12pt}

    Using these NE and PO conditions independently we can assign each region of the sphere a certain kind of game .  The NE and PO planes cut the sphere into 12 regions, with 3 different sizes. There are 8 triangular (elementary) regions covering each $\frac{1}{24}$ of the sphere, 4 bigger triangular regions covering each $\frac{1}{12}$ of the sphere (conformed by two elementary regions), and  2 square regions covering each $\frac{1}{6}$ of the sphere, conformed by 4 elementary regions.

    We should, however, take into account the arbitrariness of the labeling of strategies 0 and 1, and only distinguish diagonal (symmetric) positions where both players play the same strategy and nondiagonal (nonsymmetric) conditions where they don't.  Then we would have 9 kinds of games.\vspace{12pt}

    It can be wise, however, to discriminate whether a certain diagonal P0 position (pure or mixed) have higher or lower payoff than a certain NE position.  This give us 12 kinds of games.   This is exactly twice the number obtained by Rappoport and others for classes of 2x2 symmetric games \cite{Rapoport},\cite{Robinson}.

  \subsection{ROBINSON ORDER GRAPHS}\label{RobinsonNash}
   To organize the results in a simple way, We will use a very powerful tool presented in \cite{Robinson}.  This is a classification scheme for strictly ordered payoffs (where there are no two equal elements in the payoff matrix) where games are classified according to the ordering of the payoff elements.   In the diagram, a XY plot is drawn of A payoff vs. B payoff, locating the four points corresponding to the four positions.\vspace{12pt}

   The next step is connecting the points with arrows when a player can unilaterally shift from one to other.  The arrow points at the point where the payoff is higher for the involved player. If A's payoff is plotted in the Y axis, then A's arrows point allays upwards and B's arrows points allays to the right.  Doing so, we get an \textbf{order graph} where NE are points where arrows converge.\vspace{12pt}

   To find PO we draw arrows that point in the direction where the \textbf{other player}'s payoff is increased.   We will call these arrows \textit{Pareto arrows}.  The points fulfilling PC are the points where the Pareto arrows converge.Conversely, we will call the arrows pointing in the direction where the own payoff increases \textit{Nash arrows}.  This can be seen in figure \ref{Robinson1}\vspace{12pt}\\
  \begin{figure}[htb]\label{ExampleRobinson}
   \center
   \includegraphics{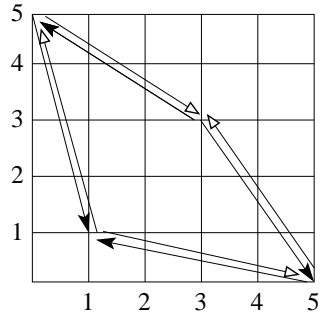}
   \caption{Robinson Diagram for the Prisoner's Dilemma}\label{Robinson1}
  \end{figure}
  If the payoff matrix $\begin{pmatrix}3 & 0 \\ 5 & 1 \end{pmatrix}$ is examined instead, the results are different.  In figure \ref{ExampleRobinson} we see Nash arrows as solid arrows, and the empty ones are Pareto arrows.   Nash Arrows are related to ordinality within one column, and Pareto Arrows to ordinality within one row.  It is the same approach used elsewhere to find Nash equilibria and Pareto optima \cite{LineaColumna}\vspace{12pt}

    A particular game of the kind we study is completely characterized with a simplified Robinson diagram, where we simply show the NE.  From the NE we can infer the direction of Nash arrows, and we get the Pareto arrows changing the direction of those arrows with a negative slope.

   Two examples of this simplified Robinson diagrams is in figure \ref{Robinson3}.   The simplified graphs in the figure allow us to distinguish the game with the  payoff matrix $\begin{pmatrix}4 & 1 \\ 5 & 0 \end{pmatrix}$ from that with a transpose payoff matrix $\begin{pmatrix}4 & 5 \\ 1 & 0 \end{pmatrix}$. The position (4,4) corresponding to the NE is shown as an empty circle in the first, as well as positions (1,5) and (5,1) in the second.  That allows us to infer the direction of the Nash arrows, all pointing at the NE.

  The Pareto arrows are inferred from the former in the following way:
  \begin{itemize}
   \item the arrows with a negative slope (those adjacent to the NE) must be reverted to be converted in Pareto arrows.
   \item The other Pareto arrows have the same direction as the correspondent Nash arrows.
  \end{itemize}
   In figure \ref{Robinson3} an example of this representation with Nash and Pareto arrows is presented.
  \begin{figure}[bht]
   \center
   \includegraphics{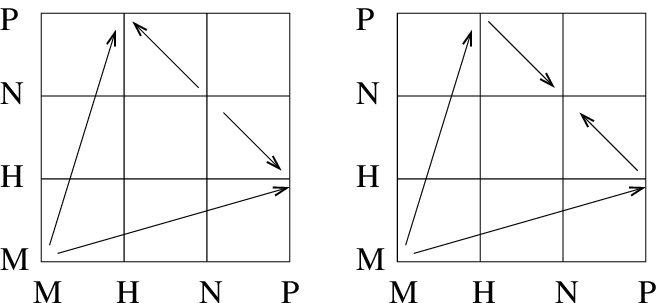} \hspace{6pt}
   \parbox{1cm}{$\longrightarrow$\vspace{1.5cm}\\}\hspace{6pt}
   \includegraphics{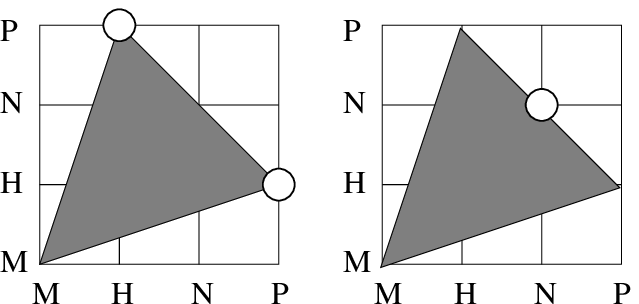}
   \caption{Simplified Robinson Diagrams}\label{Robinson3}
  \end{figure}

  \section{A LIST OF 2x2 NONZERO-SUM SYMMETRIC GAMES}
The game classes in this scheme are:
   \begin{enumerate}[1]
    \item One diagonal position is NE
    \begin{enumerate}[1]
     \item The same position is PO \hspace{12pt}{\tiny $\begin{pmatrix}4 & 2 \\ 3 & 1 \end{pmatrix}$}\hspace{5pt}\parbox{1cm}{\includegraphics{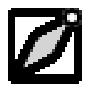}}

      These games are in two opposed pairs of adjacent elementary triangles in the cube.  The fraction is then {\large $\frac{1}{6}$} of the total.  Games in this class have an essentially unproblematic solution, for there is only one equilibrium, and it is very stable.
     \item The \textbf{other} diagonal position is PO\\
      These case also covers 4 triangles on the sphere.  The fraction is then {\large $\frac{1}{6}$} of the total.  Games in this class have an essentially unproblematic solution.
     Here we have two relevant situations: the ``generous'' PO can be more profitable than the NE itself, or the other way around.  This can be checked by the sign of $G_{A}+G_{B}$ If it is positive, then the payoff of 00 is higher, if it is negative, then it is lower.
     \begin{itemize}
      \item NC payoff is higher than the PC payoff {\tiny $\begin{pmatrix}3 & 4 \\ 1 & 2 \end{pmatrix}$}\hspace{10pt} \parbox{1cm}{\includegraphics{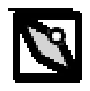}}

       This situation corresponds to one of the elementary regions conforming the rhomboidal regions: the outer one in the left graphic and the inner one in the right one. These are then {\large $\frac{1}{12}$} of all the games. 

      An example called ``\textbf{Hershey's kisses game}'' is mentioned in \cite{Kisses}.  Another is  ``\textbf{Deadlock}'' in \cite{Miracle}.  A game of this kind called ``exploitation of civil economy'' was proposed to model the decisions of scaling the conflict or negociating between the constitutional Armed Forces of Colombia and the Guerrilla groups \cite{WarGame}.

      \item NC payoff is lower than the PC payoff {\tiny $\begin{pmatrix}3 & 1 \\ 4 & 2 \end{pmatrix}$}\hspace{10pt} \parbox{1cm}{\includegraphics{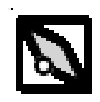}}
       This situation corresponds to the other 2 elementary regions in the two rhomboids: the inner one in the left graphic and the outer one in the right one. These also are {\large $\frac{1}{12}$} of all the games.

       In this class we find the famous ``\textbf{Prisoner's Dilemma}'' \cite{SymGames}.
     \end{itemize}
     \item The nondiagonal positions fulfills PC

       This kind of games have payoff matrices which are transpose of those of a ``\textbf{Chicken}-like'' game.   Then this kind of game also encloses $\frac{1}{12}$ of all the games. If we compare the payoff of the NE with the mixed symmetric position in the middle of the PO positions, we can assign each triangle a different kind of game:
     \begin{itemize}
      \item NE payoff is higher than the mixed diagonal PO payoff{\tiny $\begin{pmatrix}5 & 1 \\ 4 & 3 \end{pmatrix}$}\hspace{10pt} \parbox{1cm}{\includegraphics{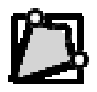}}
     Here $G_{A}>G_{AB}$ . This is shown as the left triangle in the graphic ($\frac{1}{24}$ of all the games)
      \item NE payoff is lower than the mixed diagonal PO payoff{\tiny $\begin{pmatrix}5 & 1 \\ 3 & 2 \end{pmatrix}$}\hspace{10pt} \parbox{1cm}{\includegraphics{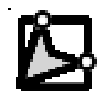}}
    Here $(-G_{A})>G_{AB}$. This is shown as the right triangle in the graphic ($\frac{1}{24}$ of all the games)
     \end{itemize}
     \item Both diagonal positions fulfill PC
      This is the case in  other two of the triangular regions (conformed by one elementary region each). This kind of games cover then $\frac{1}{12}$ of the sphere.
    \end{enumerate}
    \item Both diagonal positions fulfill NC
    \begin{enumerate}[1]
     \item One of the diagonal positions fulfill PC\\
      This is the case in two of the triangular regions (conformed by one elementary region each).This kind of games cover then $\frac{1}{12}$ of the sphere. In this class we find the so called \textbf{assurance game} \cite{Strategy}   But, again, we can distinguish two cases:
     \begin{itemize}
      \item The PC payoff is higher than the other diagonal payoff{\tiny $\begin{pmatrix}5 & 1 \\ 4 & 3 \end{pmatrix}$}\hspace{10pt} \parbox{1cm}{\includegraphics{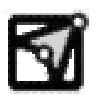}}
      These are $\frac{1}{24}$ of all the games (Upper triangle in the graphic)
      \item The PC payoff is lower than the other diagonal payoff{\tiny $\begin{pmatrix}5 & 1 \\ 4 & 2 \end{pmatrix}$}\hspace{10pt} \parbox{1cm}{\includegraphics{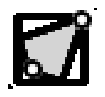}}
     These are $\frac{1}{24}$ of all the games (Lower triangle in the graphic) 
     \end{itemize}
     \item Both of the diagonal positions fulfill PC{\tiny $\begin{pmatrix}4 & 1 \\ 2 & 3 \end{pmatrix}$}\hspace{10pt} \parbox{1cm}{\includegraphics{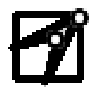}}

      This is the case in one of the square regions of the sphere (conformed by 4 elementary regions).  So, these are $\frac{1}{6}$ of all the games.   These can be problematic games, because there are two equilibria, both very stable.  Half of them, of course, will give higher payoff to one of the equilibria, and half of them to the other.  This can give the players a certain predilection.  But in the line between the two halves (shown in the figure) there are two equilibria, and we have a real dilemma for the players.  In this class we find the ``\textbf{Pareto Coordination}'' game \cite{Dictionary}. 
    \end{enumerate}
    \item Nondiagonal positions fulfill NC
    \begin{enumerate}[1]
     \item One diagonal position fulfills PC

      This is the case in two of the triangular regions ($\frac{1}{12}$ of all the games).   We find in this class the famous ``\textbf{Chicken game}``  \cite{SymGames}, \cite{Strategy},  \cite{Dictionary} also called ``\textbf{Stag-Hunt}'' \cite{Miracle}  We can, again, distinguish two possibilities:
     \begin{itemize}
      \item PO payoff is lower than the mixed diagonal NE payoff{\tiny $\begin{pmatrix}3 & 2 \\ 5 & 1 \end{pmatrix}$}\hspace{10pt} \parbox{1cm}{\includegraphics{R3251.eps}}
      This class encompasses $\frac{1}{24}$ of all the games. (Lower region in the graphic)
      \item PO payoff is higher than the mixed diagonal NE payoff{\tiny $\begin{pmatrix}4 & 2 \\ 5 & 1 \end{pmatrix}$}\hspace{10pt} \parbox{1cm}{\includegraphics{R4251.eps}}

      This class encompasses $\frac{1}{24}$ of all the games. (Upper region in the graphic)
     \end{itemize}
     \item Nondiagonal positions fulfill PC{\tiny $\begin{pmatrix}2 & 3 \\ 4 & 1 \end{pmatrix}$}\hspace{10pt} \parbox{1cm}{\includegraphics{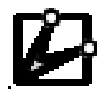}}

      This is the case in the other square region of the sphere ($\frac{1}{6}$ of all the games) These are also games with a nonproblematic solution, because there is a mixed symmetric NE which is extraordinarily stable.
    \end{enumerate}
   \end{enumerate}

  \section{CONCLUSIONS}
    The following image presents a map of the different kinds of games according to the Robinson classification, projected on the unfolded cube:
   \begin{figure}[htb]
    \center
    \includegraphics{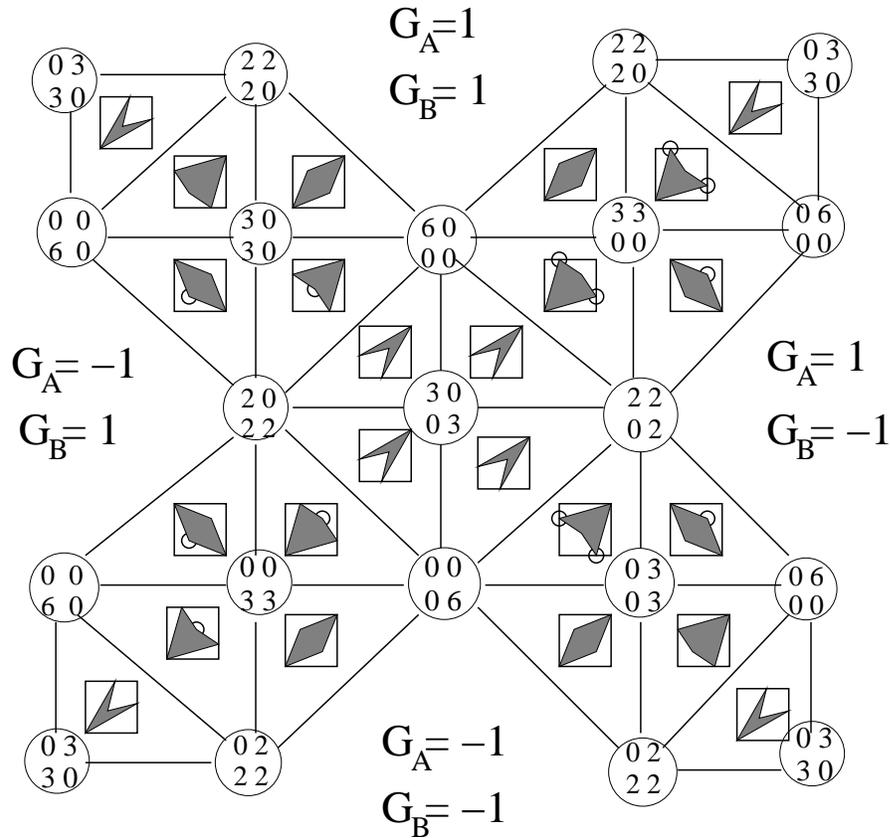}
    \caption{Zones of the Payoff-Matrix-Space on the Unit Cube}\label{classesmap}
   \end{figure}
  \subsection{HOW TO READ THE MAP}
   Each triangle in the map corresponds to a plane triangle, and all the variations of the payoff matrix are linear within it.  But the triangles can be in one out of 8 planes defined by the signs of the three coordinates.

   In every triangle a certain point corresponds to a convex combination of the payoff matrices written in the vertexes, weighted by their closeness.   Payoff matrices are not written in the normalized form, but in a easier-to-use integer form with entries from 0 to 12, to make easier to recognize the known games. 

   Examples:
   \textbf{PRISONER'S DILEMMA}
    \begin{equation}
     \begin{pmatrix}3 & 0 \\ 5 & 1\end{pmatrix} =
     \frac{9}{6}\left(\frac{2}{9}\begin{pmatrix}0 & 0 \\ 6 & 0\end{pmatrix}
     + \frac{4}{9}\begin{pmatrix}3 & 0 \\ 3 & 0\end{pmatrix}
     + \frac{3}{9}\begin{pmatrix}2 & 0 \\ 2 & 2\end{pmatrix} \right)
    \end{equation}
   The factor $\frac{9}{12}$ is to convert a game with a total payoff sum of 12 to one with a total payoff sum of 9.  Prisoner's Dilemma can thus be found in the map at $\frac{2}{9}$ from the upper-left vertex of the triangle, at $\frac{4}{9}$ from the upper-right vertex of the triangle, and at $\frac{3}{9}$ from the lower vertex of the triangle.

   \textbf{CHOLESTEROL: FRIEND OR FOE}
   This game has a payoff matrix that is between the ``cholesterol is healthy'' and the ``cholesterol is harmful''.   Each payoff matrix corresponds to ($G_A,G_B,G_{AB}$ vectors that are projected in different sides of the cube. (Good cholesterol in the (-,-,+) face and bad cholesterol in the (+,-,-) face).
   The decompositions are:
    \begin{equation}
     \begin{pmatrix}9 & 15 \\ 5 & 7\end{pmatrix} =
     \frac{16}{6}\left(\frac{3}{8}\begin{pmatrix}0 & 6 \\ 0 & 0\end{pmatrix}
     + \frac{3}{8}\begin{pmatrix}2 & 2 \\ 0 & 2\end{pmatrix}
     + \frac{2}{8}\begin{pmatrix}3 & 3 \\ 0 & 0\end{pmatrix} \right)
     + \begin{pmatrix}5 & 5 \\ 5 & 5\end{pmatrix}
    \end{equation}
   For cholesterol healthy and
    \begin{equation}
     \begin{pmatrix}-9 & -3 \\ -1 & 1\end{pmatrix} =
     \frac{6}{24}\left(\frac{2}{24}\begin{pmatrix}0 & 0 \\ 0 & 6\end{pmatrix}
     + \frac{4}{24}\begin{pmatrix}0 & 0 \\ 3 & 3\end{pmatrix}
     + \frac{18}{24}\begin{pmatrix}0 & 2 \\ 2 & 2\end{pmatrix}\right)
     + \begin{pmatrix}-9 & -9 \\ -9 & -9 \end{pmatrix}
    \end{equation}
  Nature would choose a point lying \textbf{between} this two points in the space of payoff matrices.    This line, when projected into the unit cube we used as a map, gives a trajectory of two straight segments joining with an angle. To compute this trajectory, all we need to do is to find the point where the the projections arrive at the edge where the faces of the cube join.

  From the decomposition of the payoff matrix, we can locate the extremal points of the game.   The distance to the vertexes of the triangle are given by the inverse of the coefficient (the higher coefficient, the nearer to the vertex).
 The trajectory from the two extremal payoff matrices for the game ``Cholesterol: friend or foe'' are ilustrated in figure \ref{Trajectory}.

   \begin{figure}[hbt]
    \center
    \includegraphics{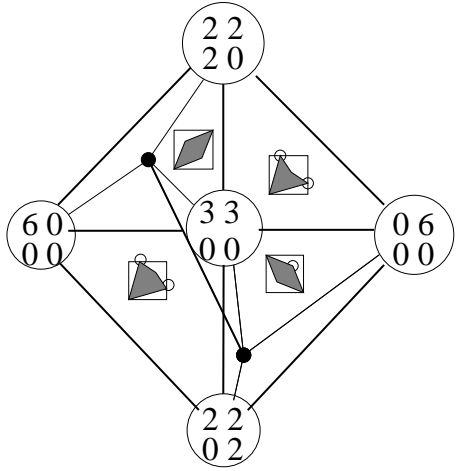}
    \caption{Position of ``Cholesterol: frien or foe'' for different p}
   \end{figure}

  In this map is all the information we need about the relations between classes of games, like these:
  \begin{itemize}
   \item The probability of getting a certain class with a random payoff matrix, is the number of cells occupied by this class, divided by 24 (the total number of cells)
   \item Distorting a payoff matrix we can change the class of the game.  The map shows us in which direction (in the G parameter space) a change from one class to another happens.
   \item The games with non-strictly-ordered payoff matrices are represented by the lines between classes, and can also be quantified by the length of the line. (one example is the ``\textbf{Friend-or-Foe game}'', a non-ordered-payoff game lying between the Prisoner's dilemma and the Chicken game \cite{FOF}\\
   Even though in the circle there seem to be two kinds of lines, long ones with unit length and short ones with length 1/$\sqrt{2}$, in the sphere all lines are 60 degrees long.
  \end{itemize}
   This map can be used as a starting point for the classification of the quantum games.   The quantum game with zero entanglement is equivalent to the classical game, so this map is useful for quantum games in that limit.   The questions that the following parts of the work must answer are the following:
   {\large \begin{itemize}
     \item Can two classes of games merge when the entanglement is introduced?
     \item Do new classes appear in quantum games that have no classical analogues in this kinds of games?
     \item Are some differences between classes emphasized or softened in the quantum games?
    \end{itemize}}

 \chapter[DETERMINISTIC STRATEGIES]{THE QUANTUM GAME WITH DETERMINISTIC STRATEGIES}
  
  In the Eisert scheme a "strategy" is an action the player performs on a quantum system, which can be entangled with the other player's system. \cite{Eisert}.  We are going to call \textit{semideterministic strategies} to those who does not involve randomness neither in the totally nonentangled nor in the totally entangled cases.

  These strategies can be represented by the identity and the Pauli matrices multiplied by i.  In what follows, we are going to call them $E$,$Z$,$Y$ and $X$:
  \begin{equation}
    I=\begin{pmatrix}1&0\\0&1\end{pmatrix}\hspace{20pt}
    Z=\begin{pmatrix}i&0\\0&-i\end{pmatrix}\hspace{20pt}
    Y=\begin{pmatrix}0&-1\\1&0\end{pmatrix}\hspace{20pt}
    X=\begin{pmatrix}0&i\\i&0\end{pmatrix}.
  \end{equation}
  They form a group with a product defined as the matrix product of the hermitian conjugate of the first times the second:
  \begin{equation}
    P:(S_i,S_j)\mapsto S_k \hspace{1cm}P(S_i,S_j)=S_i S_j.
  \end{equation}
  This group is isomorphic to the V group \cite{Vgroup}. The strategies of both players can be described as direct product of the strategies of each player, so the total group of strategies is indeed a product group.

  Some of the direct products of strategies will commute with the entangling unitary $J$, and thus will not be affected by its action, but some are mixed as if orthogonally transformed:
  \begin{equation}
   \begin{aligned}
   (I\otimes I),\hspace{16pt}(Z\otimes Z)&,\hspace{16pt}(Y\otimes Y)\hspace{16pt}(X\otimes X)\\
   (I\otimes X),\hspace{16pt}(X\otimes I)&,\hspace{16pt}(Y\otimes Z), \hspace{16pt}(Z\otimes Y).
   \end{aligned}
  \end{equation}
  The other possible products will be transformed in the following way:
  \begin{align}
   J^\dagger \begin{pmatrix}Z\otimes I\\Y\otimes X\end{pmatrix} J &=
   \begin{pmatrix} \sqrt{1-E} & \sqrt{E} \\-\sqrt{E} & \sqrt{1-E}\end{pmatrix}
   \begin{pmatrix}Z\otimes I\\Y\otimes X\end{pmatrix}\\
   J^\dagger \begin{pmatrix}I\otimes Z\\X\otimes Y\end{pmatrix} J &=
   \begin{pmatrix} \sqrt{1-E} & \sqrt{E} \\-\sqrt{E} & \sqrt{1-E}\end{pmatrix}
   \begin{pmatrix} I\otimes Z \\ X\otimes Y \end{pmatrix}\\
   J^\dagger \begin{pmatrix}Z\otimes X\\Y\otimes I\end{pmatrix} J &=
   \begin{pmatrix} \sqrt{1-E} & \sqrt{E} \\-\sqrt{E} & \sqrt{1-E}\end{pmatrix}
   \begin{pmatrix}Z\otimes X\\Y\otimes I\end{pmatrix}\\
   J^\dagger \begin{pmatrix}X\otimes Z\\I\otimes Y\end{pmatrix} J &=
   \begin{pmatrix} \sqrt{1-E} & \sqrt{E} \\-\sqrt{E} & \sqrt{1-E}\end{pmatrix}
   \begin{pmatrix} X\otimes Z \\ I\otimes Y \end{pmatrix}.
  \end{align}
  where we used the entangling transformation defined in equation \ref{entangling}.   The parameter $E$ is defined as $\sqrt{E}=2\gamma \sqrt{1-\gamma^2}$.

 \section{THE 4x4 EQUIVALENT GAME}
  Let's consider a symmetric nonzero-sum 2x2 game.  Without loss of generality we can describe it with a restricted payoff with null average payoff, and unit average square payoff:
  \begin{equation}\begin{aligned}
	 G_A=\begin{pmatrix} a & b \\ c & d \end{pmatrix}&\\ a+b+c+d=0 \hspace{2cm}& a^2+b^2+c^2+d^2=1.
  \end{aligned}\end{equation}
  The complete expression for the expected payoff for a player can be written as follows:
  \begin{equation}
   \langle G \rangle = Tr \left( J^\dagger(U_A\otimes U_B)^\dagger J \Pi_{00} J^\dagger(U_A\otimes U_B)J \hat{G} \right) 
  \end{equation}
  where
  \begin{equation}
   \Pi_{00} = \begin{pmatrix}1&0\\0&0\end{pmatrix} \otimes
   \begin{pmatrix}1&0\\0&0\end{pmatrix}\hspace{20pt}
   \hat{G_A} = \begin{pmatrix}a&0&0&0\\0&b&0&0\\0&0&c&0\\0&0&0&d\end{pmatrix}.
  \end{equation}

  If we construct an extended payoff matrix with the four strategies for each player, we get:
  \begin{equation}\label{ExtendedPayoff}
   \begin{matrix}
   & \begin{matrix} I_B & Z_B & Y_B & X_B \end{matrix}\\
   \begin{matrix} I_A \\ Z_A \\ Y_A \\ X_A \end{matrix} &
   \begin{pmatrix}a&a'&b'&b\\a'&a&b&b'\\ c'&c&d&d' \\ c&c'&d'&d\end{pmatrix}
   \end{matrix}
  \end{equation}
  where 
  \begin{equation}\begin{aligned}
   a' &= (1-E)a + E d \hspace{20pt}
   b' = (1-E)b + E c \\
   c' &= (1-E)c + E b \hspace{20pt}
   d' = (1-E)d + E a.
  \end{aligned}\end{equation}

  We can expect that constructing 4x4 payoff matrices with the same parameters of 2x2 payoff matrices plus one (the entanglement parameter) will lead us to a somehow symmetric result, ant it is in fact so.   The resulting payoff matrix has a remarkable symmetry: if we exchange strategies $I$ and $Z$ simultaneously for both players or if we exchange strategies $X$ and $Y$ simultaneously for both players, we get exactly the same payoff matrix.

  We can think of a two-stage classical game corresponding to this density matrix.   Each player chooses to play "classically" or "quantumly", and if they choose the opposite, then there is a probability $E$ that the referee transpose both columns and rows of the payoff matrix for the next stage of the game, where they use the same strategies 0 and 1.  

  We can regard the four strategies as being two choices each:
  \begin{enumerate}
   \item $I$: CLAS y  0
   \item $Z$: CUANT y 0
   \item $Y$: CUANT y 1
   \item $X$: CLAS y 1
  \end{enumerate}

   The tree representing the choices and payoffs in the classical game is shown in figure \ref{2x2tree},tougether with a \textbf{completely transposed game} with  a payoff matrix:
  \begin{equation}
   G^* = \begin{pmatrix}d & c \\ b & a \end{pmatrix}
  \end{equation}
  \begin{figure}[htb]
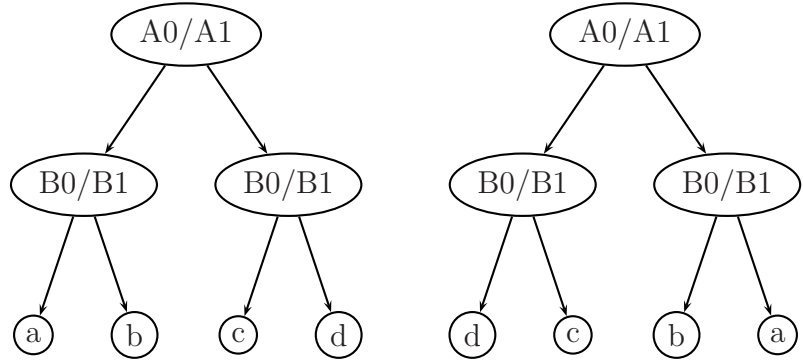

   \center
   \pstree[arrows=->]{\Toval{A0/A1}}{\pstree{\Toval{B0/B1}}{\Tcircle{a} \Tcircle{b}} \pstree{\Toval{B0/B1}}{\Tcircle{c} \Tcircle{d}}}\hspace{1cm}
   \pstree[arrows=->]{\Toval{A0/A1}}{\pstree{\Toval{B0/B1}}{\Tcircle{d} \Tcircle{c}} \pstree{\Toval{B0/B1}}{\Tcircle{b} \Tcircle{a}}}
   \caption{Extensive forms of a 2x2 game and its transposed version}\label{2x2tree}
  \end{figure}
   The semideterministic quantum game can be represented by a three-player game where one of the players is the entanglement introducing referee (figure \ref{DetTree}).  In the figure, Ac and Bc means A and B choosing to play classical moves, and A0 and A1 choosing the classical or quantum version of move 0 and move 1.

  \begin{figure}
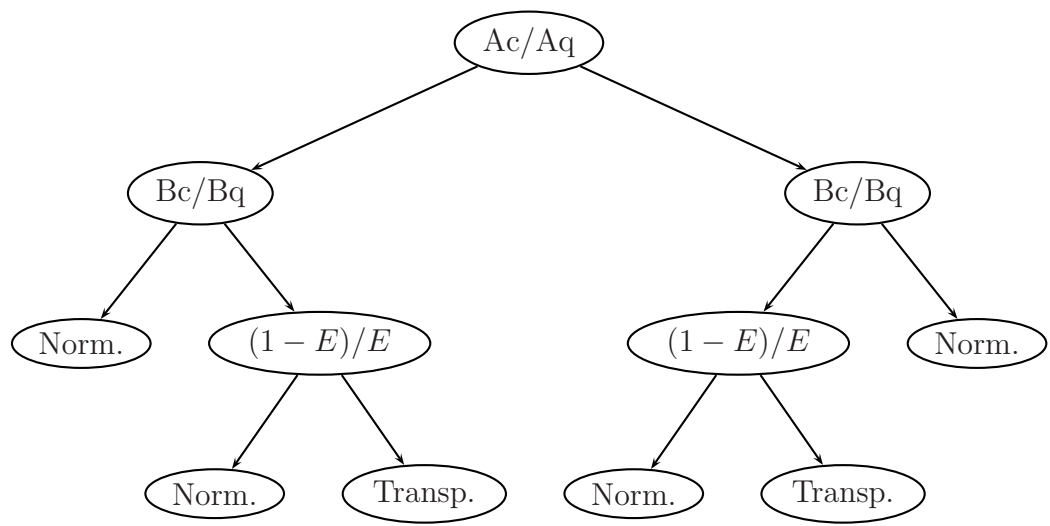

   \center
   \pstree[arrows=->]{\Toval{Ac/Aq}}{\pstree{\Toval{Bc/Bq}}{\Toval{Norm.} \pstree{\Toval{$(1-E)/E$}}{\Toval{Norm.} \Toval{Transp.}}} \pstree{\Toval{Bc/Bq}}{\pstree{\Toval{$(1-E)/E$}}{\Toval{Norm.} \Toval{Transp.}} \Toval{Norm.}}}
   \caption{Extensive form of a 2x2 deterministic quantum game}\label{DetTree}
  \end{figure}
   The deterministic game evolves as if a referee transposed the payoff matrix for the next stage of the game (with probability $E$) or left it unchanged (with probability $1-E$).   Observe that our probabilistic knowledge of what the referee does is similar to the knowledge of the action of nature in ``Cholesterol: friend or foe'' game described in section \ref{ExtensiveForm}.  Like in that game, we can reduce the game to a two-person game.

  \subsection{A PROBLEMATIC FEATURE}
  It must be noted that even though the Harsanyi procedure of including commitments in the game tree can remove the dilemma of choosing among equilibria, the extra choices introduced in the quantum game are indeed adding degeneracy to the game.  This happens because in the new payoff matrix every value appear twice, including the Nash equilibria and Pareto Optima, causing the player to have even more equilibria to choose.  This problem can be solved introducing some extra elements to the game, like treats and commitments \cite{Commitments}.

  \subsection{ORDER OF PAYOFFS IN THE EXTENDED GAME}
   The extended game is a symmetric nonzero-sum 4x4 classical game, with 16 positions corresponding to all possible strategy choices of the players.  However, the the payoff matrix has only 8 different values (each appearing twice), all of them depending only of three independent parameters: two from the classic game (section \ref{ClasParameters}), and the entanglement parameter $E$.\vspace{4pt}

   Four out of the eight parameters are convex combinations of the others with weights $E$ and $1-E$, so it the order of these will be determined by the order of the elements of the classical payoff matrix and the entanglement parameter.  For example, if $a>b$ and $d>c$, then it is always true that $a'>b'$ and $d'>c'$.  But not every ordering relations result in such a certainty.  For example, if $a>b$ and $c>d$, then the inequalities $a'>b'$ and $d'>c'$ hold or not hold depending on $E$.  It can be visualized in figure \ref{ordering}, where the vertical lines mark the values of $E$ and  $1-E$.

  \begin{figure}
   \center
   \psset{unit=18pt}
   \begin{pspicture}(0,0)(12,3)
    \pspolygon(0,0)(0,3)(5,3)(5,0)
    \uput[l](0,3){a}
    \uput[l](2,2){a'}
    \psline(0,3)(5,2)
    \uput[r](5,2){d}
    \uput[r](3,2){d'}
    \psline(0,0)(5,1)
    \uput[l](0,0){b}
    \uput[l](2,1){b'}
    \uput[r](5,1){c}
    \uput[r](3,1){c'}
    \psline(2,0)(2,3)
    \psline(3,0)(3,3)
    \uput[r](1,0){$E$}
    \uput[r](2,0){$(1-E)$}
    \pspolygon(7,0)(7,3)(12,3)(12,0)
    \uput[l](7,3){a}
    \uput[l](9,2){a'}
    \psline(7,3)(12,1)
    \uput[r](12,2){c}
    \uput[r](10,2){c'}
    \psline(7,0)(12,2)
    \uput[l](7,0){b}
    \uput[l](9,1){b'}
    \uput[r](12,1){d}
    \uput[r](10,1){d'}
    \psline(9,0)(9,3)
    \psline(10,0)(10,3)
    \uput[r](8,0){$E$}
    \uput[r](9,0){$(1-E)$}
   \end{pspicture}
   \caption{Payoffs vs. Entanglement degree}\label{ordering}
  \end{figure}
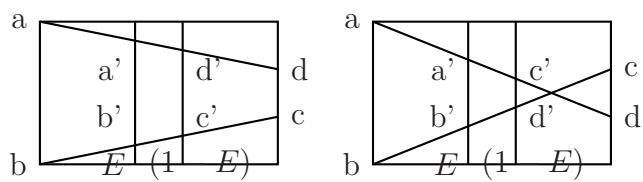
  It is clear from the drawing that in certain conditions there are qualitative changes in the game when changing the entanglement parameter $E$, for there is a change in the ordering of the different values in the payoff matrix.  This can make different entanglement regimes have different qualitative features.

 \section{AVAILABLE POSITIONS FOR THE PLAYERS}\label{Positions}
  To examine the consequences of the ordering in the extended payoff matrix we can use the Robinson order graphs, used in a previous work to classify the possible 2x2 classical games.

  In the quantum games with deterministic strategies, the eight different values mark eight nodes in the Robinson graph, corresponding to two positions each.  Then any Nash equilibrium or Pareto optimal position correspond to two positions of the payoff matrix.  We will have always an even number of Nash equilibria and Pareto optimal positions.

  \subsection{CONNECTIVITY}
   An important thing to be analyzed in the quantum game is the capability of each player to change the payoff from one possible value to another, given a certain strategy used by the other player.

   When we examine a node in the order graph corresponding to some position the connections (arrows) tell us which positions are available by unilateral actions.  In the quantum game we have more strategies than we have in the classical game, and there will be more possible payoffs to choose (given a certain strategy chosen by the other player).  Let's take another look at the payoff matrix:\
  \begin{equation*}
   \begin{matrix}
   & \begin{matrix} I_B & Z_B & Y_B & X_B \end{matrix}\\
   \begin{matrix} I_A \\ Z_A \\ Y_A \\ X_A \end{matrix} &
   \begin{pmatrix}a&a'&b'&b\\a'&a&b&b'\\ c'&c&d&d' \\ c&c'&d'&d\end{pmatrix}
   \end{matrix}
  \end{equation*}
  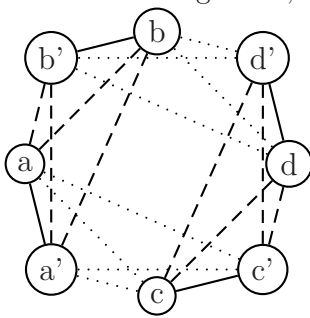
\begin{figure}[bht]
   \center
   \psset{unit=10pt}
   \begin{pspicture}(0,0)(10,10)
    \rput[cc](0,5){\circlenode{a}{a}}
    \rput[cc](1,9){\circlenode{b'}{b'}}
    \rput[cc](5,10){\circlenode{b}{b}}
    \rput[cc](9,9){\circlenode{d'}{d'}}
    \rput[cc](10,5){\circlenode{d}{d}}
    \rput[cc](9,1){\circlenode{c'}{c'}}
    \rput[cc](5,0){\circlenode{c}{c}}
    \rput[cc](1,1){\circlenode{a'}{a'}}
    \ncline{a}{a'}
    \ncline[linestyle=dashed]{a}{b}
    \ncline[linestyle=dashed]{a}{b'}
    \ncline[linestyle=dotted]{a}{c}
    \ncline[linestyle=dotted]{a}{c'}
    \ncline[linestyle=dashed]{a'}{b}
    \ncline[linestyle=dashed]{a'}{b'}
    \ncline[linestyle=dotted]{a'}{c}
    \ncline[linestyle=dotted]{a'}{c'}
    \ncline{b}{b'}
    \ncline[linestyle=dotted]{b}{d}
    \ncline[linestyle=dotted]{b}{d'}
    \ncline[linestyle=dotted]{b'}{d}
    \ncline[linestyle=dotted]{b'}{d'}
    \ncline{c}{c'}
    \ncline[linestyle=dashed]{c}{d}
    \ncline[linestyle=dashed]{c}{d'}
    \ncline[linestyle=dashed]{c'}{d}
    \ncline[linestyle=dashed]{c'}{d'}
    \ncline{d}{d'}
   \end{pspicture}
   \caption{Connections between payoff positions}\label{Connectivity}
  \end{figure}
   A given strategy of player A restricts us to a single row, and a given strategy of B restricts us to a single column.  Then we can conclude: \textit{player B can choose payoffs within a row, and player A can choose payoffs within a column}.

   In figure \ref{Connectivity} the connections associated with player A's moves (same column in the matrix) are represented by dotted lines, and player B's moves (same row in the matrix) are represented by dashed lines.  Solid lines represent double relations (A \textbf{and} B moves)}

   If we compare the connectivity graph of the quantum game with that of the classical game, we can see that each lines corresponding to each player in the classical game have become a \textit{complete subgraph} in the quantum game.
   \begin{itemize}
    \item \pstree[treemode=R]{\Tcircle{a}}{\Tcircle{b}} turns into \hspace{16pt}
     \parbox{5cm}{
     \begin{pspicture}(0,0)(5,2)
       \rput(0,1){\circlenode{a}{a}}
       \rput(2,2){\circlenode{a'}{a'}}
       \rput(3,0){\circlenode{b'}{b'}}
       \rput(5,1){\circlenode{b}{b}}
        \ncline{a}{b} \ncline{a}{a'} \ncline{a}{b'} \ncline{b}{b'} \ncline{a'}{b'} \ncline{a'}{b}
     \end{pspicture}}
    \item \parbox{3cm}{
     \pstree[treemode=R]{\Tcircle{a}}{\Tcircle{c}}} turns into \hspace{16pt}
     \parbox{5cm}{
     \begin{pspicture}(0,0)(5,2)
       \rput(0,1){\circlenode{a}{a}}
       \rput(2,2){\circlenode{a'}{a'}}
       \rput(3,0){\circlenode{c'}{c'}}
       \rput(5,1){\circlenode{c}{c}}
        \ncline{a}{c} \ncline{a}{a'} \ncline{a}{c'} \ncline{c}{c'} \ncline{a'}{c'} \ncline{a'}{c}
     \end{pspicture}}
   \end{itemize}
   and so forth.  Note that the first subgraph corresponds to moves of player B, and the second subgraph corresponds to moves of player A.\\
   The subgraph originated in $a-b$ is formed by three of the dashed lines in the complete graph, and share one solid line ($a-a'$) with the subgraph originated in $a-c$, and another solid line ($b-b'$) with the subgraph originated in $b-d$.

  \section{FROM ARROWS TO LATTICES}
   It is important to know the possibilities available for each player, but when that is clear, we still need to know what will she do in order to maximize her payoff within this set of possibilities.  What happen, then,  when we introduce strict order relations between the payoff parameters $a$,$b$,$c$,$d$?

   First of all, there are two order relations for the nodes, one a relation for player A (horizontal in the Robinson graphs) and a relation for player B (vertical in the Robinson graphs).  Each subgraph, with the corresponding order relations constitutes indeed a boolean lattice, the lattice of payoffs available for a given player for a given strategy chosen by the other player.

   The arrow that shows the way each player can improve his payoff in the Robinson graph of a classical game turns now into a lattice in the Robinson graph of the restricted quantum game, with a maximum payoff node which is going to be chosen by the player.
   \subsection{NASH EQUILIBRIA}
   Each point belongs to a lattice ``belonging'' to player A, and a lattice belonging to player B.  To be a Nash equilibrium, the point must be a supremum in both lattices.  So, we still can find Nash equilibria easily with the Robinson graphs as we did in the classical game (chapter \ref{RobinsonNash}).  The games where $a>b>c>d$ are a rather trivial example, but will allow us to show how the Robinson graphs look for a quantum restricted game:

   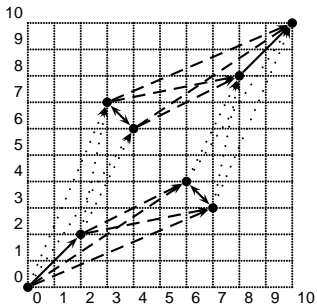
\begin{figure}[htb]
    \center
    \psset{unit=10pt}\grilla
    \begin{pspicture}(0,0)(10,10)
     \dotnode(10,10){a}
     \dotnode(3,7){b}
     \dotnode(7,3){c}
     \dotnode(0,0){d}
     \ncline[linestyle=dashed]{->}{b}{a}
     \ncline[linestyle=dotted]{->}{c}{a}
     \ncline[linestyle=dashed]{->}{d}{c}
     \ncline[linestyle=dotted]{->}{d}{b}
     \dotnode(8,8){a'}
     \dotnode(4,6){b'}
     \dotnode(6,4){c'}
     \dotnode(2,2){d'}
     \ncline[linestyle=dashed]{->}{b'}{a'}
     \ncline[linestyle=dotted]{->}{c'}{a'}
     \ncline[linestyle=dashed]{->}{d'}{c'}
     \ncline[linestyle=dotted]{->}{d'}{b'}
     \ncline{->}{a'}{a}
     \ncline{<->}{b'}{b}
     \ncline{<->}{c}{c'}
     \ncline{->}{d}{d'}
     \ncline[linestyle=dashed]{->}{b'}{a}
     \ncline[linestyle=dotted]{->}{c'}{a}
     \ncline[linestyle=dashed]{->}{d}{c'}
     \ncline[linestyle=dotted]{->}{d}{b'}
     \ncline[linestyle=dashed]{->}{b}{a'}
     \ncline[linestyle=dotted]{->}{c}{a'}
     \ncline[linestyle=dashed]{->}{d'}{c}
     \ncline[linestyle=dotted]{->}{d'}{b}
    \end{pspicture}
    \caption{Robinson Graph for a quantum game with deterministic strategies, where a>b>c>d}\label{QRobinson}
   \end{figure}
   The existence of the Nash equilibria is guaranteed for the classical game \cite{NashTheorem}.

   It is, however, not obvious how at least one supremum of a lattice corresponding to player A \textbf{must} correspond to the supremum of a lattice corresponding to player B.  But if we see the quantum deterministic-strategy game as a certain 4x4 classical game, the existence of at least Nash Equilibrium is again guaranteed by Nash Theorem.

  \subsection{PARETO OPTIMA}
   To find if a position is Pareto optimal, we can also draw the ``Pareto Lattices'' for both players.   A Pareto optimal position will be a supremum in the Pareto lattice of each player.   As in the Nash arrows and Pareto arrows, the Pareto and Nash lattices for one player have the same nodes, but some arrows can have different directions.  In the Robinson graphs, one of these lattices has a ``horizontal'' ordering (arrows pointing rightwards), the other has a ``vertical'' order (arrows pointing upwards).

 \section{REGIME TRANSITIONS}
  Every one among these payoff lattices involves two \textit{primed} parameters (like $a'$, $b'$, $c'$ and $d'$) which depend on a pair of payoff parameters (like $a$, $b$, $c$ and $d$) and the entanglement parameter $E$.  That means that for a given strict ordering of the payoff parameters they can undergo qualitative changes as we change the entanglement parameter, as a result of the order relations discussed above.

   Three different entanglement regimes can be defined then \textbf{for each lattice}: low, medium and high.  They are presented with an example, a trivial game with order relation $a>b>c>d$, and payoff matrix $(a,b,c,d)=(12,4,10,0)$.

  \subsection{THE LOW ENTANGLEMENT REGIME}
   In this regime the supremum of the lattice preserves its status.  There are no differences with the classical game, except that of the degeneracy of maxima.    In the example, the lower lattices (those who are not implied in the Nash equilibrium) are in this regime for $E<\frac{1}{3}$, and the upper lattice are always in it, because being the highest payoff, it is impossible that the supremum lose its character.
   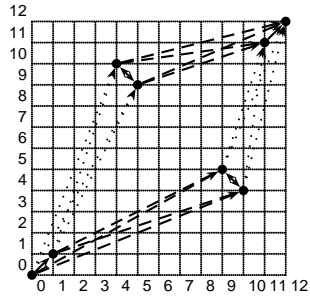
\begin{figure}[htb]
    \center
    \psset{unit=8pt}
     \begin{pspicture}(0,0)(12,12)\grilla
     \dotnode(12,12){a}
     \dotnode(4,10){b}
     \dotnode(10,4){c}
     \dotnode(0,0){d}
     \ncline[linestyle=dashed]{->}{b}{a}
     \ncline[linestyle=dotted]{->}{c}{a}
     \ncline[linestyle=dashed]{->}{d}{c}
     \ncline[linestyle=dotted]{->}{d}{b}
     \dotnode(11,11){a'}
     \dotnode(5,9){b'}
     \dotnode(9,5){c'}
     \dotnode(1,1){d'}
     \ncline[linestyle=dashed]{->}{b'}{a'}
     \ncline[linestyle=dotted]{->}{c'}{a'}
     \ncline[linestyle=dashed]{->}{d'}{c'}
     \ncline[linestyle=dotted]{->}{d'}{b'}
     \ncline{->}{a'}{a}
     \ncline{<->}{b'}{b}
     \ncline{<->}{c}{c'}
     \ncline{->}{d}{d'}
     \ncline[linestyle=dashed]{->}{b'}{a}
     \ncline[linestyle=dotted]{->}{c'}{a}
     \ncline[linestyle=dashed]{->}{d}{c'}
     \ncline[linestyle=dotted]{->}{d}{b'}
     \ncline[linestyle=dashed]{->}{b}{a'}
     \ncline[linestyle=dotted]{->}{c}{a'}
     \ncline[linestyle=dashed]{->}{d'}{c}
     \ncline[linestyle=dotted]{->}{d'}{b}
    \end{pspicture}
    \caption{Robinson Graph for Low Entanglement}
   \end{figure}
  \subsection{THE MEDIUM ENTANGLEMENT REGIME}
   In this regime, the supremum of the lattice is the primed node derived from the initial higher point in the classical lattice.  For example: if we have a $b\to a$ ($b<a$) order relation , now we have $b\to a \to a'$ ($b<a<a'$).  In this case no new Nash equilibria appears, but the payoff of the equilibrium increases with $E$.

\begin{figure}[hbt]
    \center
    \psset{unit=8pt}
     \begin{pspicture}(0,0)(12,12)\grilla
     \dotnode(12,12){a}
     \dotnode(2,8){b}
     \dotnode(8,2){c}
     \dotnode(0,0){d}
     \ncline[linestyle=dashed]{->}{b}{a}
     \ncline[linestyle=dotted]{->}{c}{a}
     \ncline[linestyle=dashed]{->}{d}{c}
     \ncline[linestyle=dotted]{->}{d}{b}
     \dotnode(10,10){a'}
     \dotnode(3,7){b'}
     \dotnode(7,3){c'}
     \dotnode(2,2){d'}
     \ncline[linestyle=dashed]{->}{b'}{a'}
     \ncline[linestyle=dotted]{->}{c'}{a'}
     \ncline[linestyle=dashed]{->}{d'}{c'}
     \ncline[linestyle=dotted]{->}{d'}{b'}
     \ncline{->}{a'}{a}
     \ncline{<->}{b'}{b}
     \ncline{<->}{c}{c'}
     \ncline{->}{d}{d'}
     \ncline[linestyle=dashed]{->}{b'}{a}
     \ncline[linestyle=dotted]{->}{c'}{a}
     \ncline[linestyle=dashed]{->}{d}{c'}
     \ncline[linestyle=dotted]{->}{d}{b'}
     \ncline[linestyle=dashed]{<-}{b}{a'}
     \ncline[linestyle=dotted]{<-}{c}{a'}
     \ncline[linestyle=dashed]{->}{d'}{c}
     \ncline[linestyle=dotted]{->}{d'}{b}
    \end{pspicture}
    \caption{Robinson Graph for Medium Entanglement}\label{MediumRobinson}
   \end{figure}
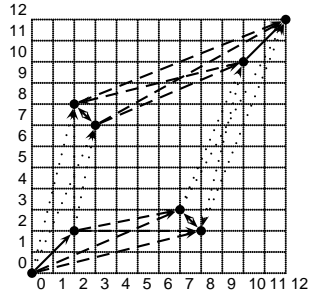

   In figure \ref{MediumRobinson}, the lower lattices are in this regime for $\frac{1}{6}<\sqrt{E}<\frac{1}{3}$. 
  \subsection{THE HIGH ENTANGLEMENT REGIME}
   In this regime, the supremum of the lattice is the primed node derived from the initial lower point in the classical lattice.  For example: if we have a $b\to a$ ($b<a$) order relation , now we have $b\to a \to b'$ ($b<a<b'$).  In the example, there appears a new Nash Equilibrium node in the Robinson graph, meaning two more Nash Equilibrium in the game.  This Nash equilibrium has a payoff of $12\sqrt{E}$.
   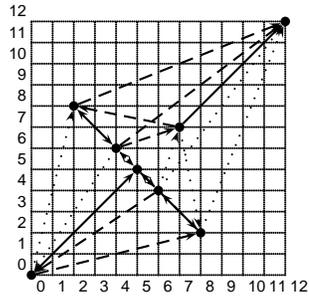
\begin{figure}[htb]
    \center
    \psset{unit=8pt}
     \begin{pspicture}(0,0)(12,12)\grilla
     \dotnode(12,12){a}
     \dotnode(2,8){b}
     \dotnode(8,2){c}
     \dotnode(0,0){d}
     \ncline[linestyle=dashed]{->}{b}{a}
     \ncline[linestyle=dotted]{->}{c}{a}
     \ncline[linestyle=dashed]{->}{d}{c}
     \ncline[linestyle=dotted]{->}{d}{b}
     \dotnode(7,7){a'}
     \dotnode(4,6){b'}
     \dotnode(6,4){c'}
     \dotnode(5,5){d'}
     \ncline[linestyle=dashed]{->}{b'}{a'}
     \ncline[linestyle=dotted]{->}{c'}{a'}
     \ncline[linestyle=dashed]{->}{d'}{c'}
     \ncline[linestyle=dotted]{->}{d'}{b'}
     \ncline{->}{a'}{a}
     \ncline{<->}{b'}{b}
     \ncline{<->}{c}{c'}
     \ncline{->}{d}{d'}
     \ncline[linestyle=dashed]{->}{b'}{a}
     \ncline[linestyle=dotted]{->}{c'}{a}
     \ncline[linestyle=dashed]{<-}{d}{c'}
     \ncline[linestyle=dotted]{<-}{d}{b'}
     \ncline[linestyle=dashed]{<-}{b}{a'}
     \ncline[linestyle=dotted]{<-}{c}{a'}
     \ncline[linestyle=dashed]{<-}{d'}{c}
     \ncline[linestyle=dotted]{<-}{d'}{b}
    \end{pspicture}
    \caption{Robinson Graph for High Entanglement}
   \end{figure}
 \section{CONCLUSIONS}
  At this point, we are able to give a number of general results, that even when they are restricted to deterministic strategies, will prove to be valid within bigger sets of strategies:
  \subsection{EQUIVALENT CLASSICAL GAME}
   The quantum nonzero-sum 2x2 symmetric game with deterministic strategies is equivalent to a two-stage symmetric game, where the second choice is a normal 2x2 game, but the first choice of both players, if anticorrelated, produces a double (column and row) transposition of the payoff matrix.

  \subsection{DEGENERACY OF THE EQUIVALENT 4x4 GAME}\label{DeterministicSym}
   The equivalent game is invariant under interchange of strategies 0 and Z or interchange of strategies X and Y made for both player.  This causes each payoff value to be repeated twice in the 4x4 payoff matrix in such a way that there are pairs of completely equivalent positions.

  \subsection{POSSIBILITY OF ENTANGLEMENT REGIMES}
   It is possible to change the qualitative features of the quantum version of some classical games by modulating the entanglement parameter.  The ordering among payoffs of one player for a fixed strategy of the other player (which we named the player's lattices) can change twice from the unentangled case to the totally entangled case.  We have named the possible entanglement regimes low, medium and high.  In the high entanglement regime, there can appear new pairs of Nash equilibria in some games.

   In all cases the position that becomes Nash equilibrium is a' if d>a or d' if a>d. The payoff for this position becomes (in the transition) higher than the higher non-diagonal classical payoff (b or c).  This allows us to write the condition like this: 
   \begin{equation}
    (1-E)\ min(a,d) + E\ max(a,d) > max(b,c)
   \end{equation}
   which means: 
   \begin{equation}
    E > \frac{max(b,c)-min(a,d)}{max(a,d)-min(a,d)}
   \end{equation}
    
   In terms of the G parameters defined in equation \ref{transformation}, these maxima and minima are easy to compute:
   
   \begin{multline}
    max(a,d)= \frac{1}{2}(|G_0+G_{AB}|+|G_A+G_B|)\\
    min(a,d)=\frac{1}{2}(|G_0+G_{AB}|-|G_A+G_B|)\\
    max(b,c)=\frac{1}{2}(|G_0-G_{AB}|+|G_A+G_B|).
   \end{multline}

    The transition value for E will be:
   \begin{equation*}
    E > 1-\frac{|G_0+G_{AB}|-|G_0-G_{AB}|}{2|G_A+G_B|};
   \end{equation*}

    Or, more briefly:
   \begin{equation}
    E > 1\frac{|min(G_0,G_{AB})|}{|G_A+G_B|}
   \end{equation}
    
   We know that $G_0$ can be set to 0, and that means that $G_{AB}$ \textbf{must be a negative number to allow the existence of an entanglement transition}.
    
  \section{RESULTS: REGIMES AND CLASSIFICATION}\label{DetermClass}
   In the following listing the different regimes appearing in different kinds of games are presented.   Each class of games is characterized by several payoff matrices with maximum payoff 1, minimum payoff 0, and two free parameters which can vary from 0 to 1.
   \begin{enumerate}[1]
    \item \textbf{One diagonal position is Nash Equilibrium}
    \begin{enumerate}[1]
     \item The same position is PO
      \begin{itemize}
       \item Classical Payoff Matrix: $\begin{pmatrix}1 & xy \\ x & 0 \end{pmatrix}$
       \item Quantum Semideterministic Payoff Matrix:
        {\small \begin{equation}
         \begin{pmatrix}
          1                 & 1-E               & (1-E)xy + Ex & xy\\
          1-E               & 1                 & xy           & (1-E)xy + Ex\\
          (1-E)x + Exy      & x                 & 0            & E\\
          y                 & (1-E)x + Exy      & E            & 0 \\
         \end{pmatrix}
        \end{equation}}
       \item Regimes: There is only one relevant transition, that takes place when $E=x(y-E(1-y))$.   The maximum value for E is 1, and this means that the transition is only possible when $2xy<1-x$.  Games in this class that do not fulfill this condition do not exhibit such transition.  In the high entanglement regime there appears a new Nash Equilibrium with payoff E, which is nevertheless less appealing than the initial one.

      It must be noted that even thought in the classical game the game with transposed payoff matrix has the same features, here the game with payoff matrix {\tiny $\begin{pmatrix}1 & x \\ xy & 0 \end{pmatrix}$} has this transition at E=x, that is, at higher payoff.  Then the transition is present in all such games

      This can be seen for a suitable example in the Robinson Graphs for low and high entanglement.  The Robinson Graph shows a new Nash Equilibrium for payoff {\tiny $\begin{pmatrix}6 & 1 \\ 3 & 0 \end{pmatrix}$} but not for {\tiny $\begin{pmatrix}6 & 3 \\ 1 & 0 \end{pmatrix}$}.
       \begin{center}\begin{figure}[htb]
        \center 
        \includegraphics{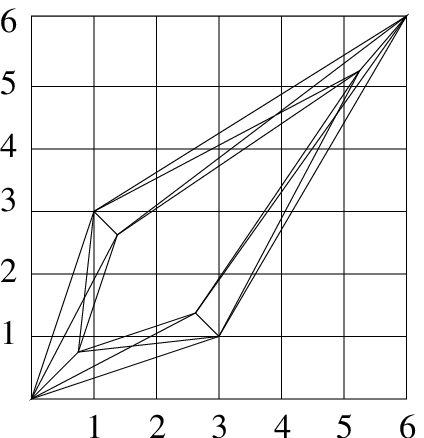}\includegraphics{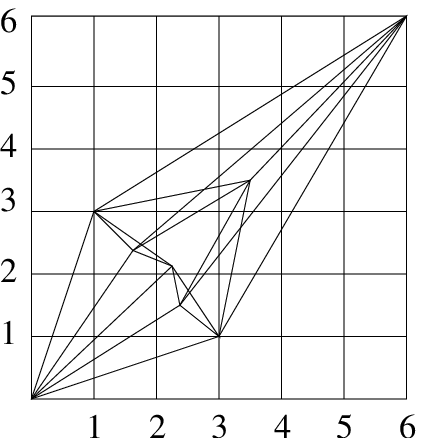}
        \caption{Diagonal position is NE and PO}
       \end{figure}\end{center}
      \end{itemize}
      \item The \textbf{other} diagonal position is PO. \hspace{8pt} There are two possibilities:
      \begin{enumerate}
       \item The Nash Equilibrium payoff is higher than the Pareto Optimal payoff.\\
      (Hershey's kisses, Deadlock)
       \begin{itemize}
        \item Classical Payoff Matrix: $\begin{pmatrix}x & 0 \\ 1 & xy \end{pmatrix}$
        \item Quantum Semideterministic Payoff Matrix:
         {\small \begin{equation}
          \begin{pmatrix}
          x            & x(1-E + yE) & 1-E         & 1\\
          x(1-E + yE)  & x           & 1           & 1-E\\
          E            & 0           & xy          & x(y(1-E)+E)\\
          0            & E           & x(y(1-E)+E) & xy\\         \end{pmatrix}
         \end{equation}}
        \item Regimes: In these games there is no transition at all.
       \end{itemize}
       \item The Pareto Optimal payoff is higher than the Nash Equilibrium payoff\\
        (\textbf{Prisoner's Dilemma})
       \begin{itemize}
        \item Classical Payoff Matrix: $\begin{pmatrix}x & 1 \\ 0 & xy \end{pmatrix}$
        \item Quantum Semideterministic Payoff Matrix:
         {\small \begin{equation}
          \begin{pmatrix}
           x            & x(1-E + yE) & E           & 0\\
           x(1-E + yE)  & x           & 0           & E\\
           1-E          & 1           & xy          & x(y(1-E)+E)\\
           1            & 1-E         & x(y(1-E)+E) & xy\\
          \end{pmatrix}
         \end{equation}}
        \item Regimes: Here we find another transition where a new equilibrium arises, in a similar way as the first case considered.  This happens at $E=x(y(1-E)+E)$.   In the known game "`Prisoner's Dilemma"' the scaled payoff matrix is defined by $x=\frac{3}{5}$ and $y=\frac{1}{3}$.  The transition is in $E=\frac{1}{3}$, the same result found by \cite{QTransition}

      For these games $G_{AB}=x+xy-1$, a positive or negative number according to the value of x an y.  Games with $G_{AB}>0$ will not have an entanglement transition, and games with  $G_{AB}<0$ will have one.

         In the image an example of either this or the above considered kind of games is illustrated:
        \begin{center}\begin{figure}[htb]
         \center
         \includegraphics{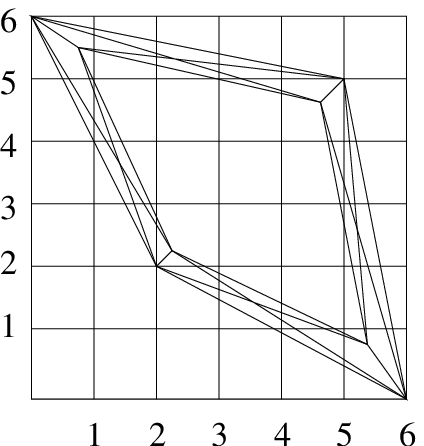}\includegraphics{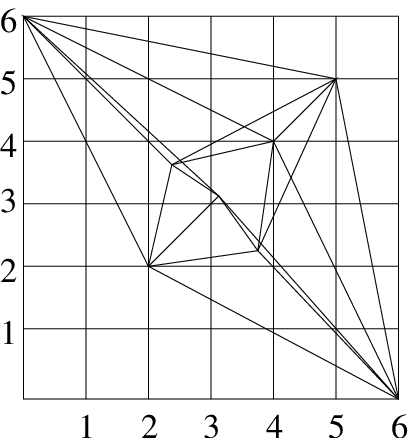}
         \caption{One diagonal position is NE and the other is PO}
        \end{figure}\end{center}
       \end{itemize}
       \item Both diagonal positions are Pareto Optimal
       \begin{itemize}
        \item Classical Payoff Matrix: $\begin{pmatrix}1 & x \\ 0 & xy \end{pmatrix}$
        \item Quantum Semideterministic Payoff Matrix:
         {\small \begin{equation}
          \begin{pmatrix}
          1            & 1-(1-xy)E   & x(1-E)      & x\\
          1-(1-xy)E    & 1           & x           & x(1-E)\\
          Ex           & 0           & xy          & xy+E(1-xy)\\
          0            & Ex          & xy+E(1-xy)  & xy\\
          \end{pmatrix}
         \end{equation}}
         \item Regimes: In these games we find another transition corresponding to the appearance of a new Nash Equilibrium.  This happens at $xy+E(1-xy)=x$
         \begin{center}\begin{figure}[htb]
          \center
          \includegraphics{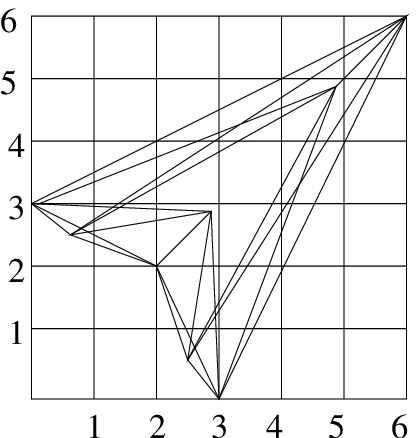}\includegraphics{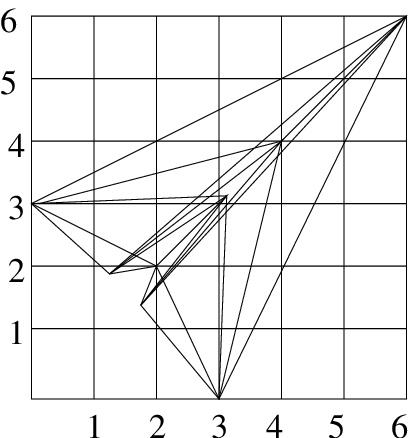}
          \caption{One diagonal position is NE and both are PO}
        \end{figure}\end{center}
       \end{itemize}
      \end{enumerate}
      \item \textbf{Both diagonal position are Nash Equilibria}
       There are two cases here:
      \begin{enumerate}
       \item One of the diagonal position is Pareto Optimal\\
        (Assurance Game, Stag-Hunt)
       \begin{itemize}
        \item Classical Payoff Matrix: $\begin{pmatrix}x & 1 \\ xy & 0 \end{pmatrix}$
        \item Quantum Semideterministic Payoff Matrix:
         {\small \begin{equation}
         \begin{pmatrix}
          x            & x(1-E)      & 1-E(1-xy)   & 1\\
          x(1-E)       & x           & 1           & 1-E(1-xy)\\
          xy+(1-xy)E   & xy          & 0           & Ex \\
          xy           & xy+(1-xy)E  & Ex          & 0\\
         \end{pmatrix}
         \end{equation}}
        \item Regimes: In these games there is no transition at all.
         \begin{center}\begin{figure}[htb]
         \center
         \includegraphics{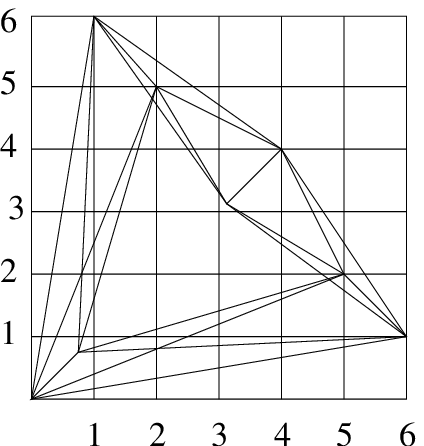}\includegraphics{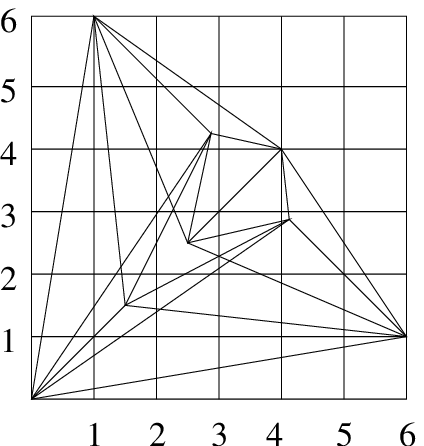}
         \caption{Both diagonal are Nash Equilibria, One is Pareto Optimal}
         \end{figure}\end{center}
        \item Both positions are Pareto Optimal
        \begin{itemize}
         \item Classical Payoff Matrix: $\begin{pmatrix}1 & xy\\ 0 & x \end{pmatrix}$
         \item Quantum Semideterministic Payoff Matrix:
          {\small \begin{equation}
          \begin{pmatrix}
          x            & x(1-E)      & 1-E(1-xy)   & 1\\
          x(1-E)       & x           & 1           & 1-E(1-xy)\\
          xy+(1-xy)E   & xy          & 0           & Ex \\
          xy           & xy+(1-xy)E  & Ex          & 0\\
          \end{pmatrix}
          \end{equation}}
         \item Regimes: In these games there is no transition at all.
         \begin{center}\begin{figure}[htb]
          \center
          \includegraphics{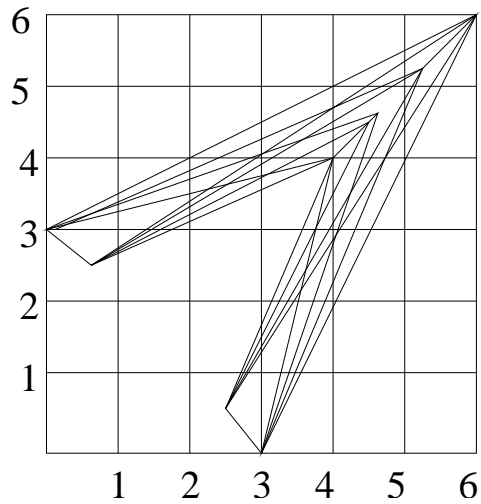}
          \caption{Both diagonal are Nash Equilibria and Pareto Optimal}
         \end{figure}\end{center}
        \end{itemize}
       \end{itemize}
      \end{enumerate}
      \item Nondiagonal positions are Nash Equilibria
      \begin{enumerate}
       \item One diagonal position is Pareto Optimal\\
        (Chicken Game)
        \begin{itemize}
         \item Classical Payoff Matrix: $\begin{pmatrix}x & xy \\ 1 & 0 \end{pmatrix}$
         \item Quantum Semideterministic Payoff Matrix:
          {\small \begin{equation}
          \begin{pmatrix}
          x            & x(1-E)      & xy+(1-xy)E   & xy\\
          x(1-E)       & x           & xy           & 1-E(1-xy)\\
          1-E(1-xy)    & 1           & 0            & Ex \\
          1            & 1-E(1-xy)   & Ex           & 0\\
          \end{pmatrix}
          \end{equation}}
         \item Regimes: In these games there is no transition
         \begin{center}\begin{figure}[htb]
          \center
          \includegraphics{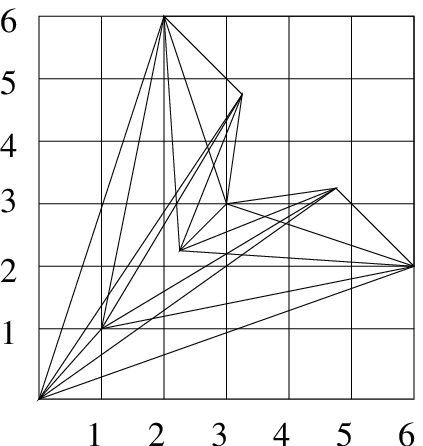}
          \caption{Both diagonal are Nash Equilibria and Pareto Optimal}
         \end{figure}\end{center}
        \end{itemize}
       \item Both diagonal positions are Pareto Optimal
        \begin{itemize}
         \item Classical Payoff Matrix: $\begin{pmatrix}xy & x \\ 1 & 0 \end{pmatrix}$
         \item Quantum Semideterministic Payoff Matrix:
          {\small \begin{equation}
          \begin{pmatrix}
          xy           & xy(1-E)     & E+x(1-E)     & x\\
          xy(1-E)      & xy          & x            & E+x(1-E)\\
          1-E(1-x)     & 1           & 0            & xyE\\
          1            & 1-E(1-x)    & xyE          & 0\\
          \end{pmatrix}
          \end{equation}}
         \item Regimes: In these games there is no transition
         \begin{center}\begin{figure}[htb]
          \center
          \includegraphics{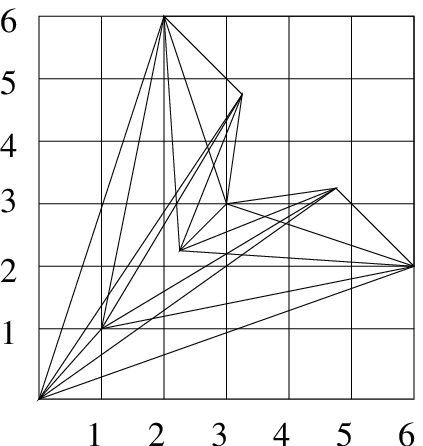}
          \caption{Both diagonal are Nash Equilibria and Pareto Optimal}
         \end{figure}\end{center}
        \end{itemize}
      \end{enumerate}
     \end{enumerate}
    \end{enumerate}

   We can locate in the game classes map (figure \ref{classesmap}) for example, the games that can have regime transitions for $E<\frac{1}{2}$  These are, according to the map, $\frac{1}{4}$ of all possible games (figure \ref{MapTrans}):
   \begin{figure}[hbt]
    \center
    \includegraphics{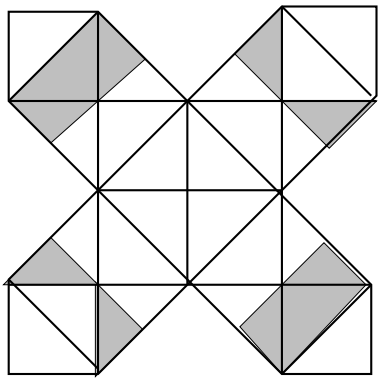}
    \caption{Entanglement Regimes in the payoff zones}\label{MapTrans}
   \end{figure}

  In this chapter we considered the smaller subspace of interesting strategies. The next step in this work, is to consider unitary strategies, and the first step is given in the next chapter, developing a methodology to obtain the pursued results in the space of unitary strategies.

 \chapter{A GEOMETRICAL APPROACH TO UNITARY STRATEGIES}
  
  In this chapter, a methodology to compute optimal response strategies for unitary strategies is presented.   It is a little of a digression from the subject of quantum games, but can be interesting for the reader that is interested in the quantum mechanics of entangled qubits.
 
 \section{TENSOR REPRESENTATION OF THE SEPARABLE DENSITY OPERATOR}
  Identity and Pauli matrices constitute a base for the space of operators acting on a bidimensional Hilbert space.   Then we can represent any linear operator acting on this space as a 4-vector with the coefficients of the basis matrix as elements.  That is:
  \begin{multline*}
\vec{O} = \frac{1}{2}(\langle 0\mid\hat{O}\mid0 \rangle +\langle 1\mid\hat{O}\mid1 \rangle,\langle 0\mid\hat{O}\mid1 \rangle +\langle 1\mid\hat{O}\mid0 \rangle,\\,i\langle 0\mid\hat{O}\mid1\rangle -i\langle 1\mid\hat{O}\mid0 \rangle,\langle 0\mid\hat{O}\mid0 \rangle -\langle 1\mid\hat{O}\mid1 \rangle),
  \end{multline*}
  Or:
  \begin{equation}
\vec{O} =\frac{1}{2}(Tr(\mathbb{I}\mathbb{O}),Tr(\sigma_x\mathbb{O}),Tr(\sigma_y\mathbb{O}),Tr(\sigma_z\mathbb{O})
  \end{equation}
  where:
  \begin{equation*}
\mathbb{O}_{i,j} = \sum(\langle\hbox{i}\mid\hat{O}\mid\hbox{i}\rangle).
  \end{equation*}
   If we study a composite system, we can represent the operators as rank 2 and dimension 4 tensors.   The tensor representation for separable and general linear operators acting on the total space is:
  \begin{align} 
(A\otimes B)_{\mu\nu} &= A_{\mu}B_{\nu}\\
O_{(spaces A and B)} &= Tr \left(\begin{pmatrix}
\mathbb{I}\\ \sigma_{z}\\ \sigma_{y}\\ \sigma_{x}
    \end{pmatrix}\begin{pmatrix}
\mathbb{I} & \sigma_{z} & \sigma_{y} & \sigma_{x}
    \end{pmatrix} \hat{O}\right). \label{RepSeparable}
  \end{align}
   Even if an operator is not separable it can be written as a sum of separable operators, and the same is valid for its tensor representation.

 \section{ENTANGLED STATES}
  A transformation generated by a direct product of Pauli matrices is completely nonlocal, in the sense that it can cause a maximal change in local information (defined as von Neumann entropy, length of Bloch vector of reduced density operator, etc).

  In the Eisert scheme the chosen entangling unitary is:
  \begin{equation}\label{TransformEnreda}
\hat{J}=\sqrt{1-\gamma^{2}}(\mathbb{I}\otimes\mathbb{I})+i\gamma(\sigma_{x}\otimes\sigma_{x}).
  \end{equation}

   We will use the notation  (ab)=$\sigma_{a}\otimes\sigma_{b}$ where $\sigma_{0}$ is identity:
  \begin{equation*}
\hat{J}=\sqrt{1-\gamma^{2}}(00)+i\gamma(xx).
  \end{equation*}

  Let us examine the effect of this transformation on the state  $\mid$ kl $\rangle$:
  \begin{equation*}
\hat{J}^{\dagger}\rho(kl)\hat{J}=(1-\gamma^{2})\rho\hbox{(kl)}+\gamma^{2}(xx)\rho\hbox{(kl)}(xx)\\+i\gamma\sqrt{1-\gamma^{2}}\left[\rho\hbox{(kl)},(xx)\right].
  \end{equation*}

  Using an explicit expression for $\rho$(kl):
  \begin{equation}
\rho\hbox{(kl)}=\frac{1}{2}((00)+(-1)^{k}(z0)+(-1)^{l}(0z)+(-1)^{k+l}(zz).
  \end{equation}

   The effect of the transformation J on the tensor can be deduced from the entangled deterministic strategies shown before:
   \begin{align*}
\hat{J}^{\dagger}(00)\hat{J}&=(00)
\\ \hat{J}^{\dagger}(n0)\hat{J}&=(1-2\gamma^{2})(n0)-i\gamma\sqrt{1-\gamma^{2}}((n\times x)x)
\\ \hat{J}^{\dagger}(0n)\hat{J}&=(1-2\gamma^{2})(0z)-i\gamma\sqrt{1-\gamma^{2}}(x(n\times x))
\\ \hat{J}^{\dagger}(nn)\hat{J}&=(nn).
   \end{align*}
  where n is $x$, $y$ or $z$.  We are particularly interested in those with $n=z$, because they appear in the density operator of the eigenstates of the payoff operator:
  \begin{align*}
\hat{J}^{\dagger}(00)\hat{J}&=(00)\\
\hat{J}^{\dagger}(z0)\hat{J}&=(1-2\gamma^{2})(z0)-i\gamma\sqrt{1-\gamma^{2}}(yx))\\
\hat{J}^{\dagger}(0z)\hat{J}&=(1-2\gamma^{2})(0z)-i\gamma\sqrt{1-\gamma^{2}}(xy))\\
\hat{J}^{\dagger}(zz)\hat{J}&=(zz).
   \end{align*}

   The tensor representation of the transformed initial state is:
   \begin{equation}\label{TensRhoEnredado}
\overleftrightarrow{\rho\hbox{(ij)}}=
    \begin{pmatrix}
1 & (-1)^{k}\sqrt{1-E} & 0 & 0\\
(-1)^{l}\sqrt{1-E} & (-1)^{k+l} & 0 & 0\\
0 & 0 & 0 & (-1)^{k+1}\sqrt{E} \\
0 & 0 & (-1)^{l+1}\sqrt{E} & 0 \\
    \end{pmatrix}
   \end{equation}
   where we defined:
   \begin{equation}\label{ParametroE}
E=4\gamma^{2}(1-\gamma^{2}).
   \end{equation}

 \section{REPRESENTATION FOR THE PAYOFF OPERATOR}
  The payoff function is represented in classical games in normal form as a matrix whose row is the state chosen by player A and the column that chosen by player B.   In the quantum game, we define the payoff as an operator, whose eigenvalues are the entries of the classical payoff matrix:
   \begin{equation}
\hat{G}=(00)G_{prom} +(z0)G_{A} +(0z)G_{B} +(zz)G_{AB}.
   \end{equation}
  where the parameters $G_{A}$, $G_{B}$ and $G_{AB}$ are the same defined before for the classical game in section \ref{ClassParameters}:
  \begin{itemize}
   \item $G_{A}$ is the difference in the own payoff when a player changes from strategy 0 to strategy 1
   \item $G_{B}$ is the difference in the other player's  payoff when a player changes from strategy 0 to strategy 1
   \item $G_{AB}$ is the difference in any player's payoff when both players change simultaneously from a coordinated strategy to an anticoordinated one.
   \item $G_{prom}$ is the average payoff for completely random playing by both players.  As we have seen before, it is completely irrelevant, and can be set equal to zero.
  \end{itemize}

  The tensor representing the non-entangled payoff operator is:
   \begin{equation}
   \overleftrightarrow{G}=
    \begin{pmatrix}
0 & G_{B} & 0 & 0\\
G_{A} & G_{AB} & 0 & 0\\
0 & 0 & 0 & 0\\
0 & 0 & 0 & 0\\
    \end{pmatrix}.
   \end{equation}

   It must be observed that this operator is not necessarily expressible as a direct product of partial operators.  In fact, the games where it can be are rather trivial both in the classical and quantum version.  The reason is that there must be some \textit{correlation} between the player A and player B payoff, for the game to be non-trivial.

   Entanglement acts on the tensor representation in a similar way to how it acts on the representation of density operator:
   \begin{equation}\label{TensGEnredado}
{\overleftrightarrow{G}}=
    \begin{pmatrix}
0 & \sqrt{1-E}G_{B} & 0 & 0\\
\sqrt{1-E}G_{A} & G_{AB} & 0 & 0\\
0 & 0 & 0 & \sqrt{E}G_{B} \\
0 & 0 & \sqrt{E}G_{A}  & 0\\
    \end{pmatrix}.
   \end{equation}

 \section{LOCAL UNITARY OPERATIONS}
  The action of the unitary operators on the tensor representation is very simple; much simpler than that of nonlocal transformations.  This can be easily seen on separable operators (tensor representation given in (\ref{RepSeparable}):
    \begin{align*}
(U_{A}^{\dagger} A U_{A}\otimes U_{B}^{\dagger} B U_{B})_{\mu\nu} &= (U_{A}^{\dagger}A U_{A})_{\mu}(U_{B}^{\dagger} B U_{B})_{\nu}\\
\hat{A} &= \langle A \rangle \mathbb{I}+\vec{A}\centerdot \vec{\sigma}\\
U_{A}^{\dagger} A U_{A} &= \langle A \rangle \mathbb{I}+\vec{A}\centerdot (U_{A}^{\dagger} \vec{\sigma} U_{A})\\
&= \langle A \rangle \mathbb{I}+\overleftrightarrow{R_{A}}\centerdot \vec{A}\centerdot \vec{\sigma}.
    \end{align*}

   The tensor $\overleftrightarrow{R_{A}}$ is simply the matrix representation of a SO(3) transformation.   To apply this result to the 4-vector representation we enlarge the dimension by one, and replacing the SO(3) matrix to  his SO(4) equivalent.
   \begin{equation}
(R_{A})_{\mu,\mu' }= 1 \oplus (R_{A})_{ij}.
   \end{equation}

   For the subspace of player B we proceed in a similar way
   \begin{equation}
(U_{A}^{\dagger} A U_{A}\otimes U_{A}^{\dagger} B U_{B})_{\mu\nu} = (R_{A})_{\mu,\mu' }(A\otimes B)_{\mu,\nu}(R_{B})_{\nu ,\nu'}.
   \end{equation}

   In the last equation all indexes are \textit{subindexes}, but there is no problem with that because the metric is identity.   It can be seen, anyway that the formula is that of a \textbf{matrix multiplication}.   There is only one detail regarding the representation of the rotation for B.  The subindex convention forces us to use \textit{the transpose of the rotation matrix}, and to always multiply $(\mathbb{R}_{B})^{T}$ on the right.
   \begin{equation}\label{TransLocalTensor}
U_{A}^{\dagger} \hat{A} U_{A}\otimes U_{A}^{\dagger} \hat{B} U_{B} = \mathbb{R}_{A}\overleftrightarrow{A\otimes B}(\mathbb{R}_{B})^{T}.
   \end{equation}
   
 \section{EXPECTED PAYOFF}
  In the tensor representation,  there is a very simple expression for the expected payoff:
  {\large
   \begin{equation}
    \langle G \rangle = \overleftrightarrow{G}:\overleftrightarrow{\rho}
   \end{equation}.
  }
   where $\overleftrightarrow{G}$ is given in (\ref{TensGEnredado}), and $\overleftrightarrow{\rho}$ is given in (\ref{TensRhoEnredado})  This contraction is equivalent to the trace of a product of matrices:
  {\large
   \begin{equation}
\langle G \rangle = (R_{A})_{\mu,\mu' }\rho_{\mu' ,\nu' }(R_{B})_{\nu' ,\nu}G_{\mu,\nu}.
   \end{equation}}

   This expression is valid for the payoff of player A.  In symmetric games, it is unimportant which player are we considering.   If we want to consider the \textit{other player} payoff, all we have to do is change $G_{A}$ by $G_{B}$ in the expression.

 \section{A SYMMETRY IN THE QUANTUM GAME}\label{symmetryQG}
  To simplify the problem, we can change the base used for the tensor representation, in a way that the representation of the payoff operator become diagonal in the $x,y,z$ subspace.   We achieve this multiplying by a rotation of 90 degrees in z by the right. (that means that we change the convention for the unitary strategies of B):
   \begin{equation}\label{ConvencionB}
    \begin{aligned}
    \begin{pmatrix}
1 & 0 & 0 & 0\\
0 & 1 & 0 & 0\\
0 & 0 & 1 & 0\\
0 & 0 & 0 & 1\\
     \end{pmatrix} &\mapsto  
    \begin{pmatrix}
1 & 0 & 0 & 0\\
0 & 0 & 1 & 0\\
0 & -1 & 0 & 0\\
0 & 0 & 0 & 0\\
     \end{pmatrix}\\
    \begin{pmatrix}
1 & 0 & 0 & 0\\
0 & 1 & 0 & 0\\
0 & 0 & 0 &-1\\
0 & 0 & 1 & 0\\
     \end{pmatrix} &\mapsto  
    \begin{pmatrix}
1 & 0 & 0 & 0\\
0 & 0 &-1 & 0\\
0 & 0 & 0 & 1\\
0 & 1 & 0 & 0\\
     \end{pmatrix}\\
    \begin{pmatrix}
1 & 0 & 0 & 0\\
0 & 0 & 0 & -1\\
0 & 0 & 1 & 0\\
0 & 1 & 0 & 0\\
     \end{pmatrix} &\mapsto  
    \begin{pmatrix}
1 & 0 & 0 & 1\\
0 & 0 & 0 & 0\\
0 & 1 & 0 & 0\\
0 & 0 &-1 & 0\\
     \end{pmatrix}\\
    \begin{pmatrix}
1 & 0 & 0 & 0\\
0 & 1 & 0 & 0\\
0 & 0 & 0 & -1\\
0 & 0 & 1 & 0\\
     \end{pmatrix} &\mapsto  
    \begin{pmatrix}
1 & 0 & 0 & 0\\
0 & 1 & 0 & 0\\
0 & 0 & 1 & 0\\
0 & 0 & 0 & 1\\
     \end{pmatrix}.
    \end{aligned}
   \end{equation}

   In this new base, the tensors will be:
{\small
   \begin{align*}
   \overleftrightarrow{\rho} &=
    \begin{pmatrix}
1 & \sqrt{1-E} & 0 & 0\\
\sqrt{1-E} & 1 & 0 & 0\\
0 & 0 & \sqrt{E} & 0\\
0 & 0 & 0 & -\sqrt{E}\\
    \end{pmatrix}\\
{\overleftrightarrow{G}} &=
    \begin{pmatrix}
0 & \sqrt{1-E}G_{B} & 0 & 0\\
\sqrt{1-E}G_{A} & G_{AB} & 0 & 0\\
0 & 0 & \sqrt{E}G_{B} & 0\\
0 & 0 & 0 & -\sqrt{E}G_{A} \\
    \end{pmatrix}.
   \end{align*}
}
   It is also possible to make a further change of base to make density operator be represented by the identity in the maximally entangled limit.  We achieve this multiplying by a 180 degree rotation about x on the left:
{\small
   \begin{equation*}
R_{x,180degrees} = 
    \begin{pmatrix}
1 & 0 & 0 & 0\\
0 & -1 & 0 & 0\\
0 & 0 & -1 & 0\\
0 & 0 & 0 & 1\\
     \end{pmatrix}.
   \end{equation*}
   \begin{subequations}\label{TensoresDiagonales}
    \begin{align}
\overleftrightarrow{\rho} &=
     \begin{pmatrix}
1 & \sqrt{1-E} & 0 & 0\\
-\sqrt{1-E} & -1 & 0 & 0\\
0 & 0 & -\sqrt{E} & 0\\
0 & 0 & 0 & -\sqrt{E}\\
     \end{pmatrix}\label{OpDensidad}\\
{\overleftrightarrow{G}} &=
     \begin{pmatrix}
0 & \sqrt{1-E}G_{B} & 0 & 0\\
-\sqrt{1-E}G_{A} & -G_{AB} & 0 & 0\\
0 & 0 & -\sqrt{E}G_{B} & 0\\
0 & 0 & 0 & -\sqrt{E}G_{A} \\
     \end{pmatrix}.\label{Ganancia}
    \end{align}
   \end{subequations}
}

    If we divide the tensor into blocks corresponding to subspaces \textit{(0Z)} and \textit{(XY)}, it becomes obvious that the tensor representation for the density operator is a multiple of the identity in the \textit{(XY)} block, and therefore is not affected by coordinated rotations in z:
    \begin{equation}
     \begin{pmatrix}
      \mathbb{I}_{2x2} & 0\\
      0 & \mathbb{R}
     \end{pmatrix}
     \begin{pmatrix}
      \mathbb{I}_{2x2} & 0\\
      0 & \sqrt{E}\mathbb{I}_{2x2}
     \end{pmatrix}
     \begin{pmatrix}
      \mathbb{I}_{2x2} & 0\\
      0 & \mathbb{R}^T
     \end{pmatrix}= 
     \begin{pmatrix}
      \mathbb{I}_{2x2} & 0\\
      0 & \sqrt{E}\mathbb{I}_{2x2}
     \end{pmatrix}.
    \end{equation}

     This means that there is a symmetry: \textbf{coordinated rotations in z do not affect the expected payoff} in this base.

{\large
   \begin{equation}\label{simetriaZ}
\mathbb{G}\mathbb{R}_{A} \mathbb{R}_{z}\rho\mathbb{R}_{z}^{T}(\mathbb{R}_{B})^{T}=\mathbb{G}\mathbb{R}_{A} \rho(\mathbb{R}_{B})^{T}
   \end{equation}
}

   This means that if we change the strategies of the players by multiplying by the mentioned matrices, we do not affect the payoff:
   \begin{equation}
    (R_x \mathbb{R}_A)\otimes((\mathbb{R}_B)^TR'_z)\equiv \mathbb{R}_A\otimes(\mathbb{R}_B)^T
   \end{equation}
   where $R_x$ is a 180 degree rotation about x and $R'_z$ is a 90 degree rotation about z

   To study with detail the expression for the expected payoff, it is convenient to translate the problem to matrices of dimension 3.   We achieve this dividing the 4-matrices in  1 dimensional and 3 dimensional blocks:
{\small
   \begin{subequations}\label{TensoresBloques}
    \begin{align}
\overleftrightarrow{\rho} &=
     \begin{pmatrix}
1 & \sqrt{1-E}\langle z \vert\\
-\sqrt{1-E}\vert z \rangle & -\mathbb{T}\\
     \end{pmatrix}\label{OpDensidad}\\
{\overleftrightarrow{G}} &=
     \begin{pmatrix}
0 & \sqrt{1-E}G_{B} \langle z \vert\\
-\sqrt{1-E}G_{A} \vert z \rangle  & -\mathbb{G}_{diag}\\
     \end{pmatrix}.\label{TGanancia3d}
    \end{align}
   \end{subequations}
}
   where \hspace{16pt} $\mathbb{T}=\begin{pmatrix}1 & 0 & 0\\ 0 & \sqrt{E} & 0\\ 0 & 0 & \sqrt{E} \end{pmatrix} $ y $\mathbb{G}=\begin{pmatrix}G_{AB}&0&0\\0&G_{A}\sqrt{E}&0\\0&0&G_{B}\sqrt{E}\end{pmatrix}$\vspace{12pt}

   When we compute the tensor product block by block we find the following expression for the expected value of payoff:
   \begin{equation*}
\langle G \rangle = (1-E)\bigl( G_{A} \langle z \vert \mathbb{R}_{A} \vert z \rangle
 + Tr(G_{B} \vert z \rangle \langle z \vert (\mathbb{R}_{B})^{T}) \bigr)
+Tr(\mathbb{R}_{A}\mathbb{T}(\mathbb{R}_{B})^{T}\mathbb{G}).
   \end{equation*}
  This can be written simply as the trace of a matrix:
   \begin{equation*}
\langle G \rangle = Tr((1-E)\bigl( G_{A} \mathbb{R}_{A} \vert z \rangle\langle z \vert -(G_{B} \vert z \rangle \langle z \vert (\mathbb{R}_{B})^{T} \bigr)
+\mathbb{R}_{A}\mathbb{T}(\mathbb{R}_{B})^{T}\mathbb{G}).
   \end{equation*}

   This expression can be simplified using the fact that $\vert z \rangle$ is an eigenvector of $\mathbb{G}$ with eigenvalue $G_{AB}$:
   \begin{multline}\label{ganancia3d}
\langle G \rangle = Tr(\bigl(\mathbb{R}_{A}-(1-E)\frac{G_{B}}{G_{AB}}  \vert z \rangle\langle z \vert \bigr)\mathbb{T}\bigl((\mathbb{R}_{B})^{T}+(1-E)\frac{G_{A}}{G_{AB}} \vert z \rangle\langle z \vert \bigr) \mathbb{G})\\ + (1-E)^{2}\frac{G_{A}G_{B}}{G_{AB}}.
   \end{multline}
 \section{EXTREMAL STRATEGIES}
   The extremum condition on a player's parameters can be computed constructing the rotations as exponentials of a sum of products of real parameters times the antisymmetric generators of SO(3):
   \begin{multline*}
    \mathbb{R}_{A}(\theta + \epsilon) = \mathbb{R}_{A}(\epsilon)\mathbb{R}_{A}(\theta)\\
    \mathbb{R}_{A}(\epsilon) = exp(\vec{\epsilon}\centerdot \vec{M}_{generators}).
   \end{multline*}

    Expanding around $\theta$:
   \begin{equation*}
    \mathbb{R}_{A}(\theta + \epsilon) = (\mathbb{I} +\vec{\epsilon}\centerdot \vec{M}_{generators}+O(\epsilon^{2}))\mathbb{R}_{A}(\theta).
   \end{equation*}

    The gradient of the expected payoff can be computed replacing the expansion in (\ref{ganancia3d}) and deriving:
   \begin{subequations}\label{GradG}
    \begin{align}
    \vec{\Delta}_{A}\langle G_{A} \rangle =  Tr(\vec{\mathbb{M}}_{generators}\mathbb{R}_{A}\mathbb{T}\bigl((\mathbb{R}_{B})^{T}+(1-E) \frac{G_{A}}{G_{AB}} \vert z \rangle\langle z \vert \bigr) \mathbb{G}_{A})\label{GradGA}\\
    \vec{\Delta}_{B}\langle G_{B} \rangle =  Tr(\vec{\mathbb{M}}_{generators}\mathbb{R}_{B}\mathbb{T}\bigl((\mathbb{R}_{A})^{T}+(1-E) \frac{G_{A}}{G_{AB}} \vert z \rangle\langle z \vert \bigr) (\mathbb{G}_{A})^{T}).\label{GradGB}
    \end{align}
   \end{subequations}
    For the second expression we used the transpose of (\ref{ganancia3d}).\\
    Let us define:
   \begin{equation}\label{MatrizA}
\mathbb{A}=\mathbb{T}\bigl((\mathbb{R}_{B})^{T}-(1-E)\frac{G_{A}}{G_{AB}} \vert z \rangle\langle z \vert \bigr) \mathbb{G}_{A}, \text{ and}
   \end{equation}
   \begin{equation}\label{MatrizB}
\mathbb{B}= \mathbb{T}\bigl((\mathbb{R}_{A})^{T}+(1-E)\frac{G_{A}}{G_{AB}} \vert z \rangle\langle z \vert \bigr)(\mathbb{G}_{B})^{T}.
   \end{equation}

    Matrix $\mathbb{A}$ depends solely on B's parameters and matrix $\mathbb{B}$ depends solely on A's parameters.

    The expression for the gradient is then:
   \begin{align}\label{GradienteG}
\vec{\Delta}_{A}\langle G_{A} \rangle &= Tr(\vec{\mathbb{M}}_{generators}\mathbb{R}_{A}\mathbb{A})\\
\vec{\Delta}_{B}\langle G_{B} \rangle &= Tr(\vec{\mathbb{M}}_{generators}\mathbb{R}_{B}\mathbb{B}).
   \end{align}

    The generating matrices of SO(3) constitute a complete basis for 3x3 antisymmetric matrices, therefore equating this gradient to zero amounts to say that the matrix multiplying the vector of generating matrix is \textbf{symmetric}\\
    Conclusion:

{\large\textbf{For a critical expected payoff, it is fulfilled that}
    \begin{equation}
      \mathbb{R}_{A}\mathbb{A}=\mathbb{A}^{T}\mathbb{R}_{A}^{T}\hspace{8pt}and\hspace{8pt}\mathbb{R}_{B}\mathbb{B}=\mathbb{B}^{T}\mathbb{R}_{B}^{T}
    \end{equation}}

     It must be noted that semideterministic strategies are represented by diagonal unitary matrices that commute or anticonmute with the conventional rotation introduced in Z.  These are then trivialy mutual critical strategies.

 \section{SOME EXTREMAL STRATEGIES}
  The form of the solutions to the extremum condition is rather simple, however difficult, in general, to compute.
   \begin{align*}
    \mathbb{R}_{A}^{eq} &=(\mathbb{A}^{T}\mathbb{A})^{1/2}\mathbb{A}^{-1}\\
    \mathbb{R}_{B}^{eq} &=(\mathbb{B}^{T}\mathbb{B})^{1/2}\mathbb{B}^{-1}.
   \end{align*}
   It is useful rewrite them in the following way:
   \begin{subequations}\label{soluciones}
    \begin{align}
     \mathbb{R}_{A}^{eq}\mathbb{A} &= (\mathbb{A}^{T}\mathbb{A})^{1/2}\label{SolA}\\
     \label{SolB}
     \mathbb{R}_{B}^{eq}\mathbb{B} &= (\mathbb{B}^{T}\mathbb{B})^{1/2}.
    \end{align}
   \end{subequations}

    The matrix $\mathbb{A}$ can be written as the product of three matrices, two orthogonal and a diagonal:
   \begin{equation*}
    \begin{split}
     \mathbb{A} =\mathbb{O}_1\mathbb{D}_{A}\mathbb{O}_2\\
     (\mathbb{A}^{T}\mathbb{A})^{1/2} = ((\mathbb{O}_2)^{T}\mathbb{D}_{A}\mathbb{O}1^{T}\mathbb{O}_1\mathbb{D}_{A}\mathbb{O}_1)^{1/2} = (\mathbb{O}_1)^{T}(\mathbb{D}_{A}\mathbb{D}_{A})^{1/2}\mathbb{O}_2.
    \end{split}
   \end{equation*}
   The extremum condition is then:
   \begin{equation}\label{diagonalA}
    \begin{split}
    \mathbb{R}_{A}\mathbb{O}_1\mathbb{D}_{A}\mathbb{O}_2 = (\mathbb{O}_2)^{T}(\mathbb{D}_{A}\mathbb{D}_{A})^{1/2}\mathbb{O}_2\\
    \mathbb{O}_2\mathbb{R}_{A}\mathbb{O}1\mathbb{D}_{A}\mathbb{O}_2(\mathbb{O}_2)^{T} = (\mathbb{D}_{A}\mathbb{D}_{A})^{1/2}\\
 \mathbb{O}_2\mathbb{R}_{A}\mathbb{O}_1\mathbb{D}_{A}= (\mathbb{D}_{A}\mathbb{D}_{A})^{1/2}.
    \end{split}
   \end{equation}

    Then $O_2\mathbb{R}_{A}O_1$ who must be diagonal in a solution.

    The SO(3) diagonal matrices can be written in the following way:
   \begin{equation}\label{RotacionDiagonal}
    \begin{pmatrix}
    s_{1}s_{2}&0&0\\0&s_{2}&0\\0&0&s_{1}
    \end{pmatrix}
   \end{equation}
    where $s_{1}=\pm1$ and $s_{2}=\pm1$.

    Given a strategy $R_{B}$ chosen by B, there are 4 \textit{critical strategies} that A can choose:
   \begin{equation}\label{SolucionesA}
    R_{A}(s_{1},s_{2})= \hspace{10pt}(O_2)^T\hspace{10pt}
    \begin{pmatrix}
    s_{1}s_{2}&0&0\\0&s_{2}&0\\0&0&s_{1}
    \end{pmatrix}\hspace{10pt} (O_2)^T
   \end{equation}
   where $O_1$ and $O_2$ ``diagonalize'' $\mathbb{A}$:  
   \begin{equation*}
    (O_2)^T\hspace{10pt} \mathbb{A} \hspace{10pt} (O_1)^T \hspace{10pt}= Diagonal.
   \end{equation*}

    It is important to characterize the extremal strategies, that is, find out if they are maxima, minima or saddle point.   Taking the expression (\ref{GradienteGanancia}) and deriving again, we find the Hessian:
    \begin{equation}
	   H_{ij} = Tr(\mathbb{M}_{i}\mathbb{M}_{j}\mathbb{R}_{A}\mathbb{A}).
    \end{equation}

     The product of the generators of SU(3) can be computed very easily using the Levi-Civita tensor:
    \begin{multline*}
	   (\mathbb{M}_{i})_{jk} = \epsilon_{jik}\\
	   (\mathbb{M}_{i}\mathbb{M}_{j})_{kl} = \epsilon_{kil}\epsilon_{kjl} = \delta_{ik}\delta_{jl} - \delta_{ij}\delta_{kl}.
    \end{multline*}

     The Hessian is the trace of a product, taken on the indexes \textit{j} and \textit{k}, and the expression left for the Hessian is:
    \begin{equation}\label{FormaHessiana}
	   H_{ij} = \bigl(\mathbb{R}_{A}\mathbb{A}-\mathbb{I}Tr(\mathbb{R}_{A}\mathbb{A})\bigr)_{ij}.
    \end{equation}

     If we replace $\mathbb{A}$ and $\mathbb{R}_A$ by their decomposed expression, we get:
    \begin{equation}\label{FormaHessiana}
	   \mathbb{H} = (O_2)^T\bigl(\mathbb{D}_R \mathbb{D}_A-\mathbb{I}Tr(\mathbb{D}_R \mathbb{D}_A)\bigr)O_2.
    \end{equation}

     Now we can conclude that the diagonal matrix $\mathbb{D}_A$ has the same eigenvalues of $\mathbb{A}$ in absolute value, with the signs $s_1 s_2$, $s_2$ y $s_1$ respectively.  These signs characterize the critical response $\mathbb{R}_A$.\vspace{10pt}

     The eigenvalues of the Hessian are then sums of two of the eigenvalues of $\mathbb{A}$ with signs $-s_2$ and $-s_1$ for the first, $-s_1$ and $-s_1s_2$ for the second, and $-s_2$ and $-s_1s_2$ for the third.

     Whatever the signs of the eigenvalues of $\mathbb{A}$, we can get any sign for the curvatures.   We can get maximum, minimum or saddlepoints with this criterion.   However, when the matrix $\mathbb{A}$ is \textbf{degenerated} some of the curvatures can be null.   This will be studied later.

 \chapter{CRITICAL RESPONSES IN QUANTUM 2x2 GAMES}
  
 In chapter \ref{OptimalNash} we introduced the concept of Optimal responses, and in the last chapter, the concept of critical responses, in the field of unitary local transformations of entangled qubits.  In this chapter, we apply this to Quantum Game Theory.
 \section{REVIEW OF SOME CONCEPTS}
  As we had shown before \cite{Quadratic}, quantum games share some common features with quadratic games, differing in the fact that the former include a quadratic constraint restricting the strategy space.

  The quantum games can thus be considered as belonging to the class of games with countably infinite strategies whose payoff is a smooth function of some parameters labeling the strategies.   We can, nonetheless, use a concept developed for general games with countably infinite strategies in which the payoff is a smooth function of the parameters that label the strategies. to characterize a quantum game: the concept of critical response.  This is based in the concept of \textbf{optimal response} (Definition \ref{DefOptResp}) used in the theory of infinite games \cite{NashResponse}.

  A strategy which gives a maximum, minimum or saddle-point payoff for a given fixed strategy of the other player is called a \textbf{critical response} to the other player's strategy.  A \textbf{Nash equilibrium} (Definition \ref{DefNashEq}) can thus be defined as a pair of strategies that are mutual optimum (maximal) responses in the own payoff function, and a \textbf{Pareto optimum} (Definition \ref{DefParetoOp}) as a pair of strategies which are mutual optimum (maximal) responses in \textit{the other player's payoff function} \cite{NashResponse}.

  \subsection{THE PROBLEM OF STABILITY}
   These two concepts (Nash equilibrium and Pareto optimum) give us situations which are stable in the sense that  players with a certain kind of rationality will not depart from such positions; those who maximize their own payoff will not depart from Nash equilibria and those who maximize the other player's payoff will not depart from Pareto optima.

   But, what happens when players choose with ``trembling hands'' \cite{Trembling}, that is, when there is some variability of strategies, some limited rationality, curiosity, or other factor that makes players use some strategies different from the ``rational'' choices?   Will Nash equilibria and Pareto Optima be stable under slight departures from rationality?

   This is a very simple problem when we deal with a few pure strategies and a set of mixed strategies which are convex combinations of the former.  But, as we will see, it turns out to be not so simple when we deal with an infinite set of strategies. 

 \section{CONVEXITY AND CONCAVITY}
   The most important class games with infinite strategies is the class of \textit{continuous games}.  The defining characteristic of these games is that the strategies can be labeled with vectors of real numbers such that the payoff is, for both players, a continuous function of the label vectors of each one of them.
   Within the class of continuous games there are two classes that have been largely studied: \textit{convex} games and \textit{concave} games \cite{Continuous}.  These are defined by the condition:
 \begin{equation}\label{ConcaveConvex}\begin{aligned}
  s\left(\alpha G(\vec{a},\vec{b}) + (1-\alpha) G(\vec{a'},\vec{b})\right) &\geqslant s\left(G(\alpha \vec{a} + (1-\alpha) \vec{a'},\vec{b})\right)\\
  s\left(\alpha G(\vec{a},\vec{b}) + (1-\alpha) G(\vec{a},\vec{b'})\right) &\geqslant s\left(G(\vec{a}, {\alpha \vec{b} + (1-\alpha) \vec{b'}})\right)
 \end{aligned}\end{equation}
 Where G is the payoff, $\vec{a},\vec{b}$ are the vectors that label the strategies, and $\alpha$ is a real number between 0 and 1.

   \begin{definition}
    A \textbf{convex game} is a game whose payoff fulfills the conditions \ref{ConcaveConvex} with $s=-1$
   \end{definition}
   \begin{definition}
    A \textbf{cocave game} is a game whose payoff fulfills the conditions \ref{ConcaveConvex} with $s=1$
   \end{definition}
 Later we will see that a quantum game is neither completely convex nor completely concave, but can be described either way under certain restrictions of the strategies.
 \section{THE TENSOR FORMULA FOR THE EXPECTED PAYOFF}
   In chapter 3 (THE EISERT SCHEME) a tensor formula for the expected payoff was developed based on the $\chi$ matrix representation of the quantum operations:
 \begin{equation}\label{ChiPayoff}
  \langle \$ \rangle = \sum_{i,j,k,l}(\chi_A)_{i,j} (\chi_B)_{k,l} P^{i,j,k,l}
 \end{equation}
  where $\chi$ is a 4x4 hermitian matrix containing the parameters of each player.

   For any unital map acting on qubit density operators, the $\chi$ matrix can be written as a convex combination of matrices corresponding to unitary transformation.  The expected payoff can therefore as the sum of a number of weighted ``unitary payoffs'', just as the classical expected payoff can be seen as a sum of weighted ``pure-strategy payoffs''.

   As it has been stressed repeatedly, the core of any fundamentally quantum behaviour lies in the interference of amplitudes \cite{FeynmanRules}, and this is manifest only \textbf{within} the unitary terms.  This is, for the moment, an assumption, but will prove later to be true. The assumption is this: \textbf{A critical response is always unitary}

  \section{MUTUALLY CRITICAL STRATEGIES}
  The main aim of this work is to classify the quantum games according to Nash Equilibria and Pareto optimal positions, and the critical strategies can be used as pathways to them. As it was already stated, a Nash Equilibrium is formed by two mutually optimal strategies, and a Pareto optimal is formed by two mutually optimal(in a pareto sense, with transpose payoff matrix) strategies as well.

  Supose Alice and Bob are in a Nash equilibrium position. A slight unilateral departure from this position by Alice will cause Bob to react in a certain way.  It is possible that Alice's best response to Bob is returning towards the equilibrium position (stable equilibrium) or to move away from it (unstable equilibrium).

  \begin{definition}
   A \textbf{stable equilibrium} is a position from which a slight departure in the space of strategies of one player causes, by optimal response, a strategy from the other player that compells the departing player to get back to the initial equilibrium strategy.
  \end{definition}
  
  The optimal strategy map gives us the tools to examine this feature of Nash Equilibrium, by computing a \textit{procuct map}.

  \subsection{PRODUCT MAPS}
   A certain critical strategy is a map from the set of strategies of one player to the set of strategies of the other:
   \begin{equation}\begin{aligned}
    \tilde{C}_B(s1_B,s2_B):\chi_A &\mapsto \chi_B\\
    \tilde{C}_A(s1_A,s2_A):\chi_B &\mapsto \chi_A
   \end{aligned}\end{equation}.

   We can compose the two maps in one that takes from the set of strategies of one player into itself:
   \begin{equation}
    \tilde{C}^2(s1_A,s2_A,s1_B,s2_B)=\tilde{C}_B(s1_B,s2_B)\tilde{C}_A(s1_A,s2_A):\chi_A \mapsto \chi_A.
   \end{equation}

   The Nash Equilibrium strategies will be fixed points of this map.  The map can also tell when a Nash equilibrium is stable or unstable.  Some tools to do it are given in this work, but the stability analysis falls beyond the scope of this study.

  \subsection{THE CRITICAL RESPONSE MAPS IN SO(3)}
   We have seen in (THE EISERT SCHEME) that any $\chi$ matrix can be expressed:
     \begin{equation}\label{ChiDecomposed}
      \chi_A = \sum_k \lambda_k \begin{pmatrix}((a_k)_0)^* \\ ((a_k)_z)^* \\ ((a_k)_y)^*  \\ ((a_k)_x)^*\end{pmatrix}
           \begin{pmatrix}(a_k)_0 & (a_k)_z & (a_k)_y  & (a_k)_x\end{pmatrix}
     \end{equation}
    where the summation is over some set $\{a_k\}$ constituting a complete basis for the $\mathbb{C}^4$ space (the space of complex vectors with dimension 4)

    Let us decompose the $\chi$ matrix for player A, and compute the payoff expression:  if A plays a unitary strategy, the payoff expression (\ref{ChiPayoff}) is:
 \begin{equation}\label{ChiPayoff}
  \langle \$ \rangle = \sum_{k,l}(\chi_B)_{k,l}\sum_n \left((\lambda_A)_n\sum_{i,j}((a_n)_i)^*(a_n)_j  P^{i,j,k,l}\right).
 \end{equation}
   where $a_k$ are unitary complex vectors, and $(\lambda_A)_n$ are real numbers between 0 and 1 that sum up to 1.   This is simply a convex combination of quadratic forms.

   Let us fix an arbitrary strategy $\chi_B$, and find the critical vectors $a_n$ without varying the $(\lambda_A)_n$.  In these conditions the critical responses of A are given by some $a_i$ that are \textbf{eigenvectors} of the matrix:
     \begin{equation}
       (\lambda_A)_n\sum_{k,l} (\chi_B)_{k,l} P^{i,j,k,l}
     \end{equation}
     Once we found some critical vectors $\tilde{a}_i$, the expression for the payoff is:
     \begin{equation}
       \sum_n(\lambda_A)_n\sum_{k,l} (\chi_B)_{k,l} P^{i,j,k,l}(\tilde{a}_i)^* \tilde{a}_j.
     \end{equation}

     The fact that all the critical vectors are identical allows us to write the expression in the following way:
     \begin{equation}
       \left(\sum_{k,l} (\chi_B)_{k,l} P^{i,j,k,l}(\tilde{a}_i)^* \tilde{a}_j\right)\sum_n(\lambda_A)_n.
     \end{equation}

   The parameters $\lambda_n$ turn out to be irrelevant, for the  completeness condition forces $\sum_n(\lambda_A)_n$ to be 1.  Then the critical responses are always \textbf{``one-term''} quantum operations, defined only by a complex 4-dimensional unitary vector.

   A unitary operation, on the other hand, is defined by a 4-dimensional unitary vector $a_i$ where $a_0$ is real and $a_z$,$a_y$ and $a_x$ are imaginary.

  \subsection{ONE-TERM QUANTUM OPERATIONS}
   It is very important at this point to check what kind of quantum operations can be described by only one complex 4-dimensional vector, that is, by a $\chi$ matrix corresponding to a projector in the $\mathbb{C}^4$ space.
   \begin{equation}
    \chi_{one-term} = \begin{pmatrix}(a_0)^*\\(a_z)^*\\(a_y)^*\\(a_x)^*\end{pmatrix} \centerdot \begin{pmatrix}a_0&a_z&a_y&a_x\end{pmatrix}.
   \end{equation}

   The action of this operation on a density matrix $\rho$ is given by
   \begin{equation}
    \hat{\chi} \rho = \left((a_0)^*\mathbb{I} + (a_z)^*\sigma_z + (a_y)^*\sigma_y + (a_x)^*\sigma_x)\right)\rho\left(a_0\mathbb{I} + a_z\sigma_z + a_y\sigma_y + a_x\sigma_x)\right).
   \end{equation}

   This operation must be \textbf{trace-preserving} to be physically meaningful.  This means that the coefficient of the identity must remain as $\frac{1}{2}$. This implies:
   \begin{equation}\begin{aligned}
    \text{for }&\rho = \frac{1}{2}(\mathbb{I}+\vec{r}\centerdot\vec{\sigma})\\
    &|a_0|^2 +||\vec{a}||^2 - 2i(Im\left[(\vec{a}^{\ *}\times \vec{r})\centerdot \vec{a}\right]) + 2Re\left[(a_0)^*(\vec{r}\centerdot \vec{a})\right] = 1\\
    \text{By unitarity: }& |a_0|^2 +||\vec{a}||^2 = 1\\
    \text{therefore }&i(Im\left[(\vec{a}^{\ *}\times \vec{r})\centerdot \vec{a}\right]) = Re\left[(a_0)^*(\vec{r}\centerdot \vec{a})\right]\\
     \text{Which can be written: }& i(Im\left[\vec{a}\right]\times \vec{r})\centerdot Re\left[\vec{a}\right] = Re\left[(a_0)^*\vec{a}\right]\centerdot\vec{r}
   \end{aligned}\end{equation}
   where $\vec{r}=(r_z,r_y,r_x)\ \epsilon \  \mathbb{R}^3$, $\vec{\sigma}=(\sigma_z,\sigma_y,\sigma_x)$ and $\vec{a}=(a_z,a_y,a_x)\ \epsilon \  \mathbb{C}^3$\vspace{6pt}

   For unitary transformations $Im\left[a_0\right]=0$ and $Re\left[\vec{a}\right]=0$, and the condition is automatically fulfilled regardless of the vector $\vec{r}$.

   There is a vector relation \cite{Identity} that states:
   \begin{equation}
    (\vec{u}\times\vec{v})\centerdot\vec{w}=\vec{u}\centerdot(\vec{v}\times{w})
   \end{equation}
   We can use it to rewrite the trace-preservation condition:
   \begin{equation}\begin{aligned}
    i(Re\left[\vec{a}\right]\times Im\left[\vec{a}\right])\centerdot \vec{r} = (Im\left[a_0\right]Im\left[\vec{a}\right]-Re\left[a_0\right]Re\left[\vec{a}\right])\centerdot\vec{r}\\
    (iRe\left[\vec{a}\right]\times Im\left[\vec{a}\right]- Im\left[a_0\right]Im\left[\vec{a}\right]+Re\left[a_0\right]Re\left[\vec{a}\right])\centerdot\vec{r}=0.
   \end{aligned}\end{equation}

   The first term within the parentheses is necessarily perpendicular to both of the other two terms, therefore the only possibility to fulfill the condition regardless of the vector $\vec{r}$ is that one of the parts (real or imaginary) of the vector $\vec{a}$ is null, as well as the complementary part of $a_0$.

   We can conclude that there are only two kinds of one-term operations that preserve trace:
   \begin{enumerate}
    \item Those where $Re\left[\vec{a}\right]=0$ and $Im\left[a_0\right]=0$ (unitary operations)
    \item Those where $Im\left[\vec{a}\right]=0$ and $Re\left[a_0\right]=0$ (antiunitary operations)
   \end{enumerate}
  If the only physically meaningful one-term quantum operations are unitary or antiunitary, we can conclude:
  \begin{center}
   {\large The critical-response strategies are always unitary}
  \end{center}

  We have to remark, however, that when there is some degeneracy in the matrix $\sum_{k,l}(\chi_B)_{k,l}P^{i,j,k,l}$ there are mixed strategies that can be considered critical-response strategies.

  \subsection{REVISITING GEOMETRIC REPRESENTATION}
   In chapter 6 we studied unitary strategies, and found some useful expressions for critical responses (called also extremal strategies).  These expressions were found using a geometric representation of the operators acting on the 2-dimensional Hilbert space: the so called ``Bloch vector representation'' \cite{BlochVector}. In this representation a density operator is represented by a 3-dimensional vector with length from 0 to 1, and unitary operations belonging to SU(2) correspond to SO(3) rotations that act on that vector.

   Given a certain unitary strategy chosen by A ($\mathbb{R}_A$) and  another chosen by B ($\mathbb{R}_B$), the payoff can be computed with the formula:
    \begin{multline}\label{ganancia3d}
\langle G \rangle = Tr(\bigl(\mathbb{R}_{A}\mathbb{R}_{1}-(1-E)\frac{G_{B}}{G_{AB}}  \vert z \rangle\langle z \vert \bigr)\mathbb{T}\bigl(\mathbb{R}_{2}(\mathbb{R}_{B})^{T}+(1-E)\frac{G_{A}}{G_{AB}} \vert z \rangle\langle z \vert \bigr) \mathbb{G})\\ + (1-E)^{2}\frac{G_{A}G_{B}}{G_{AB}}
   \end{multline}
   Where:
   \begin{equation}
    \mathbb{T}= \begin{pmatrix}
       1 & 0 & 0 \\ 0 & \sqrt{E} & 0\\ 0 & 0 & \sqrt{E}
      \end{pmatrix}
    \text{ and }
     \mathbb{G}=\begin{pmatrix}
       G_{AB} & 0 & 0 \\ 0 & \sqrt{E}G_B  & 0\\ 0 & 0 & \sqrt{E}G_A
      \end{pmatrix}
   \end{equation}
   The rotations $\mathbb{R}_1=${\tiny$\begin{pmatrix}-1&0&0\\0&-1&0\\0&0&1\end{pmatrix}$} and $\mathbb{R}_2=${\tiny$\begin{pmatrix}1&0&0\\0&0&1\\0&1&0\end{pmatrix}$} are put there to cast $\mathbb{T}$ and $\mathbb{G}$ in positive diagonal form.

   The payoff parameters, according to a payoff matrix {\tiny $\begin{pmatrix}a&b\\c&d\end{pmatrix}$} are:
   \begin{equation}
    G_A = a + b - c - d \hspace{1cm} G_B = a - b + c - d \hspace{1cm} G_{AB} = a - b - c + d.
   \end{equation}

   To examine the unilateral changes that player A can perform, it is useful to define a matrix $A$:
   \begin{equation}
    \mathbb{A}=\mathbb{T}\bigl(\mathbb{R}_{2}(\mathbb{R}_{B})^{T}-(1-E)\frac{G_{A}}{G_{AB}} \vert z \rangle\langle z \vert \bigr) \mathbb{G}
   \end{equation}

   With this matrix we can compute critical strategies and their correspondent Hessian matrix. Here we reproduce equation \ref{FormaHessiana}, where the expression for the Hessian matrix is given:
    \begin{equation*}
	   H_{ij} = \bigl(\mathbb{R}_{A}\mathbb{A}-\mathbb{I}Tr(\mathbb{R}_{A}\mathbb{A})\bigr)_{ij}
    \end{equation*}

  \subsection{THE CRITICAL RESPONSE MAP IN SO(3)}
   There are four possible critical responses (a minimum, two saddle points and a maximum, according to the number of negative eigenvalues of the Hessian), and each corresponds to a certain assignations for the signs of the eigenvectors of a square root 3x3 matrix.   This can be viewed as a map which takes from the space of unitary strategies of one player to the space of unitary strategies of the other.
   \begin{equation}
    \mathbb{R}_{A}^{critical} =\left[(\mathbb{A}^{T}\mathbb{A})^{1/2}\right]_{s1,s2}\mathbb{A}^{-1}(\mathbb{R}_1)^{t}
   \end{equation}
   where:
   \begin{equation}
    \left[(\mathbb{A}^{T}\mathbb{A})^{1/2}\right]_{s1,s2} = \mathbb{O}_1
    \begin{pmatrix}
     (s1\ s2)\lambda_{12} & 0 & 0 \\ 0 & s2\  \lambda_2 & 0 \\ 0 & 0 & s1\  \lambda_1
    \end{pmatrix}
    \mathbb{O}_2
   \end{equation}
   For $\lambda_{12},\lambda_2,\lambda_1$ positive, and $\mathbb{O}_1,\mathbb{O}_2 \ \epsilon \  SO(3)$ \\
   If we can diagonalize the $\mathbb{A}$ matrix it is easy to compute the critical response:
   \begin{multline}\label{CriticalComputed}
    \mathbb{A} = (\mathbb{O}_2)^t(\mathbb{O}_1)\mathbb{D}_A(\mathbb{O}_2)\\
    \mathbb{R}_{A}^{critical} = (\mathbb{O}_2)^t\begin{pmatrix}
      s1\ s2 & 0 & 0 \\ 0 & s2 & 0 \\ 0 & 0 & s1
     \end{pmatrix}(\mathbb{O}_1)^t(\mathbb{O}_2)(\mathbb{R}_1)^t
   \end{multline}
   The Hessian matrix for a certain response of A can be computed according to:
   \begin{equation}
     \mathbb{H}_A = \bigl(\mathbb{R}_{A}\mathbb{R}_1\mathbb{A}-\mathbb{I}Tr(\mathbb{R}_{A}\mathbb{R}_1\mathbb{A})\bigr)
   \end{equation}
   \subsubsection{A ONE-PARAMETER SUBSET}
   As an example, we can parametrize two parameter rotations for which it is possible to find an explicit expression for the critical response map.
   To illustrate how this can be compute, let us begin with rotations along the Z axis:
   \begin{equation}\begin{aligned}
    \mathbb{R}_2(\mathbb{R}_B)^t = \begin{pmatrix}
      1 & 0 & 0 \\ 0 & cos(\theta) & sin(\theta) \\ 0 & -sin(\theta) & cos(\theta)
    \end{pmatrix}\\
    \mathbb{A}= \mathbb{R}_2(\mathbb{R}_B)^{t}\begin{pmatrix}
      (1-E)G_A & 0 & 0 \\ 0 & E\ G_B & 0 \\ 0 & 0 & E\ G_A
    \end{pmatrix}
   \end{aligned}\end{equation}
   Because $\mathbb{R}_2(\mathbb{R}_B)^t$ commutes with $\mathbb{T}$.\\
   The expression of the critical response is then rather simple:
   \begin{equation}
    \mathbb{R}_A = \begin{pmatrix}
      s1\ s2 & 0 & 0 \\ 0 & s2 & 0 \\ 0 & 0 & s1
    \end{pmatrix}\mathbb{R}_B(\mathbb{R}_2)^t(\mathbb{R}_1)^t
   \end{equation}
   \subsubsection{THE OTHER ONE-PARAMETER SUBSETS}
    It is also possible to get a explicit formula for the critical response to a rotation around the X axis:
   \begin{equation}\begin{aligned}
    \mathbb{R}_2(\mathbb{R}_B)^t = \begin{pmatrix}
      cos(\theta) & sin(\theta) & 0 \\ -sin(\theta) & cos(\theta) & 0 \\ 0 & 0 & 1
    \end{pmatrix}\\
    \mathbb{A}= \begin{pmatrix}
      cos(\theta)G_{AB}+(1-E)G_A & sin(\theta)\sqrt{E}G_B & 0 \\ -sin(\theta)\sqrt{E}G_{AB} & cos(\theta)E\ G_B & 0 \\ 0 & 0 & E\ G_A
    \end{pmatrix}
   \end{aligned}\end{equation}
    The problem of finding critical responses implies here to diagonalize a 2x2 block with two SO(2) transformations: one, applied by left, to make the matrix symmetric, and then another one to make it diagonal.

    A general matrix can be made symmetric multiplying by a certain orthogonal matrix by left:
    \begin{multline}
      \left[ \frac{1}{\sqrt{(t+w)^2+(v-u)^2}}
     \begin{pmatrix}
      t+w & v-u \\ -v+u & t+w
     \end{pmatrix}\right] \begin{pmatrix} t & u \\ v & w \end{pmatrix}\\
     = \frac{1}{t+w} \begin{pmatrix}
      t^2 + v^2 + tw - uv & tu + vw \\ tu + vw &  u^2 + w^2 + tw - uv
     \end{pmatrix}
    \end{multline}
   The symmetrical matrix can be diagonalized by another SO(2) transformation:
   \begin{multline}
    \frac{1}{N^2}\begin{pmatrix} v' & -(w' + \sqrt{(v')^2+(w')^2}) \\ w' + \sqrt{(v')^2+(w')^2} & v' \end{pmatrix}
     \begin{pmatrix}
      u' + v'  & w' \\ w' &  u'- v'
     \end{pmatrix}\\\begin{pmatrix} v' & w' + \sqrt{(v')^2+(w')^2} \\ -(w' + \sqrt{(v')^2+(w'^2)^2}) & v' \end{pmatrix}\\
    = \begin{pmatrix} u' + \sqrt{(v')^2+(w')^2} & 0 \\ 0 & u' - \sqrt{(v')^2+(w')^2} \end{pmatrix}
   \end{multline}
    where N is a normalization factor for the unitary:
   \begin{equation}
    N^2 = 2\left((v')^2 + (w')^2 + w'\sqrt{(v')^2+(w')^2}\right)
   \end{equation}
   What we need for the computation of the critical response are only the unitaries used.  With the expression for $\mathbb{A}$ in terms of a diagonal matrix and two orthogonal transformations we compute easily according to \ref{CriticalComputed}.

   Applying the unitaries described above, we get an explicit (however a little complex) expression for the critical responses.  For rotations around X:
   \begin{equation}
    \mathbb{R}_{A}^{critical}=\frac{1}{N_1(N_2)^2}\begin{pmatrix}
     -s1\ s2 \alpha\  \gamma + s2\ \beta \ \delta  & s1\ s2\ \alpha \ \delta - s1\ \beta \ \gamma & 0\\
     s1\ s2 \ \beta \gamma - s2\ \alpha \ \delta & s1\ s2\ \beta \delta + s2\ \alpha \ \gamma & 0\\
     0 & 0 & s1
    \end{pmatrix}
   \end{equation}
   Where:
   \begin{equation}\begin{aligned}
    \gamma &= cos(\theta)(G_{AB}+E\ G_B)+(1-E)\ G_A\\
    \delta &= -sin(\theta)\sqrt{E}\ (G_{AB}+ G_B)\\
    \alpha &= \gamma\ (\gamma - 2cos(\theta)E\ G_b) + \delta\ (\delta + 2sin(\theta)\sqrt{E}\ G_B)\\
    \beta &= sen(\theta)\sqrt{E}(1-E)G_B(cos(\theta)G_{AB}+G_A)\\
    N_1 &= \sqrt{\gamma^2+\delta^2}\\
    (N_2)^2 &= \sqrt{\alpha^2+\beta^2}.
   \end{aligned}\end{equation}

   For rotations around axis Y the expression is similar:
   \begin{equation}
    \mathbb{R}_{A}^{critical}=\frac{1}{N_1(N_2)^2}\begin{pmatrix}
     -s1\ s2 \alpha\  \gamma + s1\ \beta \ \delta  & 0 & s1\ s2\ \alpha \ \delta - s1\ \beta \ \gamma\\
     0 & s2 & 0 \\
     s1\ s2 \ \beta \gamma - s1\ \alpha \ \delta & 0 & s1\ s2\ \beta \delta + s1\ \alpha \ \gamma
    \end{pmatrix}
   \end{equation}
   2here:
   \begin{equation}\begin{aligned}
    \gamma &= cos(\theta)(G_{AB}+E\ G_A)+(1-E)\ G_A\\
    \delta &= -sin(\theta)\sqrt{E}\ (G_{AB}+ G_A)\\
    \alpha &= \gamma\ (\gamma - 2cos(\theta)E\ G_b) + \delta\ (\delta + 2sin(\theta)\sqrt{E}\ G_A)\\
    \beta &= sen(\theta)\sqrt{E}(1-E)G_A(cos(\theta)G_{AB}+G_A)\\
    N_1 &= \sqrt{\gamma^2+\delta^2}\\
    (N_2)^2 &= \alpha^2+\beta^2.
   \end{aligned}\end{equation}

   \subsection{SOME TWO-PARAMETER SUBSPACES}
   It is possible combine this last two expressions with that for rotations around axis Z, and compute also critical responses to two-parameter strategies that can be written in the following way:
   \begin{equation}
    \mathbb{R}\mathbb{R}_1 = \mathbb{R}_z\mathbb{R}_x \hspace{24pt}\text{or}\hspace{24pt}\mathbb{R}\mathbb{R}_1 = \mathbb{R}_z\mathbb{R}_y.
   \end{equation}

   \subsubsection{THE COMPLETE GROUP}
   To compute the critical response to any SO(3) rotation means diagonalizing 3x3 nonsymmetric matrices.  This is, of course, possible, but the resulting expressions are too complicated, and are not therefore shown here.

 \chapter{EXPLORATION OF THE STRATEGY SPACE}
  
\section{ACCESSIBLE AREAS ON THE PAYOFF SPACE}
 Even when the spaces of strategies in Quantum 2x2 games are relatively big (3 parameters per player) the space of player's payoff is part of the definition of the game, and remains bidimensional.  Thus it is very useful to track the results of the quantum strategies on that arena, as it was done in the case of semideterministic strategies in chapter \ref{Positions}.
 \subsection{THE INITIAL CLASSICAL GAME}
  In the initial Classical game, we get the Robinson Graphs mentioned in chapter 4, (like figure \ref{ExampleRobinson}).  There is some delimited area of the payoffs space that is suitable to represent payoffs for both players.
       \begin{center}\begin{figure}[htb]
        \center 
        \includegraphics{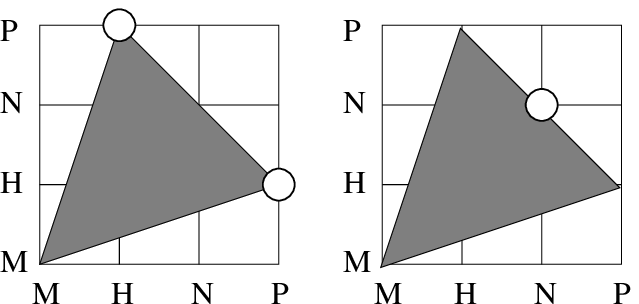}
        \caption{Accessible Payoff Area (the circle marks the Nash Equilibrium)}
       \end{figure}\end{center}
  The critical responses (there are only two of them) take to the upper and lower borders of this area for one player, and to the left and right border for the other player.
 \subsection{THE EXTENDED (SEMIDETERMINISTIC) GAME}
  In the extended (semideterministic) game, four extra points appear in the Robinson Graph; we can call them \textit{quantum points}.  In this case, the lines that in the classical game demarcated the border of the accessible area have turn into lattices, and critical responses to a mixed strategy of the other player take each player to points on the lines drawing the lattice.  If we draw a point for each critical response to every possible mixed semideterministic strategy, we draw the lines in the Robinson Graph.
       \begin{center}\begin{figure}[htb]
        \center 
        \includegraphics{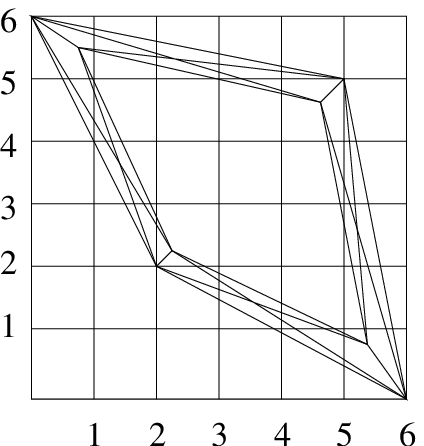}
        \caption{Lines of Critical Responses to Mixed Semideterministic Strategies}
       \end{figure}\end{center}

 \subsection{THE GAME WITH UNITARY STRATEGIES}
  The space of unitary strategies has the same dimension than the space of mixed semideterministic strategies. But the similarity between them goes no further.  When departing from a semideterministic strategy in the unitary space, the trajectory in the payoff space is far more interesting than that generated with mixed strategies.\vspace{6pt}

  The quadratic nature of the payoff function is reflected in concavity and convexity relations, and in curve shapes in the payoff space.  When one player wanders in her unitary strategy space and the other answers with a critical response, a trajectory in the payoffs space is followed.  Next we are going to see how curved can this trajectories be.\vspace{6pt}

  The famous game of the \textbf{Prisoner's Dilemma} is a good example to show this, because it has an entanglement regime transition.   The payoff matrix found in literature for this game is {\tiny$\begin{pmatrix}3&0\\5&1\end{pmatrix}$} \cite{Prisoner}.

  Let us check first the figure generated by points in the payoffs space corresponding to one player choosing random unitary strategies, and the other using critical response to the former:
       \begin{center}\begin{figure}[htb]
        \center
        \includegraphics{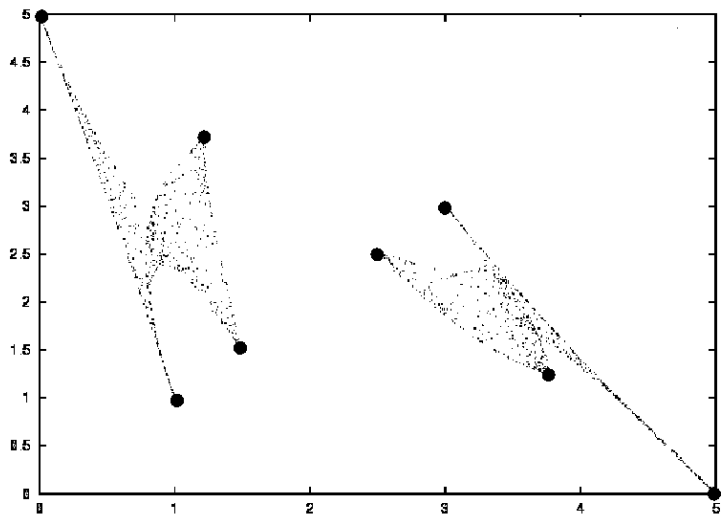}
        \caption{Points of Critical Responses to Random Unitary Strategies (low entanglement)}
       \end{figure}\end{center}
  To include mixed unitary strategies will cause the appearance of new points \textbf{between the existing ones} in the figure, thus filling two convex areas. (left and right in the example).\vspace{6pt}

  The Entanglement transitions are determined by the positions in the payoff space of the extremal points of the lattices (in the figure, the black circles), and are not thus affected by the introduction of general unitary strategies.\vspace{6pt}

  It is important to mention that the convex areas from these critical responses to unitary strategies are only slightly different to those occupied by the semideterministic lattices.\vspace{6pt}

\begin{figure}[hbt]\label{DPHighEnt}
        \center
        \includegraphics{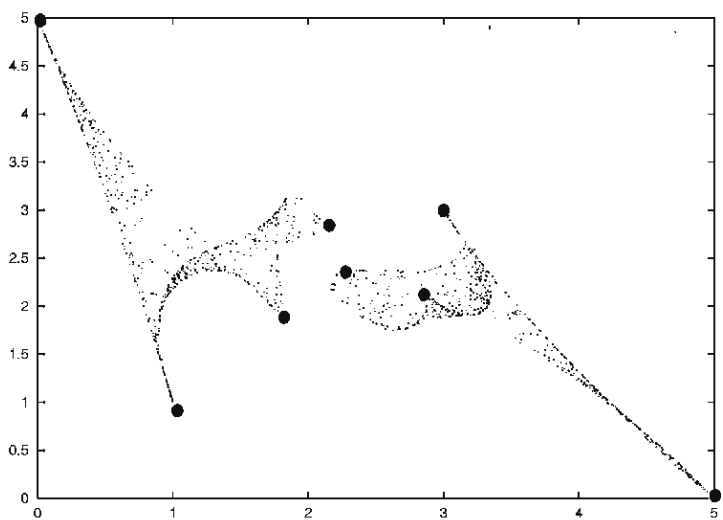}
        \caption{Points of Critical Responses to Random Unitary Strategies (high entanglement)}
       \end{figure}

   We have chosen the Prisoner's Dilemma because it has an entanglement regime transition.  What is the consequence of this in the payoff graph? The curves drawn by critical responses to unitary strategies can be seen in figure \ref{DPHighEnt}\vspace{6pt}

   For high entanglement the curvatures are even more pronounced, but that the convex area is still almost the same as that delimited by the semideterministic lattice.\vspace{6pt}
  Here it is also valid that the unitary strategies do not introduce new elements beyond those mentioned in chapter \ref{DetermClass}, just as in the low entanglement regime.

 \section{SYMMETRY}
  In section \ref{SymmetryQG} a symmetry of the quantum game was found for the unitary strategy group, and therefore for unital strategy set\footnote{A quantum operation is unital when it maps identity to identity. A strategy is unital when it corresponds to a unital quantum operation}.   But what about the complete strategy set?

  According to formula \ref{ChiPayoff} every term in the expected payoff depends on the elements of the basis in the following way:
    \begin{equation}\label{ChiPayoff}
     Tr \left(J\left((F_i\otimes F_k)^\dagger J^\dagger (\vert 00 \rangle \langle 00 \vert)J(F_j\otimes F_l)\right)J^\dagger \ \hat{\$} \right).
    \end{equation}
  As we found in section \ref{DeterministicSym}, when we examined expected payoffs consisting only of a few of these terms, there are some changes we can make on the terms without changing the expected payoff.  There is one of these equivalences for each repeated entry on the 4x4 payoff matrix in equation \ref{ExtendedPayoff}

  \parbox{6cm}{\begin{itemize}
   \item 00 $\leftrightarrow$ ZZ
   \item Z0 $\leftrightarrow$ 0Z
   \item Y0 $\leftrightarrow$ XZ
   \item X0 $\leftrightarrow$ YZ
  \end{itemize}}\hspace{1cm}
  \parbox{6cm}{\begin{itemize}
   \item 0X $\leftrightarrow$ ZY
   \item ZX $\leftrightarrow$ 0Y
   \item YX $\leftrightarrow$ XY
   \item XX $\leftrightarrow$ YY
  \end{itemize}}

  There is a lot of continuous transformations on the individual $\chi$ matrices that act in that way on the deterministic strategies.  The symmetry of the unitary game, on the other hand, suggests us to choose the following:
  {\footnotesize \begin{equation}
   \chi_A \to 
   \begin{pmatrix}
    cos(\theta) &  -isin(\theta) & 0 & 0\\
    -isin(\theta) & cos(\theta) & 0 & 0\\
     0 & 0 & cos(\theta) &  -sin(\theta)\\
     0  & 0 & sin(\theta) & cos(\theta)
   \end{pmatrix}\chi_A
   \begin{pmatrix}
    cos(\theta) & isin(\theta) & 0 & 0\\
    isin(\theta) & cos(\theta) & 0 & 0\\
     0 & 0 & cos(\theta) &  sin(\theta)\\
     0  & 0 & -sin(\theta) & cos(\theta)
   \end{pmatrix}
  \end{equation}}
 for A and
  {\footnotesize \begin{equation}
   \chi_B \to 
   \begin{pmatrix}
    cos(\phi) &  i sin(\phi) & 0 & 0\\
    i sin(\phi) & cos(\phi) & 0 & 0\\
     0 & 0 & -sin(\theta) &  cos(\theta)\\
     0  & 0 & -cos(\theta) & -sin(\theta)
   \end{pmatrix}\chi_B
   \begin{pmatrix}
    cos(\phi) & -isin(\phi) & 0 & 0\\
    -isin(\phi) & cos(\phi) & 0 & 0\\
     0 & 0 & -sin(\theta) & -cos(\theta)\\
     0  & 0 & cos(\theta) & -sin(\theta)
   \end{pmatrix}
  \end{equation}}
  for B, where $\phi=\theta+\pi/4$

 \chapter{CONCLUSIONS AND PERSPECTIVES}
   \section{CONCLUSIONS}
  The main results of this work are those obtained in chapter \ref{ConcDeterm}.  Here a summary of them is presented:

  \subsection{EQUIVALENT CLASSICAL GAMES}
   As it was stated by van Enk in \cite{ClassicalRules}, there is an extended classical game that has all the Nash Equilibria and Pareto Optima present in the quantum game.   However, there is one reason not to consider this extended game as \textit{equivalent} to the quantum game:
  \begin{center}
   {\large The unitary strategies are not equivalent to the mixed semideterministic ones.  The displacements in the payoffs space generated by changes in the unitary strategy parameters are curve and qualitatively different to those generated by changes in the probabilities that define mixed strategies.}
  \end{center}

  Even when the structure of Nash equilibria can be the same in the extended classical game\footnote{We call extended classical game to a classical game with a referee that introduces entanglement to simulate a semideterministic quantum game} the dynamics of a player moving in the complete strategy space can be very different to the dynamics of a player in the extended game.

  \subsection{SYMMETRY OF THE QUANTUM 2X2 SYMMETRIC GAMES}\label{ConcluSymm}
   The quantum game with semideterministic strategies is invariant under interchange of strategies 0 and Z or interchange of strategies X and Y  made for both player, and this was found to be a consequence of a symmetry of all possible strategy with respect to some \textbf{coordinate} operations generated by $\sigma_z\otimes\mathbb{I}$ and $\mathbb{I}\otimes\sigma_z$(chapter \ref{symmetryQG}).

   This symmetry causes the following features in Quantum Games:
   \begin{enumerate}
    \item A quantum game has always an even number of Nash Equilibria.
    \item Several equilibria will be equivalent in terms of payoff\\
     This will cause them to be less stable, because of the dilemma of choosing equilibrium points \cite{Commitments}
   \end{enumerate}

  \subsection{CRITICAL RESPONSES}
   From computations with general quantum strategies it was found that critical responses must be unitary strategies, at least for the 2x2 case.  It must be checked as a different case the two-player game with more strategies (for example (2x3, 3x3, etc).

   A geometric methodology were developed to compute the critical response map for unitary strategies,  and explicit expressions were computed for  one and two-parameter spaces.   This expressions become very simple for semi-deterministic strategies, and it is easy to show that there will allways be at least two Nash Equilibria within this set of strategies.

   The existence of at least two semi-deterministic Nash Equilibria can be proved thinking of the semi-deterministic quantum game as a 4x4 classical game, with the symmetry mentioned in \ref{ConcluSymm}

  \subsection{POSSIBILITY OF ENTANGLEMENT REGIMES}
   Some quantum games (but not all of them) show several different entanglement regimes, according to the parameter of entanglement (controlled by the referee in the Eisert scheme).

   In section \ref{DetermClass} the entanglement regime transitions are enumerated.  The main result was that there are comparatively few games showing interesting entanglement regime transitions ($\frac{1}{4}$ according to figure \ref{MapTrans}).  The games with a high value of the entanglement parameter usually show problematic features, like multiple maxima with similar characteristics.

   It was found that the analysis of the semi-deterministic strategies is enough to check the existence of entanglement regimes, and that the inclussion of unitary or other strategies does not affect this feature of the game.

 \section{PERSPECTIVES}
   A number of suggestions for future works arises, that can be classified as follows:
  \begin{itemize}
   \item On 2x2 symmetric games:
   \begin{itemize}
    \item To study the stability properties of Nash Equilibria (both intrinsic  and entanglement-generated) by means of the product of critical response maps.
    \item To check situations of asymmetric information, commitments and other refinements in quantum games
    \item To devise ways of using unitary strategies to mimic complex behaviours of interacting agents in a 2x2 game
    \item Introduce effects of decoherence and classical loss of information in this kind of games
   \end{itemize}
   \item Extensions\\
    There are some natural suggestions, that are already being followed by some workers on quantum games
   \begin{itemize}
    \item Study games with larger sets of strategies (2x3, 3x3, etc.)
    \item Study different kinds of entanglement in games with more players (2x2x2, 2x2x2x2, 3x3x3, etc)
    \item Study quantum games on networks
    \item Study quantum games with continuum of players
   \end{itemize}
  \end{itemize}

 \bibliography{JuegosCuanticos}
\end{document}